\let\accentvec\vec
\documentclass[usenatbib]{mn2e} 

\let\vec\accentvec
\let\epsilon\varepsilon
\let\theta\vartheta
\usepackage{graphicx}
\usepackage{enumerate}
\usepackage[english]{babel}
\usepackage[T1]{fontenc}
\usepackage{hyperref}
\usepackage{amsmath}
\usepackage{txfonts}
\newcommand{\sophia}{\textit{SOPHIA}}

\newcommand{\mrk}{Mrk~421}

\newcommand{\veritas}{\textit{VERITAS}}
\newcommand{\fermi}{\textit{Fermi}}
\newcommand{\fermilat}{\textit{Fermi}-LAT}
\newcommand{\lat}{LAT}

\usepackage{lineno}

\newcommand*\patchAmsMathEnvironmentForLineno[1]{%
  \expandafter\let\csname old#1\expandafter\endcsname\csname #1\endcsname
  \expandafter\let\csname oldend#1\expandafter\endcsname\csname end#1\endcsname
  \renewenvironment{#1}%
     {\linenomath\csname old#1\endcsname}%
     {\csname oldend#1\endcsname\endlinenomath}}% 
\newcommand*\patchBothAmsMathEnvironmentsForLineno[1]{%
\patchAmsMathEnvironmentForLineno{#1}%
\patchAmsMathEnvironmentForLineno{#1*}}%
\AtBeginDocument{%
\patchBothAmsMathEnvironmentsForLineno{equation}%
\patchBothAmsMathEnvironmentsForLineno{align}%
\patchBothAmsMathEnvironmentsForLineno{flalign}%
\patchBothAmsMathEnvironmentsForLineno{alignat}%
\patchBothAmsMathEnvironmentsForLineno{gather}%
\patchBothAmsMathEnvironmentsForLineno{multline}%
}
\begin{document}
%
%\linenumbers

   \title[A hadronic origin for UHBLs]{A hadronic origin for ultra-high-frequency-peaked\\ BL Lac objects}

   \author[M. Cerruti et al.]{M. Cerruti,$^1$
                    A. Zech,$^2$  C. Boisson,$^2$
          S. Inoue$^{3,4}$
          \\
$^1$ Harvard-Smithsonian Center for Astrophysics, 60 Garden Street, Cambridge MA, 02138, USA\\
                matteo.cerruti@cfa.harvard.edu\\
   $^2$ LUTH, Observatoire de Paris, CNRS, Universit\'{e} Paris Diderot, PSL Research University; 5 Place
Jules Janssen, 92190 Meudon, France\\
             andreas.zech@obspm.fr\\
$^3$   Institute for Cosmic Ray Research, 5-1-5 Kashiwanoha, Kashiwa, Chiba 277-8582, Japan  \\
$^4$  Max-Planck-Institut f\"{u}r Physik, F\"{o}hringer Ring 6, 80805 M\"{u}nchen, Germany 
             }   

% \date{}
 
 \maketitle
 
    \begin{abstract}
  %context
   {Current Cherenkov telescopes have identified a population of ultra-high-frequency peaked BL Lac objects (UHBLs), also known as extreme blazars, that exhibit exceptionally hard TeV spectra, including 1ES\,0229+200, 1ES\,0347-121, RGB\, J0710+591, 1ES\,1101-232, and 1ES\,1218+304. Although one-zone synchrotron-self-Compton (SSC) models have been generally successful in interpreting the high-energy emission observed in other BL Lac objects, they are problematic for UHBLs, necessitating very large Doppler factors and/or extremely high minimum Lorentz factors of the emitting leptonic population. In this context, we have investigated alternative scenarios where hadronic emission processes are important, using a newly developed (lepto-)hadronic numerical code to systematically explore the physical parameters of the emission region that reproduces the observed spectra while avoiding the extreme values encountered in pure SSC models.
Assuming a fixed Doppler factor $\delta=30$, two principal parameter regimes are identified, where the high-energy emission is due to: 1) proton-synchrotron radiation, with magnetic fields $B \sim 1-100\ \textrm{G}$ and maximum proton energies $E_{p;max} \lesssim 10^{19}$\,eV; and 2) synchrotron emission from p-$\gamma$-induced cascades as well as SSC emission from primary leptons, with $B \sim 0.1-1$ G and $E_{p;max} \lesssim 10^{17}$\,eV. This can be realized with plausible, sub-Eddington values for the total (kinetic plus magnetic) power of the emitting plasma, in contrast to hadronic interpretations for other blazar classes that often warrant highly super-Eddington values.}
    \end{abstract}

   \section{Introduction}
   
   %definition of blazars and description of blazar SED
   Blazars are a class of active galactic nuclei (AGN) characterized by predominantly non-thermal spectra at all wavelengths, from radio to $\gamma$-rays, relatively weak (or absent) optical/UV emission lines, rapid variability and a high degree of polarization \citep{Stein76, Moore81}. These observational properties can be consistently interpreted in the context of the unified AGN model \citep{Urry95}, assuming that blazars are radio-loud AGN, with 
   relativistic jets pointed nearly towards the observer. Their spectral energy distribution (SED) is thus dominated by the Doppler-boosted non-thermal emission from the jet. Multi-wavelength observations have 
   shown that blazar SEDs generally exhibit two bumps, one peaking at low energies (infrared to X-rays), and one peaking in $\gamma$-rays \citep[see e.g.][]{FermiSED}. 
   
   The origin of the low-energy bump is commonly ascribed to synchrotron emission from relativistic electrons in an emitting region located inside the jet, approaching relativistically towards the observer.
   
  %description of leptonic and hadronic models
   The origin of the high-energy bump is still under debate. In leptonic models, the high-energy emission is supposed to be inverse Compton emission from relativistic electrons that up-scatter either low-energy synchrotron photons emitted by the same population of electrons \citep[synchrotron-self-Compton model, SSC, e.g.][]{Konigl81}, or photons originating from outside the jet \citep[external-inverse-Compton, EIC, e.g.][]{Sikora94} such as the thermal emission from the accretion disk, the dusty torus, the broad-line region (BLR), or from stellar clusters located near the blazar emitting region. In hadronic models the high energy bump is instead assumed to originate from protons accelerated to ultra-high energies in the jet, via either the synchrotron photons radiated by the protons, or emission from secondary particles such as electron-positron pairs or muons generated in p-$\gamma$ interactions of the protons with low-energy internal and/or external photon fields \defcitealias{Mucke01}{M01}\citep[e.g.][hereafter \citetalias{Mucke01}]{Mannheim93,Dermer01, Mucke01}. For a recent modeling attempt of \fermilat\ blazars in both leptonic and hadronic scenarios, see \citet{Markus13}.
   
In an alternative scenario, protons or neutrons escaping from the emitting region trigger intergalactic cascades with the extragalactic background light (EBL) or cosmic microwave background (CMB) \citep[e.g.][]{Essey10, Dermer12, Murase12}

Proton-proton interactions \citep[important in denser environments, such as supernovae remnants or micro-quasars, see e.g.][respectively]{Ackermann13, Romero03} are commonly neglected in blazar hadronic models, because the particle density in the emitting region is considered too low for them to be important \citep[see however][]{Reynoso11}. An exception to this are models based on the interaction of relativistic protons inside the jet with stellar envelopes or gas clouds \citep[e.g.][]{Barkov10}.

\begin{table*}
       $$
   \begin{array}{ p{75 pt} p{30 pt}  p{180 pt}  p{110 pt} p{45pt}}
            \hline
             \noalign{\smallskip}
              object & z& SSC parameters & VHE variability  & Log($M_\bullet / M_\odot$)\\
            \noalign{\smallskip}
            \hline   
            \noalign{\smallskip}
          1ES\,0229+200  &  0.140 &  $\delta=$50, B$=$0.4\,G, R$=54\times10^{15}$\,cm, $\gamma_{min}=$5$\times$10$^{5}$    \newline \citep{Tavecchio09}; \newline  $\delta=$40, B$=$0.032\,G, R$=$10$^{18}$\,cm, $\gamma_{min}=$4$\times 10^{5}$    \newline   \citep{Kaufmann11};  \newline $\delta>$53, B=0.8-3.3\, mG, R=$(5-30)\times10^{15}$\,cm, \newline $\gamma_{min}=(2.5-4.5)\times$10$^{4}$  \ \citep{0229_Veritas}; & indication for variability on time scales of years \citep{0229_Veritas} &	$9.16 \pm 0.11$\\
                      \noalign{\smallskip}
           1ES\,0347-121 & 0.188 & $\delta=$25, B$=$0.035\,G,  R$=$3.2$\times$10$^{16}$\,cm, $\gamma_{min}=$10$^3$ \newline  \citep{0347_HESS}  \newline  $\delta=$61, B$=$1.3\,mG,  R$=$1.6$\times$10$^{17}$\,cm, $\gamma_{min}=2\times$10$^4$ \newline \citep{Tanaka14}; &   no known variability  & $8.02 \pm 0.11$\\  
            \noalign{\smallskip}
                       RGB\, J0710+591  & 0.125 & $\Gamma=$30, B$=0.036$\,G, R$=2\times10^{16}$\,cm, $\gamma_{min}=6\times10^4$  \newline \citep{0710_veritas};    & no known variability & $8.25 \pm 0.22$\\
             \noalign{\smallskip} 
            1ES\,1101-232 & 0.186 & $\delta=$25, B$=$0.1\,G, R$\approx$10$^{16}$\,cm, $\gamma_{min}=$10$^3$  \newline   \citep{1101_HESS}; &  no known variability & \textit{9}\\
            \noalign{\smallskip}
           1ES\,1218+304 & 0.184 &$\delta=$80, B$=$0.04\,G,  R$=$3$\times$10$^{15}$\,cm  \newline  \citep{Rueger10}; \newline  $\delta=$44, B$=$0.12\,G,  R$=$3$\times$10$^{15}$\,cm  \newline  \citep{Weidinger10}; & variability on the scale of days\newline \citep{1218_Veritas}  & $8.04 \pm 0.24$\\
           \noalign{\smallskip}
            \hline
         \end{array} 
      $$
        \caption[]{Characterization of the five UHBLs with one-zone SSC models. The model parameters are: the Doppler factor ($\delta$) or bulk Lorentz factor ($\Gamma$); the magnetic field B; the size of the emission region R; the minimum Lorentz factor $\gamma_{min}$ of the primary electron distribution. The references for the redshift measurements are \citet{Woo05} for both 1ES\,0229+200 and 1ES\,0347-121, \citet{Falomo03} for RGB\, J0710+591,  \citet{Remillard89} for 1ES\,1101-232 , and \citet{Ahn12} for 1ES\, 1218+304. The masses of the super-massive black-holes are taken from \citet{0347_HESS} (for 1ES\,0347-121) and \citet{Meyer12} (for the remaining objects). For 1ES\,1101-232, in absence of a reliable mass estimation, we considered a standard average value of $10^9 M_\odot$. }
        \label{tab:uhbl}
  \end{table*}    

Ninety percent of the AGN detected so far with ground-based Cherenkov telescope arrays, i.e. at very high energies (VHE) above $\sim$100\,GeV, are blazars of the BL Lac type\footnote{\url{http://tevcat.uchicago.edu}}. They can be further classified according to the frequency of the first SED peak \citep{Padovani95}  into high-frequency-peaked BL Lac objects (HBLs, peaking in X-rays), or intermediate and low-frequency-peaked BL Lac objects (IBLs and LBLs, peaking in optical-infrared). Due to their hard spectra, HBLs currently account for the largest fraction of ``TeV-loud'' AGN (75$\%$). Out of the 44 HBLs detected so far in the TeV range, five
sources --- 1ES\,0229+200, 1ES\,0347-121,  RGB\,J0710+591, 1ES\,1101-232 and 1ES\,1218+304 --- are characterized by particularly hard spectra with the high-energy peak above $\sim$1\,TeV, which is consistent with their very hard and weak signal in the {\it Fermi}-LAT band. Another common characteristic is the lack of observable variability on short time-scales of the very-high-energy flux \citep[with the exception of 1ES\,1218+304, which showed VHE flaring activity on the scale of days,][]{1218_Veritas}.
Although the BL Lac object H\,1426+428 belongs to this sub-class as well, since it was detected at VHE only during flaring activity without extensive multi-wavelength coverage \citep{1426_hegra, Arache, 1426_whipple2, 1426_whipple, Leonardo09, wystan} we decided to exclude it from our sample. 

The modelling of these ultra-high-frequency peaked BL Lacs \citep[UHBLs,  also known as extreme BL Lacs, EHBLs, see e.g.][for a more recent review]{Costamante01, Gunes13} with one-zone SSC scenarios, which usually yield good results for HBLs, is proving difficult in most cases, as can be seen from Table~\ref{tab:uhbl}, where some recent attempts
at SSC modeling of the sources are summarized. 
A good representation of the complete SEDs requires either extremely high values of the minimum Lorentz factor of the electron distribution (1ES\,1101-232, 1ES\,0347-121), or an elevated Doppler factor (1ES\,1218+304), or both (1ES\,0229+200, RGB\, J0710+591). 
Very large Doppler factors imply either a very fast movement of the plasma blob and/or a very small viewing angle with respect to the jet axis and could thus be difficult to 
reconcile with radio observations of the movement of knots inside jets \citep[see e.g.][]{Lister13} or with the statistics of observed blazars and radio-galaxies \citep{Henri06}.

A very high value of $\gamma_{min}$, and thus a very narrow stationary electron energy distribution, requires specific and finely-tuned conditions to occur. It might arise if electrons can somehow be injected into the emission region with a narrow energy distribution, and their subsequent cooling is inefficient, or if the cooling of particles accelerated at a shock is compensated by stochastic turbulent re-acceleration \citep{Katarzynski06, Lefa11, Asano14}.

Another possibility to achieve very hard VHE spectra with high minimum electron energies is given by models including external Compton upscattering of ambient photon fields \citep{Lefa11}, which are however generally thought to dominate in powerful flat-spectrum-radio-quasars (FSRQs) and LBLs \citep[see e.g.][]{EyleenMeyer12}, but not in HBLs, in view of the absence of a detectable emission from the accretion disk or the BLR in these sources. The above mentioned intergalactic cascades
may also provide a hard spectral component at the highest energies to fit the spectra of UHBLs, as was shown for example for 1ES\,0229+200 by \citet{Murase12}.\\ A hadronic origin of the TeV spectrum of 1ES\,1101-232 has recently been proposed by \citet{Cao14}. They ascribe the spectrum measured with the H.E.S.S. Cherenkov telescopes from this source to secondary 
$\gamma$-ray emission from neutral pion decay following interactions of sub-PeV protons with synchrotron radiation inside the source. This interpretation, however, requires an extremely large power in protons of a few 10$^{53}$\,erg\,s$^{-1}$, around six orders of magnitudes higher than the Eddington luminosity of the super-massive black hole (SMBH) powering the AGN.    

% purpose of the paper
In this paper, we systematically investigate a global interpretation of the SEDs of the five UHBLs within a (lepto-)hadronic framework. By ascribing the low-energy and high-energy bumps in the SEDs to different particle populations, 
sharing however a common acceleration and emission region, extreme values for the Doppler factor and the minimum electron Lorentz factor can be avoided. We show that two distinct sets of solutions with acceptable 
power requirements can be found in this way. In general, hadronic scenarios have the added benefit of providing a potential link to the outstanding question of the sources of ultra-high energy cosmic rays \citep[UHECRs, see][for a recent review]{Kumiko11}, as well as possibly the sources of PeV neutrinos recently discovered \citep{IceCube}.

  %outline of the paper
In Sec.~\ref{sec2} we present a new stationary code, which computes both the leptonic and hadronic components and permits the study of a much wider parameter space than the standard SSC or hadronic codes. In particular, leptonic or hadronic scenarios can be evaluated in a consistent framework, by simply varying the physical parameters of the emitting region (i.e. its particle content and its magnetic energy density). Interesting mixed --- ``lepto-hadronic'' --- scenarios, in which the high energy bump has comparable contributions from both SSC and proton-induced cascade emission, naturally arise in this framework. In its current form, the code is focused on the interpretation of emission 
from (U)HBLs: in particular, we do not consider external photon fields which are thought to be important in flat-spectrum radio quasars (FSRQs) and probably in LBLs.\\  
 In Sec.~\ref{sec3} we present, as an example, a first application of our code to the well-studied HBL Mrk\,421 and compare our results to those of a previously published,  independent hadronic code.\\ 
 In Sec.~\ref{sec4} we focus on the systematic application of our code to the five UHBLs. By scanning the parameter space of all acceptable values of the strength of 
 the magnetic field and the size of the emission region for a given Doppler factor, we arrive at two separate regions of solutions
 with distinct ranges for these parameters. The first set of solutions is dominated by proton-synchrotron emission, while the other is dominated by synchrotron emission from pair cascades triggered by hadronic interactions plus SSC emission from the primary leptons.\\
The implication on the source physics of the solutions found with this parameter scan are discussed in Sec. \ref{sec5}. We focus specifically on the viability of the solutions with respect to the required jet power. We also study the constraints our solutions pose on the acceleration processes, the expected flux variability and a potential connection with UHECRs.
   
   \section{Description of the code}
   \label{sec2}
   
   \subsection{Leptonic processes}
   \label{lep}
   The leptonic part of the code we present here is a direct evolution of the SSC code developed by \citet{Kata01} (hereafter\defcitealias{Kata01}{K01}\citetalias{Kata01}): a spherical emitting region of radius $R$, moving with Doppler factor $\delta$ in a relativistic jet with angle $\theta$ to the line of sight, is filled with a tangled, homogeneous magnetic field with amplitude $B$ and a stationary population of primary electrons described by $N_e(\gamma_e)$.\footnote{This population can be interpreted as consisting of electrons and positrons, but we will refer to it  simply as \textit{''electrons''} in the following.} Here $\gamma_e = E_e/mc^2$ is the Lorentz factor of the electrons, $N_e(\gamma_e)\ d\gamma_e$ representing the number of electrons per unit volume with a Lorentz factor between $\gamma_e$ and $\gamma_e + d\gamma_e$.  A break in the stationary electron distribution is expected as a consequence of the cooling of the particles via synchrotron emission and inverse Compton scattering, see e.g. \citet{Susumu}. The electron energy distribution is thus described by a broken power-law function, defined by the two slopes $\alpha_{e;1}$ and $\alpha_{e;2}$, the Lorentz factors $\gamma_{e;min}$, $\gamma_{e;break}$ and $\gamma_{e;max}$, and a normalization factor $K_e$.\\
   
   The synchrotron emission is evaluated using the standard relativistic formulae for the emissivity and self-absorption \citep[see for example][]{Rybicki}; the inverse Compton emission is evaluated using the Compton kernel given by \citet{Jones68}, which correctly describes the Comptonized spectrum in both the Thomson and the Klein-Nishina regimes.\\
   
   High-energy photons are absorbed by the pair-production process ($\gamma+\gamma'\rightarrow e^+ + e^-$) after interaction with the low-energy synchrotron photons inside the emitting region, which can modify the observed $\gamma$-ray spectrum \citep[see e.g.][]{Aha08}. In general, a more significant effect is $\gamma$-$\gamma$ pair production with the infrared photons of the extragalactic background light (EBL) experienced by the high-energy photons traveling from the source towards the Earth \citep{Salamon98}. Both effects are included in the code described by \citetalias{Kata01}.\\     
       
   Several modifications and improvements have been applied to the original leptonic code:
   \begin{itemize}
   \item The stationary distribution of primary electrons has been modified, replacing the sharp cut-off at the maximum energy ($N_e(\gamma_e)=0$ if $\gamma_e > \gamma_{e;max}$)  with a more natural exponential cut-off:
   \begin{equation}
   N_e(\gamma_e) = \left\{
\begin{array}{rl}
K_e\ \gamma_e^{-\alpha_{e;1}}\ e^{-\gamma_e/\gamma_{e;max}} & \text{if } \gamma_{e;min} \leq \gamma_e < \gamma_{e;break}\\
\gamma_{e;break}^{\alpha_{e;2} - \alpha_{e;1}}\ K_{e}\ \gamma_e^{-\alpha_2}\  e^{-\gamma_e/\gamma_{e;max}} & \text{if } \gamma_{e;break} \leq \gamma_e 
\end{array} \right.
   \end{equation} 
    where $K_{e}$ is the normalization factor of the electron distribution, in cm$^{-3}$.\\   
   \item The computation of the synchrotron emission has been improved in precision by performing the complete integration over the pitch angle between the charged particle and the magnetic field, while in the original code this integral was approximated by a simple analytical function \citepalias[see][appendix A]{Kata01}, evaluated in a well defined energy range. This modification is unavoidable for a hadronic extension of the code that includes proton synchrotron emission, as will be discussed below. The synchrotron emissivity $j(\nu)$ (in erg s$^{-1}$ cm$^{-3}$ Hz$^{-1}$ sterad$^{-1}$) is now evaluated as: 
   \begin{equation} 
   \label{eq1}
   j(\nu) = \frac{1}{8\pi}\int^{\pi}_{0}{d\theta\ \sin{\theta}} \int^{\infty}_{\gamma_{e;min}}{d\gamma_e\ N_e(\gamma_e)\ P(\nu,\gamma_e,\theta)}
   \end{equation}
   where $P(\nu,\gamma_e,\theta)$ is the power emitted by each electron (in erg s$^{-1}$ Hz$^{-1}$ sterad$^{-1}$), equal to:  
   \begin{equation} 
   \label{eq2}
   P(\nu,\gamma_e,\theta) = \frac{\sqrt{3} e^3 B \sin{\theta}}{m_e c^2} \frac{\nu}{\nu_c} \int^{\infty}_{\frac{\nu}{\nu_c}}{dx\ K_{5/3}(x)} 
   \end{equation}
   where $\nu_c = (3 e B / (4\pi m_e c))\ \gamma_e^2 \sin{\theta}$ and $K_j(x)$ is the modified Bessel function of the second kind of order $j$. To reduce the computing time of the code, the integral $I(a)=\int^{\infty}_{a}{dx\ K_{5/3}(x)}$ has been tabulated, and then a linear interpolation as a function of $a$ is performed.\\
   \item The evaluation of the absorption due to internal $\gamma-\gamma$ pair production has been modified, replacing the simple cross-section $\delta$-function approximation with the formula given by \citet{Aha08}, yielding a better accuracy: 
   \begin{equation}
   \begin{split}
   \sigma_{\gamma\gamma} = & \frac{3\ \sigma_T}{2\ s^2}\left[ \left(s + \frac{1}{2}\ln{s} - \frac{1}{6} + \frac{1}{2 s} \right)\ \ln{(\sqrt{s}+\sqrt{s-1})} \right.+ \\
   & \left. - \left(s + \frac{4}{9} - \frac{1}{9 s} \right)\ \sqrt{1-\frac{1}{s}} \right]
   \end{split}
   \end{equation}
   where $s$ represents the normalized, non-dimensional product of the energies of the primary and target photons ($E$ and $E'$): $s = E E' / (m_e^2 c^4)$.\\
   \item The secondary population of leptons coming from $\gamma-\gamma$ pair production is computed using the injection function $Q_e(\gamma_e)\ d\gamma_e$ (defined as the number of injected pairs per unit volume and time, with a Lorentz factor between $\gamma_e$ and $\gamma_e + d\gamma_e$) given by \citet{Aha83}: 
   \begin{equation}
   \begin{split}
   Q_e(\gamma_e) = & \frac{3\sigma_T c}{32}\ \int_{\gamma_e}^{\infty}{d\epsilon\ \frac{n_{\epsilon}(\epsilon)}{\epsilon^3}} \int_{\frac{\epsilon}{4\gamma_e(\epsilon-\gamma_e)}}^{\infty}{d\epsilon'\ \frac{n_{\epsilon'}(\epsilon')}{\epsilon'^2}}\ \cdot \\
   & \cdot\ \left[ \frac{4\epsilon^2}{\gamma_e(\epsilon-\gamma_e)} \ln{\left( \frac{4\gamma_e \epsilon' (\epsilon-\gamma_e)}{\epsilon} \right)} -8\epsilon \epsilon' + \right. \\
   & \left. +\frac{2\epsilon^2(\epsilon\epsilon'-1)}{\gamma_e(\epsilon-\gamma_e)}-\left(1-\frac{1}{\epsilon\epsilon'}\right)\left( \frac{\epsilon^2}{\gamma_e(\epsilon-\gamma_e)} \right)^2 \right]
   \end{split}
   \end{equation}
   where  $\epsilon = E / (m_ec^2)$ and $\epsilon'= E' / (m_ec^2)$ are the normalized, non-dimensional energies of the two photons (with $\epsilon \gg \epsilon'$), and $n_{\epsilon,\epsilon'}(\epsilon,\epsilon')$ is the number density of photons of high ($\epsilon$) and low ($\epsilon'$) energy.  \\
    The injection function $Q_e(\gamma_e)$ has been used as a source term in the continuity equation in order to determine the stationary state of the population of produced pairs $N'_e(\gamma_e)$: 
    \begin{equation}
    \label{contequation}
    \frac{\partial}{\partial t}N'_e(\gamma_e) = \frac{\partial}{\partial \gamma_e}\left[\gamma_e\ \frac{N'_e(\gamma_e)}{\tau_c(\gamma_e)}\right] + Q_e(\gamma_e) - \frac{N'_e(\gamma_e)}{\tau_{ad}}
    \end{equation} 
    where we chose $\tau_{ad} = 2R/c$ as the adiabatic time scale \citepalias[the factor $0.5 c$ is typical for this type of objects, see][]{Mucke01} and 
    \begin{equation}
    \tau_c(\gamma_e) = \frac{3m_ec}{4(u_B+u_{soft})\sigma_T} \frac{1}{\gamma_e} 
    \end{equation}
     is the radiative cooling time, including both synchrotron losses ($u_B$ being the magnetic energy density) and inverse-Compton losses (in the Thomson regime; $u_{soft}$ being the synchrotron photon energy density). 
    The stationary state has been computed using the integral expression given by \citet{Susumu}:
    \begin{equation} 
    \label{intsolution}   N'_e(\gamma_e)=e^{-\gamma^{\ast}_e/\gamma_e}\frac{\gamma^{\ast}_e\tau_{ad}}{\gamma_e^2}\int_{\gamma_e}^{\infty}{d\zeta\ Q_e(\zeta)e^{+\gamma^{\ast}_e/\zeta}}
    \end{equation}
    with 
    \begin{equation}
    \label{gammacool}
    \gamma^{\ast}_e = \frac{3m_ec^2}{8(u_B+u_{soft})\sigma_TR}
    \end{equation}
     representing the Lorentz factor at which $\tau_c(\gamma_e) = \tau_{ad}$. The associated stationary synchrotron emission is then being computed.\\
 %  \item The absorption by the EBL can be evaluated using the models developed by \citet{Stecker98, Kneiske04, Franceschini, Kneiske10}. All the results presented in this paper have been computed using the model by \citet{Franceschini}, which agrees with the latest constraints obtained by very-high-energies (VHE; E>100 GeV) observations \citep{Aha06,HESSEBL}.    
  \item The absorption by the EBL is computed using the model by \citet{Franceschini}, which agrees well with the latest constraints obtained by very-high-energy (VHE; E>100 GeV) observations \citep{Aha06,HESSEBL}. 
   \end{itemize}
   
   \subsection{Hadronic processes}
   The emitting region is assumed to be filled with a stationary population of relativistic protons, in addition to the electrons. In analogy with the electron population, the proton distribution $N_p(\gamma_p = E_p/m_pc^2)$ is described by a broken power law function:
   \begin{equation}
   N_p(\gamma_p) = \left\{
\begin{array}{rl}
K_p\ \gamma_p^{-\alpha_{p;1}}\ e^{-\gamma_p/\gamma_{p;max}} & \text{if } \gamma_{p;min} \leq \gamma_p < \gamma_{p;break}\\
\gamma_{p;break}^{\alpha_{p;2} - \alpha_{p;1}}\ K_{p}\ \gamma_p^{-\alpha_2}\  e^{-\gamma_p/\gamma_{p;max}} & \text{if } \gamma_{p;break} \leq \gamma_p 
\end{array} \right.
   \end{equation}   
    The normalization of the proton spectrum is $K_p = \eta K_e$, with the $\eta$ factor representing the ratio between the number density of protons and electrons at $\gamma_p=\gamma_e=1$.\\
   
   The proton synchrotron emission is evaluated in the same way as for the electrons (cf. Equations \ref{eq1} and \ref{eq2}), by replacing the electron mass with
   the proton mass.
   % given the fact that the synchrotron emissivity is computed using the exact formula, the only difference is the use of $m_p$ instead of $m_e$, and of $N_p(\gamma_p)$ instead of $N_e(\gamma_e)$ (with the integration performed, of course, over $\gamma_p$).
   When proton synchrotron photons are assumed to be responsible for the high energy bump, they suffer $\gamma$~-~$\gamma$ absorption from both internal photons and the EBL. The $\gamma$~-~$\gamma$ absorption and the emission from the population of secondary pairs is evaluated in analogy with what is done for the inverse Compton emission for the leptonic part of the code.\\
   
   The proton population in the emitting region interacts with the low energy photons through photo-meson processes 
   \begin{equation}
   \begin{split}
   p + \gamma  \rightarrow &\ p' + n^0\pi^0 + n^+\pi^+ + n^-\pi^-+\ldots \\
   \textrm{or}\\
   p + \gamma  \rightarrow &\ n + n^0\pi^0 + n^+\pi^+ + n^-\pi^-+\ldots
%              & \ \ \ \ \ \ \ \ \ \ \  \rightarrow e^- + p' + \bar{\nu_e} + \mu^+ + \nu_{\mu}         
   \end{split}
   \end{equation}  
   and through electron-positron pair production (Bethe-Heitler process)   
   \begin{equation}
   p + \gamma  \rightarrow p' + e^+ + e^-\\    
   \end{equation}
   Photo-meson production has been evaluated using the publicly available Monte-Carlo code \sophia\ \citep{Sophia} which computes $N_{iter}$ interactions of a proton of energy $E_p$ with a given low-energy photon field, and provides as output the distributions (expressed as $E\frac{dN}{dE} = \gamma\frac{dN}{d\gamma}$) of the stable and long-lived particles ($e^{\pm}$, $\gamma$, $p$, $n$, $\nu_{e,\mu}$ and $\bar{\nu}_{e,\mu}$ \footnote{As the study of the neutrino emission from blazars is beyond the purposes of this paper, they will not be discussed in detail here.}).\\ The only photon target field we consider is given by the synchrotron emission from the primary electron population in the jet, which represents by far the dominant component at low energies in (U)HBLs. We slightly modified the \sophia\ code so that it can accept as input photon-field any numerical function (and not only a power-law or a black-body function as in the original version of the code).  
   We call \sophia\ for 50 different proton energies, spaced by $\Delta \log(\gamma_p)=0.1$, using for each call $N_{iter}=10^4$. For the $i$-th call, the proton Lorentz factor is thus:
   \begin{equation}
   \log{(\gamma_{p;i})} = \log{(\gamma_{p;max})} - i\cdot0.1
   \end{equation} 
   
   The \sophia\ code cannot properly evaluate the energy distributions of charged secondary particles in a magnetized environment, where they can be modified by synchrotron cooling. Following \citetalias{Mucke01}, we thus modify the energy of the proton before interaction to account for the synchrotron losses that can occur before the proton-photon collision: 
   \begin{equation}
   \label{synclosses}
   E'_p \simeq \frac{E_p}{1+\frac{r_{p,syn}(E_p)}{r_\pi(E_p)}}
   \end{equation} 
where $r_{p,syn}$ represents the inverse of the proton synchrotron loss time, given by: 
\begin{equation}
\label{synrate}
r_{p,syn}(E_p) = \frac{1}{\tau_{p,syn}(E_p)} = \frac{4}{3}\left(\frac{m_e}{m_p}\right)^2 \frac{\sigma_T u_B c}{(m_pc^2)^2}\ E_p
\end{equation}
 and $r_{\pi}$ represents the mean pion-production interaction rate, which is evaluated using a routine integrated in \sophia\ \citepalias[following again][]{Mucke01}.
  
    The \sophia\ output spectra are expressed as $\gamma\frac{dN}{d\gamma}$ for each type of particle. To use them, we need to convert these spectra into injection functions, which will be then used to obtain stationary distributions. We first divide these spectra by $\gamma$ and multiply them by $\int_{\gamma'_{p;i}}^{\gamma'_{p;i+1}}{d\gamma_p\ N'_p(\gamma_p)/\tau_{ad}}$, i.e. the rate of \textit{injected} proton density in the emitting region with energies between $\gamma'_{p;i}$ and $\gamma'_{p;i+1}$. Here $N'_p(\gamma'_p)$ represents the proton distribution without spectral break and $\gamma'_p=E'_p/m_pc^2$ is defined according to equation \ref{synclosses}. Finally, the particle spectra are multiplied by $r_\pi / r_{tot}$, the ratio of the pion-production interaction rate over the total interaction rate, given by $r_\pi + r_{BH} + 1/\tau_{ad}$, where $r_{BH}$ represents the Bethe-Heitler interaction rate \citep[see equation 2.2 in][]{Chodorowski92}.  
 In this stationary framework, the protons are assumed to be confined in the emitting region, and multiple $p$-$\gamma$ interactions may occur, which are evaluated as suggested in \citetalias{Mucke01}. The stationary distributions of electrons and positrons are then computed following Equation \ref{intsolution}, while the stationary distribution of photons is obtained simply by multiplying their injection rate by $\tau_{ad}$.\\ 
     
   Photons from the $\pi^0$ decay and the synchrotron emission from $e^\pm$ coming from the $\pi^\pm$ channel can reach energies up to $ m_pc^2\gamma_{p;max}$. Interacting with the low-energy photon field,  they trigger electro-magnetic cascades, mediated by synchrotron emission and $e^\pm$ pair production. The stationary state of the cascade emission is evaluated as follows: the first generation of pairs injected into the emitting region is treated in the same way as in the leptonic part of the code, by computing the injection rate, the stationary state of pair distribution and the associated synchrotron emission. In this case however, the synchrotron photons from pairs are still energetic enough to produce a second generation of pairs, which in their turn can produce a third generation, and so on. We iterate the process until the  i-th generation of pairs gives a negligible contribution to the sum of the previous generations. As a general rule, for the physical parameters used in the following applications, the spectrum computed including five generations of pairs already provides a good description of the cascade. During the computation of the cascade spectrum, the low energy photon field is considered as being represented only by the synchrotron emission of  primary electrons, neglecting the emission from the cascade itself (i.e. we make the assumption that the cascade is not self-sustained, and we verify \textit{a posteriori} that this condition is respected).\\  
   
   The \sophia\ code considers as output only stable or long-lived particles. However, in highly magnetized environments, the synchrotron emission from kaons, pions and muons before decay can be non-negligible and radiative losses can also affect the resulting spectra of electrons and positrons. Following \citetalias{Mucke01}, we modified the spectra of kaons, pions and muons before decay, taking into account their synchrotron losses. \\

   Synchrotron emission from muons can significantly contribute to the overall SED for magnetic fields of the order of tens of Gauss \citep[see e.g.][]{Rachen00}. We extract the muon ($\mu^\pm$) spectra from \sophia\ \textit{before} their decay into electrons and positrons, and we treat them in the same way as all the other \sophia\ outputs. The only difference occurs in the evaluation of the steady state distribution: for non-stable particles such as muons, we need to add the decay term (equal to $-N(\gamma)/\gamma\tau_{dec}$) in Equation \ref{contequation}. The integral solution (equation \ref{intsolution}) is then modified as follows: 
   \begin{equation} 
   \begin{split}
    \label{intsolutionmuons}   N_\mu(\gamma_\mu)=&\exp{\left[-\frac{\gamma^{\ast}_\mu}{\gamma_\mu} -\gamma^{\ast}_\mu \frac{\tau_{ad}}{2\gamma_\mu^2\tau_{dec}} \right]}\ \frac{\gamma^{\ast}_\mu \tau_{ad}}{\gamma_\mu^2}\cdot  \\
    &\cdot \int_{\gamma_\mu}^{\infty}{d\zeta\ \  Q_\mu(\zeta)\exp{\left[+\frac{\gamma^{\ast}_\mu}{\zeta} +\gamma^{\ast}_\mu \frac{\tau_{ad}}{2\zeta^2\tau_{dec}} \right]}}
       \end{split} 
    \end{equation}
    where $Q_\mu(\gamma_\mu)$ is represented by the \sophia\ output, and $\gamma_\mu^\ast$ is the Lorentz factor at which $\tau_c(\gamma_\mu)=\tau_{ad}$ (as in equation \ref{gammacool}).\\
    
   It should be noted that a fast alternative to the direct use of the \sophia\  Monte Carlo code is the approach by \citet{Kelner}, in which an analytical parametrization of the secondary particle distributions produced in the $p$-$\gamma$ interactions is given. However, an analytical expression for the contribution from muons does not yet exist, and in highly magnetized environments, synchrotron losses can significantly affect the distribution of secondary electrons and positrons.\\ 
   
  A process that competes with the photo-meson channel, but is dominant at lower energies, is Bethe-Heitler pair production (which is not included in the \sophia\ package). The pairs injected into the emitting region through this process have been computed using the analytical formulae by \citet{Kelner} \citep[in which the Bethe-Heitler cross-section is expressed following the work by][]{Blumenthal70}. The pairs injected into the emitting region are energetic enough to trigger an electro-magnetic cascade, which is computed in the same way as for the photo-meson induced cascades. \\
  It should be noted that our approach is different from the one used by \citetalias{Mucke01}, who simulated the Bethe-Heitler pair production via the Monte-Carlo code described in \citet{Protheroe96}.\\
   
   To summarize, the hadronic component is given by seven different contributions: the synchrotron emission from protons and muons and their associated synchrotron emission from $\gamma$-$\gamma$ secondary pairs constitute four distinct contributions; the three remaining ones are given by the synchrotron emission from the cascades triggered by photons produced via the $\pi^0$ decay, by $e^\pm$ produced via the $\pi^\pm$ decay and by $e^\pm$ produced via the Bethe-Heitler process.  
   
   \subsection{Physical constraints and systematic parameter scan} 
   \label{phycons}
   
  In hadronic scenarios, the fact of considering an additional proton population in the emitting region leads to six more free parameters with respect to the simple SSC scenario: the three Lorentz factors $\gamma_{p;min}$, $\gamma_{p;break}$, and $\gamma_{p;max}$, the indices of the proton power-law distribution $\alpha_{p;1,2}$, and the normalization factor $\eta$.\\
  
  The value of $\alpha_{p;1}$ can be constrained assuming that electrons and protons share the same acceleration mechanism in magnetic fields whose spectrum of turbulence is characterized by a single power-law at all scales, and thus have injection functions with the same index. 
 It is fixed to $\alpha_{e;1}$, the index of the electron distribution before the break. This assumes that the break in the electron distribution is induced by radiative cooling and that the distribution before the break is representative of the injection spectrum. In the following approach, we consider that the particle populations are cooled mainly by synchrotron emission, and thus characterized by a spectral break of $1$. The data in the optical and X-ray bands are used to constrain the index of the electron population $\alpha_{e;2}$, which is then used to derive $\alpha_{e;1}$ and $\alpha_{p;1,2}$.\\
  
  The value of $\gamma_{p;min}$ does not affect the modelling as long as it is low enough, so it can be considered as a fixed parameter. It impacts however the value of the particle energy density in the emitting region (especially if the proton-distribution slope is softer than 2.0). To be conservative, and in order not to bias our solutions by a systematic reduction of the total proton energy density, we fix it at $\gamma_{p;min}=1$.\\
  
\begin{figure*}
	   \centering
		\includegraphics[width=220pt,height=140pt]{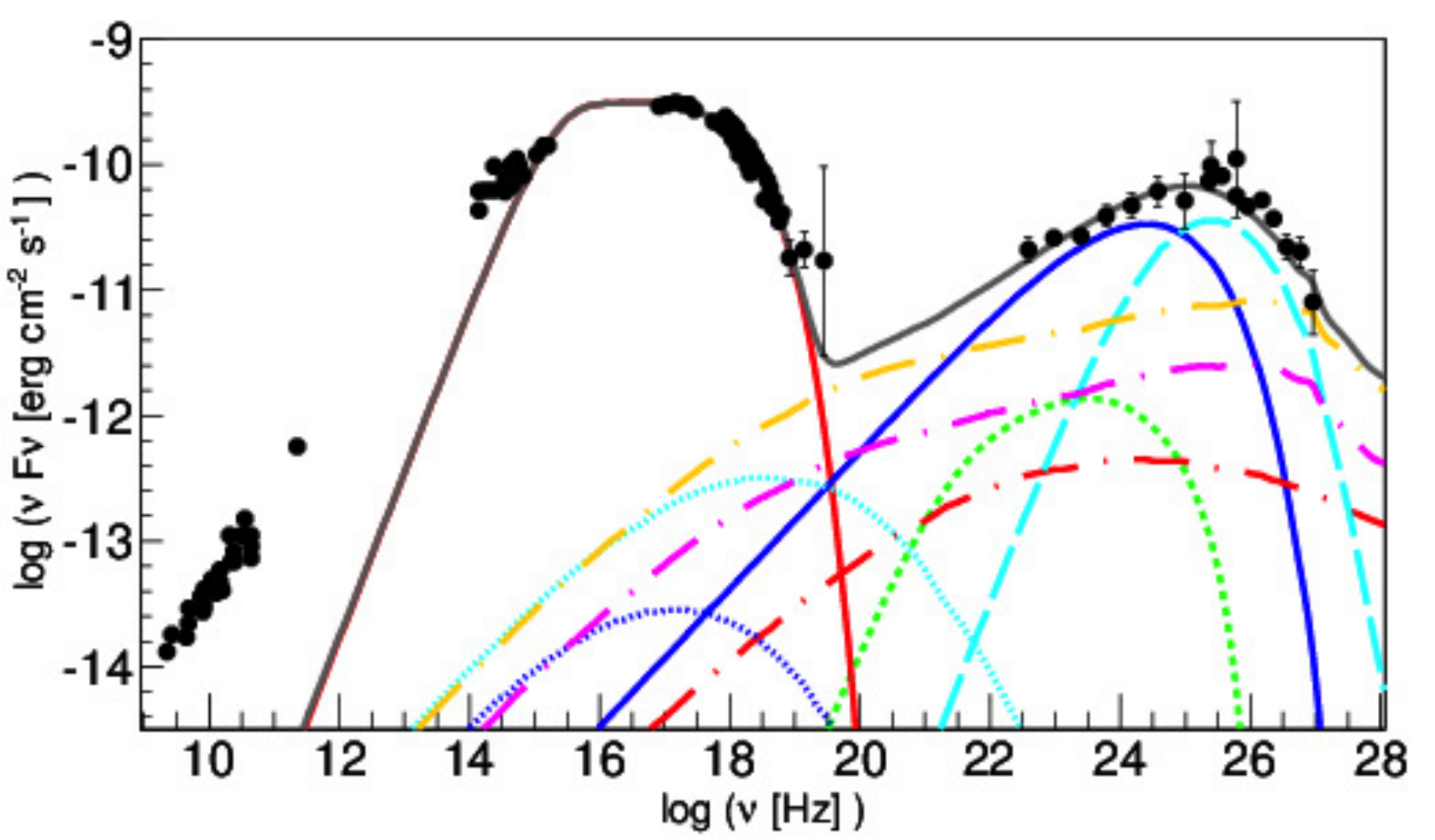}
		\includegraphics[width=220pt,height=138pt]{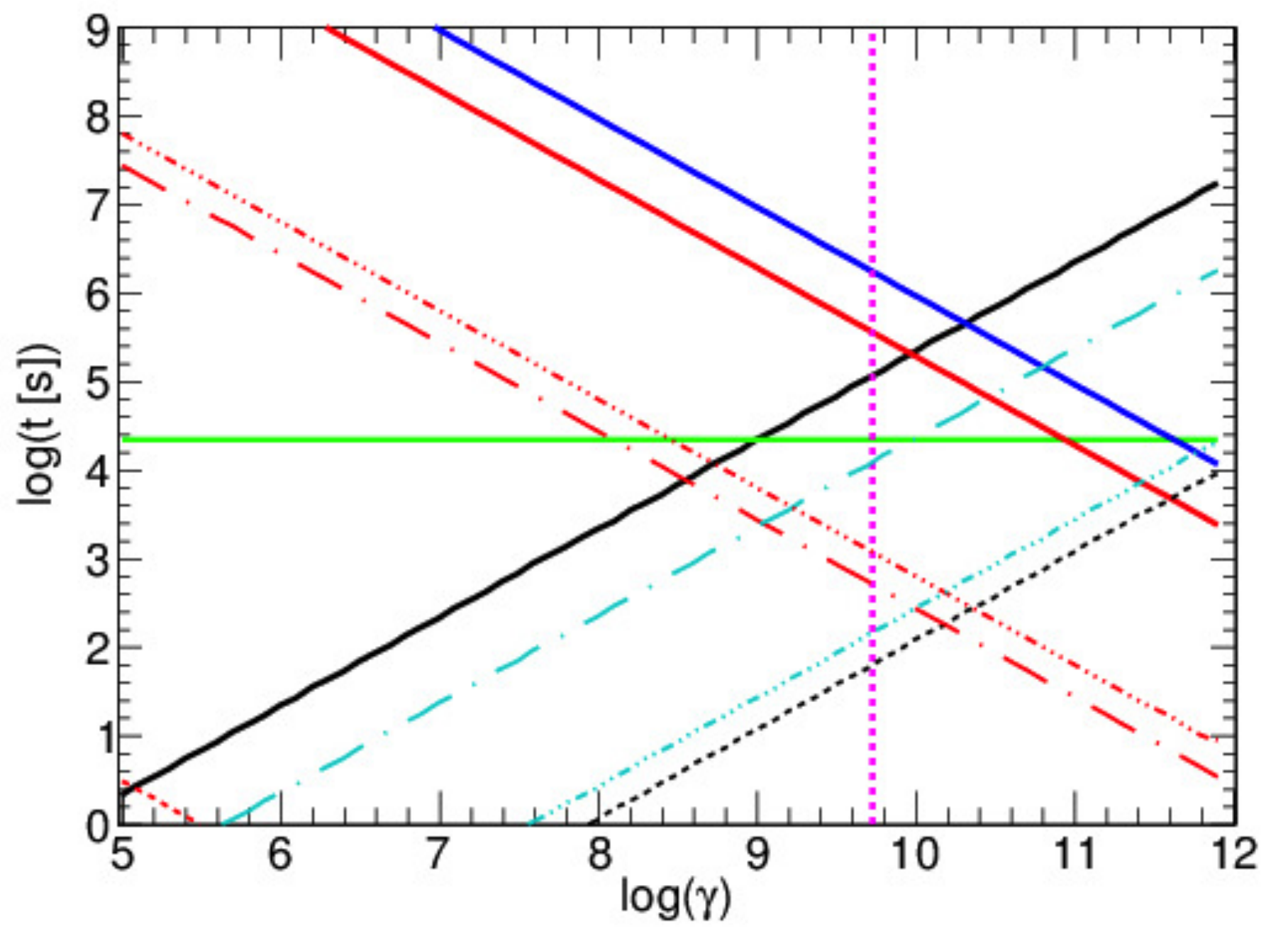}
		\includegraphics[width=220pt,height=70pt]{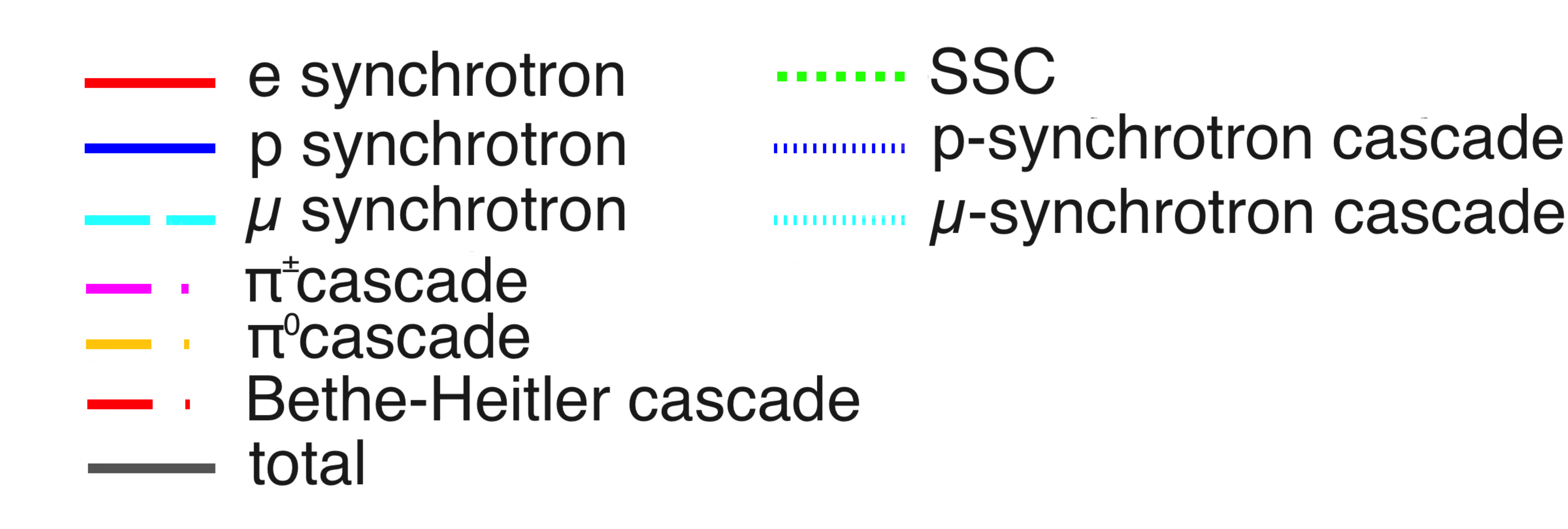}
		\includegraphics[width=220pt,height=70pt]{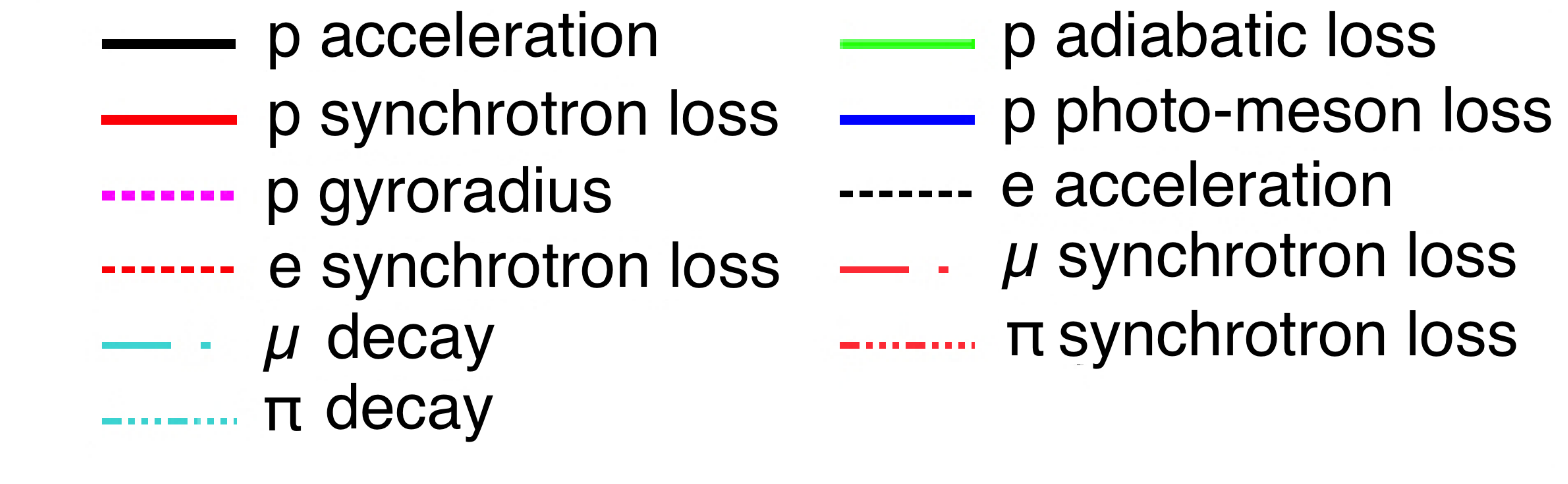}
	  \caption{\textit{Left:} hadronic modeling of the spectral energy distribution of \mrk\ \citep[data points taken from][]{Mrk421}. Data have been corrected for the EBL absorption, which is thus not included in the model. \textit{Right:} acceleration and cooling time-scales for the different particles in the emitting region. For the parameter values, see Section \ref{sec3}. \label{figmrk}}
   \end{figure*}

The value of $\gamma_{p;max}$ is constrained by physical considerations on acceleration and cooling time-scales. In particular, assuming that the acceleration takes place at diffusive shocks, the expression of the acceleration time-scale \citep[see e.g.][]{ODrury83, Protheroe04, Rieger07} can be expressed as 
  \begin{equation}
  \label{equation18}
  \tau_{acc}=\frac{1}{\psi}\frac{m_p\ c}{e\ B}\ \gamma_p
  \end{equation}
where $\psi$ is an efficiency factor characterizing the acceleration rate, and has been fixed at  a physically plausible value of $1/10$ \citepalias[see][]{Mucke01}.
The cooling terms are represented by the adiabatic losses ($\tau_{ad}=2\ R/c$), the synchrotron losses (the inverse of the synchrotron cooling rate, see Equation \ref{synrate}) and the photo-meson losses \citep[expressed analytically following ][]{Sikora}. 
The maximum proton energy depends on the most rapid cooling mechanism. It is determined from the equality $\tau_{acc}(\gamma_{p;max}) = \min[ \tau_{ad} ; \tau_{syn}(\gamma_{p;max}) ;
\tau_{pm}(\gamma_{p;max}) ]$ with $\tau_{syn}$ and $\tau_{pm}$ the characteristic time-scales for synchrotron and photo-meson losses, respectively. \\ % break AZ
The plot on the right panel of Fig. \ref{figmrk} shows an example of all the relevant time-scales in the emitting region. Under the assumption that the photo-meson losses are always the slowest, which holds for the application to AGN without strong external photon fields discussed in the present paper, we have to consider two different regimes, $\tau_{ad} \leq \tau_{syn}$ or $\tau_{ad} > \tau_{syn}$, for all $\gamma \le
\gamma_{p;max}$.

\subsubsection{Adiabatic cooling dominated regime ($\tau_{ad} \leq \tau_{syn}$)}

In this regime,  the fastest cooling mechanism for all $\gamma_{p} \le
\gamma_{p;max}$ is the adiabatic one, and no break is expected in the proton energy distribution.  The equation  $\tau_{acc} (\gamma_{p;max})=\tau_{ad}$  is thus used to define $\gamma_{p;max}$:
   \begin{equation}
   \label{equation20}
 \gamma_{p;max}=6.44\times10^{9}  \frac{B}{1 \rm{G}}\ \frac{R}{10^{17} \rm{cm}}
  \end{equation}

The condition on the cooling terms translates into a relation between the size of the emitting region and the magnetic field: 
     \begin{equation}
     \label{equation19}
  \frac{R}{10^{17} \rm{cm}} \leq 10.13 \frac{B}{1 \rm{G}}^{-3/2}
  \end{equation} 
 
  The peak frequency of the proton synchrotron component can be derived from the synchrotron emissivity and the expression (Equation \ref{equation20}) for $ \gamma_{p;max}$ as \citep[see e.g.][]{Tavecchio98, constraints}:
        \begin{equation}
        \label{equation21}
 \frac{\nu_{peak, p}}{10^{27} \rm{Hz}} =  \frac{1.25 \times 10^{-3}} {1+z} \frac{(3-\alpha_{p;1})}{1.5} \frac{\delta}{10} \left( \frac{B}{1 \rm{G}}\right)^3 \left(\frac{R}{{10^{17} \rm{cm}}}\right) ^2 
  \end{equation}
   
   where $z$ is the redshift of the source and the numerical coefficient takes into account the exponential cut-off at $ \gamma_{p;max}$. Interestingly, considering Equation \ref{equation19}, $\nu_{peak, p}$ has to be lower than a given value which is a function only of the injection index $\alpha_{p;1}$ and the Doppler factor:
           \begin{equation}
           \label{equation22}
 \frac{\nu_{peak, p}}{10^{27} \rm{Hz}} \leq \frac{0.128} {1+z} \frac{(3-\alpha_{p;1})}{1.5} \frac{\delta}{10}  
  \end{equation}
  or, in terms of the peak energy,
             \begin{equation}
           \label{equation23}
 \frac{E_{peak, p}}{1 \rm{TeV}} \leq \frac{0.529} {1+z} \frac{(3-\alpha_{p;1})}{1.5}  \frac{\delta}{10}
  \end{equation}
  This inequality shows the capability of hadronic models to reproduce the observed peak of the UHBL $\gamma$-ray emission at TeV energies.\\
   
  An additional constraint comes from the observed variability time-scale $\tau_{var}$ and the usual causality argument:
           \begin{equation}
           \label{equation24}
\frac{R}{10^{17} \rm{cm}}\leq \frac{2.59}{1+z} \frac{\delta}{10} \frac{\tau_{var}}{10\ \rm{days}}
  \end{equation}
   
   We can thus systematically study the parameter space for a given value of $\delta$ by iterating over $R$, starting from the maximum allowed value, (Equation \ref{equation24}), and over $B$, which is constrained by Equation \ref{equation21}, while keeping the value of $\nu_{peak, p}$ consistent with the data. 
   The value of $\gamma_{p;max}$ is then computed following Equation \ref{equation20}.
  The luminosity of the high-energy component of the SED fixes the normalization of the proton distribution (the parameter $\eta=K_p/K_e$), which is the last free parameter of the hadronic component.\\
  
  Now turning to the leptonic component, the break in the electron distribution is fixed by the equality of the adiabatic and the synchrotron time-scale ($\tau_{ad} = \tau_{syn}(\gamma_{e;break})$ ), and can be expressed as: 
  \begin{equation}
  \label{equation25b}
  \gamma_{e;break}=75\  \left( \frac{B}{1 \rm{G}}\right)^{-2} \left(\frac{R}{10^{17} \rm{cm}} \right)^{-1} 
  \end{equation}
  As can be seen, in the proton-synchrotron scenario (characterised by $B\gtrsim 1$ G) the magnetic field is sufficiently high such that the entire leptonic population is cooled by synchrotron emission and is thus described by a simple power-law function with index $\alpha_{e;2} = \alpha_{p;1} + 1$. \\ % break AZ
   The remaining free parameters are the normalization of the electron distribution, which is constrained by the luminosity of the low-energy component of the SED, and the maximum energy of the electrons $\gamma_{e;max}$.  In our model, the latter is constrained by the observed peak of the low-energy bump in the SED (and not by the equality of the acceleration and cooling time-scales, as discussed in Section \ref{sec5}):
  \begin{equation}
  \label{equation25}
  \gamma_{e;max}=1.64\times10^5\ (1+z)  \left(\frac{\nu_{peak, e}}{10^{18} \rm{Hz}}\right)^{1/2}  \left(\frac{\delta}{10} \right)^{-1/2} \left( \frac{B}{1 \rm{G}}\right)^{-1/2}
  \end{equation} 

\subsubsection{Synchrotron cooling dominated regime  ($\tau_{ad} > \tau_{syn}$)}

In this second regime, the synchrotron cooling time-scale is shorter than the adiabatic one for all $\gamma_{p} \le
\gamma_{p;max}$. The first consequence is that the stationary proton distribution is described by a broken power-law, with index $\alpha_{p;2}=1+\alpha_{p;1}$ above $\gamma_{p;break}$. The maximum proton energy is determined by the equation  $\tau_{acc}(\gamma_{p;max}) = \tau_{syn}(\gamma_{p;max})$, which yields: 
   \begin{equation}
   \label{equation26}
 \gamma_{p;max}=6.53\times10^{10}  \left( \frac{B}{1 \rm{G}}\right)^{-1/2}
  \end{equation}
The relation $\tau_{ad}=\tau_{syn}(\gamma_{p;break})$  is used to define the proton break energy:
    \begin{equation}
   \label{equation27}
 \gamma_{p;break}=6.61\times10^{11}  \left( \frac{B}{1 \rm{G}}\right)^{-2} \left( \frac{R}{10^{17} \rm{cm}}\right)^{-1}
  \end{equation}
 In this case we expect $\gamma_{p;break} <  \gamma_{p;max}$, and indeed this inequality leads to
      \begin{equation}
     \label{equation28}
  \frac{R}{10^{17} \rm{cm}} > 10.13 \frac{B}{1 \rm{G}}^{-3/2}
  \end{equation}
  
   which is complementary to Eq. \ref{equation19}, as expected.\\ 
   The peak frequency of the proton synchrotron component is associated to $\gamma_{p;break}$ only if $\alpha_{p;1} \in (2.0, 3.0)$. In this case,
           \begin{equation}
        \label{equation21b}
 \frac{\nu_{peak, p}}{10^{27} \rm{Hz}} =  \frac{13.15} {1+z} \frac{(3-\alpha_{p;1})}{1.5} \frac{\delta}{10} \left( \frac{B}{1 \rm{G}}\right)^{-3} \left(\frac{R}{{10^{17} \rm{cm}}}\right) ^{-2} 
  \end{equation}
   By imposing the condition from Equation \ref{equation28}, we find again that the upper bound for $\nu_{peak, p}$ is given by  the inequality in Equation \ref{equation22}.\\
   On the other hand, if $\alpha_{p;1} < 2.0$, the peak frequency is associated with $\gamma_{p;max}$. In this case $\nu_{peak, p}$ is constant, and equal to 
              \begin{equation}
          \label{equation29new}
\frac{\nu_{peak, p}}{10^{27} \rm{Hz}} = \frac{0.128} {1+z} \frac{(3-\alpha_{p;2})}{1.5}  \frac{\delta}{10} 
\end{equation}
which is a factor $1 - 1/(3-\alpha_{p;1})$ lower than the one in Equation \ref{equation22}.\\
  There exists thus a maximum peak frequency of the proton synchrotron emission, corresponding to the equality in Equation \ref{equation22}, for which $R$ and $B$ are related via the equality in Equations \ref{equation19} and \ref{equation28}. This is the case where $\tau_{acc}(\gamma_{p;max}) = \tau_{ad} = \tau_{syn}(\gamma_{p;max})$.\\
      For a given value of $\delta$, we can again study systematically the parameter space, iterating over $R$ and $B$,  with $\nu_{peak, p}$ constrained by the data. 
      Here, we are applying again the constraint on $\tau_{var}$ (Equation~\ref{equation24}) and the constraints on the electron distribution (Equations~\ref{equation25b} and
      \ref{equation25}), though generally, in this regime, $\gamma_{e;break}$ occurs below or very close to $\gamma_{e;min}$.

      \subsubsection{Additional constraints}  
       One may think that another constraint on the maximum energy of protons in the emitting region  could be provided by their gyro-radii in the assumed homogeneous magnetic field: particles with gyro-radius larger than $R$ would escape the emitting region, and should not be considered in the framework of a stationary emission model. For relativistic particles, the expression of the gyro-radius is 
     \begin{equation}
     \label{equation30}
     R_{gyro} = \frac{m\ c^2}{e\ B} \gamma
     \end{equation} 
       However, for the protons in the emitting region, by substituting the values of $\gamma_{p;max}$ in Equation \ref{equation30}, it is easy to show that $R_{gyro}$ is always smaller than $R$, i.e. even the most energetic protons are confined in the plasma blob, given the constraints from acceleration and loss time-scales.  \\
      
      For a given Doppler factor $\delta$, the equations described above can be used to determine the complete set of solutions for a proton-synchrotron scenario by systematically scanning different values of the magnetic field $B$ and the size of the emission region $R$. This scenario provides a satisfactory description of the blazar SED only for values of $\nu_{peak, p}$ in agreement with the observed peak frequency of the high-energy bump in the SED. Removing this constraint on $\nu_{peak, p}$, and studying the parameter space for all other values of $R$ and $B$, we discovered an interesting part of the parameter space, in which $B$ is too low for a dominant proton-synchrotron component and the denser emission region leads to a significant contribution from secondary pairs from p-$\gamma$ interactions. This set of secondary solutions exists only in
      the regime where $\tau_{ad} \leq \tau_{syn}$, and is systematically studied by iterating over $R$ and $B$ for lower values of $\nu_{peak, p}$ (see Equation \ref{equation21}) and normalizing the secondary-pair synchrotron emission in order to match the observed gamma-ray emission. The resulting lepto-hadronic scenario is very distinct from the proton-synchrotron scenario and resembles more the ``proton-induced cascade'' model by \citet{Mannheim93}.\\
      In this lepto-hadronic scenario, the primary leptonic population is not completely cooled down by synchrotron emission, and a cooling break should be present in their stationary distribution. In this case, the electron particle population is thus defined by a broken power-law with $\alpha_{e;1}=\alpha_{p;1}$, $\alpha_{e;2}=\alpha_{e;1} + 1$, and $\gamma_{e;break}$ is given by Equation \ref{equation25b}.
      
 The lepto-hadronic solutions occupy a region in the parameter space with low B ($\sim$0.1 to a few G) and higher particle densities than in the proton-synchrotron solutions. 
In this scenario, proton synchrotron radiation can lead to a small spectral bump at intermediate energies between the X-ray and \fermilat\ bands and the primary proton
spectrum is not significantly cooled. Both SSC emission and photons from proton-induced cascades contribute to the high-energy bump.

   \section{Application to \mrk}
   \label{sec3}
   
  As a first application of our code we have studied the SED of the well known HBL \mrk, which was the very first extragalactic source detected at TeV energies \citep{Punch92}. Its VHE spectrum has been well studied with the current generation of Cherenkov telescopes \citep{Mrk421HESS, Mrk421MAGIC, Mrk421VERITAS}. Up to now, the most complete simultaneous SED of \mrk\  in a low state has been published by \citet{Mrk421}, including data from \fermilat\ and \textit{MAGIC}. Apart from the usual SSC model, the authors applied also a hadronic model to this SED, using the code by \citetalias{Mucke01}. We have applied our code to this data set to compare the result with an existing model.\\
   
   The low-energy component of the SED can be satisfactorily associated with a completely cooled leptonic population with power-law index $\alpha_{e;2}=2.9$ and a simple exponential cut-off. Following the physical constraints presented in Section \ref{sec2}, we impose thus $\alpha_{p;1}=1.9$. To compare our model directly with the one presented in \citet{Mrk421}, we adopt the same value for the magnetic field, $B=50$ G. The main difference with respect to their model comes from the constraint on $\gamma_{p;max}$ and, in particular, from the hypothesis on the acceleration mechanism (equation \ref{equation18}). \citetalias{Mucke01}  considered a more efficient acceleration term due to oblique shocks, yielding higher values for $\gamma_{p;max}$. To compensate for this effect we use a higher Doppler factor (25 instead of 12). We find a similar solution imposing $\nu_{peak, p} = 3\times10^{24}$ Hz in equation \ref{equation21}, which results in $R=3.3\times10^{14}$ cm and $\gamma_{p;max}=1.06\times 10^9$ \citep[compared respectively to $4\times10^{14}$cm and $2.3\times10^9$ in][]{Mrk421}. In Figure \ref{figmrk} we present the results of the modeling: the high energy bump of the SED is ascribed to different components, with the muon synchrotron emission dominating the VHE emission, and cascades dominating above $10^{27}$ Hz. This example  illustrates the importance of a detailed simulation of secondary particles from p-$\gamma$ interactions. It can also be seen that synchrotron emission from muons cannot be neglected.\\
     
  Besides the small differences in the parameter values imposed by our physical constraints, our code fully reproduces the result presented in \citet{Mrk421}. In addition we are able to correctly estimate the emission by the cascades triggered by Bethe-Heitler pair production, which is negligible with respect to the emission by cascades associated with the photo-meson interaction. Although negligible in this case, as we show later, the SSC component can play an important role in the lepto-hadronic scenario. It should be noted that for this first application, we have not carried out a systematic scan of the parameter space, as will be done for the investigation of the UHBLs in the following section. \\

      \section{Modelling of ultra-high-frequency-peaked BL Lacs}
   \label{sec4}
   
    \begin{table}
       $$
   \begin{array}{ p{58 pt} p{13 pt}  p{38 pt}  p{38 pt}  p{14 pt} p{20 pt} }
            \hline
             \noalign{\smallskip}
              object & $\sigma$ & $\Gamma$& $\Phi_0$& $E_{dec}$ & $E_{max}$ \\
            \noalign{\smallskip}
            \hline   
            \noalign{\smallskip}
          1ES\, 0229+200  &  9.1 & 1.90$\pm$0.16 &1.67$\pm$0.31 & 5.03	& 171 \\
                      \noalign{\smallskip}
          1ES\, 0347-121 & 6.8 & 1.70$\pm$0.14  & 1.09$\pm$0.25 &   5.16 & 27\\ 
            \noalign{\smallskip}
                       RGB\, J0710+591 & 16.4 & 1.56$\pm$0.09  & 1.45$\pm$0.18    & 7.24 & 175 \\
             \noalign{\smallskip} 
            1ES\, 1101-232 & 9.4 & 1.91$\pm$0.16  & 3.60$\pm$0.62 &  3.50 & 48 \\
            \noalign{\smallskip}
           1ES\, 1218+304 & 42.0 &   1.68$\pm$0.03  & 17.16$\pm$0.87& 4.48 & 366\\
           \noalign{\smallskip}
            \hline
         \end{array} 
      $$
        \caption[]{Results from the analysis of \fermilat\ data. For all objects the best-fit model is given by a power-law function with index $\Gamma$. The differential flux normalization $\Phi_0$ is in units of ($10^{-14}$ cm$^{-2}$ s$^{-1}$ MeV$^{-1}$), and computed at the decorrelation energy  $E_{dec}$ (in Gev). We also provide the energy ($E_{max}$, in GeV) of the most energetic photon consistent with the source position. The significance of the detection is expressed in standard deviations ($\sigma$) above the background. }
        \label{tab:newfermi}
  \end{table}

   We have modelled the SEDs of the five UHBLs listed in Table \ref{tab:uhbl} with our code in a systematic way. The model parameters have been constrained following the relations detailed in Section \ref{phycons}.  For a given value of the Doppler factor (fixed in the following at $\delta=30$), we scan the parameter space in $R$ and $B$, while the remaining free parameters are all determined by physical assumptions and by observations. The maximum value of $R$ is determined from equation \ref{equation24} by assuming $\tau_{var}=10$ days, except for the modelling of 1ES\, 1218+304, for which a variability time-scale of one day has been adopted, based on observational evidence \citep{1218_Veritas}. Under the assumption that leptons and protons emitting at all energies are confined in the same emitting region, $\tau_{var}$ should be considered as the shortest variability time-scale at all wavelengths. Beside 1ES\, 1218+304,  for 1ES\, 0229+200  \citep{0229_Veritas} and RGB\, J0710+591\footnote{See \url{http://www.swift.psu.edu/monitoring/source.php?source=RGBJ0710+591}} a $\tau_{var}$ of the order of ten days has been observed in soft X-rays. Regardless of the exact value of $\tau_{var}$, radii larger than $10^{18}$ cm can also be excluded on the basis of the required power of the emitting region (see Fig. \ref{fig2} and Section 5), as well as the opening angle of relativistic jets measured in the radio band. For all sources we have arbitrarily imposed a minimum size of the emitting region $R_{min}= 10^{14-15}$ cm, which corresponds to $\sim 10$ gravitational radii for a SMBH of $10^{8-9} M_\odot$ (see Table \ref{tab:uhbl} for the exact value of $M_\bullet$ we considered). This is based on the expectation that the size of the emitting region will likely be at least of order the scale of the base of the jet. \\
      
   Multi-wavelength data have been taken from \citet[][for 1ES\, 0229+200]{0229_Veritas}; \citet[][for 1ES\, 0347-121]{0347_HESS}; \citet[][for RGB\, J0710+591]{0710_veritas}; \citet[][Fig.6, upper panel, for 1ES\, 1101-232]{1101_HESS}; and \citet[][for 1ES\, 1218+304]{Rueger10}.   In this work, publicly available data from the \fermilat\ taken from August 4, 2008 to February 15, 2014 (MJD 54682-56703) were analyzed using the standard \fermi\ analysis software, version v9r32p5, available from the Fermi Science Support Center (FSSC)\footnote{\url{http://fermi.gsfc.nasa.gov/ssc/}}. Events with energy between 100 MeV and 500 GeV were
selected from the Pass 7 data set. Only events passing the SOURCE class filter and located within
a square region of side length 20$^\circ$ centered on the source position were selected. Cuts on the zenith angle
($<$ 100$^\circ$) and rocking angle ($<$ 52$^\circ$) were applied to the data. The post-launch \textit{P7SOURCE$\_$V15}
instrument response functions (IRFs) were used in combination with the corresponding Galactic
and isotropic diffuse emission models. The model of the region includes the diffuse components
and all sources listed in the Second Fermi-LAT Catalog \citep[2FGL][]{2FGL} located within
a  20$^\circ$ circle centered on the source. The spectral parameters of the sources were
left free during the fitting procedure. A power-law correction in energy with free normalization
and spectral slope was applied to the Galactic diffuse component. Events were analyzed using the
binned maximum likelihood method as implemented in \textit{gtlike}. 
   The results of our \fermilat\ analysis are shown in Table \ref{tab:newfermi}.\\
    
    Among the five sources under study, only RGB\, J0710+591, 1ES\, 1101-232, and 1ES\, 1218+304 are included in the 2FGL catalog. Our analysis is fully consistent with the 2FGL for both 1ES\, 1101-232 and 1ES\, 1218+304, while for RGB\, J0710+591 only the spectral index is consistent but not the flux normalization (the differential flux from the 2FGL, estimated at the same energy provided in Table \ref{tab:newfermi}, is $(2.26 \pm 0.34)\times10^{-14}$ cm$^{-2}$ s$^{-1}$ MeV$^{-1}$). This discrepancy may by due to the known variability of the source at GeV energies\footnote{See \url{http://tevcat.uchicago.edu/LATCat2/spectra/lat_1281.html}.} \\
    The GeV emission from 1ES\, 0229+200 was first studied by \citet{Dermer11}, who analyzed \lat\ data from August 2008 to September 2010 without detecting the source. With a larger data-set (3.25 years of \fermilat\ observations), and considering only energies from 1 to 300 GeV,  \citet{Vovk12} have claimed the first GeV detection of the source. They estimated a $1-300$ GeV spectral index of $1.36\pm0.25$ and a flux normalization of $(2.15\pm1.45)\times10^{-15}$ cm$^{-2}$ s$^{-1}$ MeV$^{-1}$ at our decorrelation energy.\\ % break AZ
  The discrepancy in the spectral index estimation between our work and the results presented by \citet{Vovk12} could be related both to the different energy band in which the fit was performed, as well as the different time range. In order to perform a better comparison, we analyzed the same time interval analyzed by \citet{Vovk12}, but using all the \fermilat\ photons from 100 MeV to 500 GeV. The resulting index is $2.14\pm0.19$, with a differential flux of $(4.80\pm1.25)\times10^{-14}$ cm$^{-2}$ s$^{-1}$ MeV$^{-1}$ at $E_0 = 2.69$ GeV, consistent with the results from the 2008-2014 analysis, but again not in agreement with the results from \citet{Vovk12}.\\ % break AZ
The remaining possibility is that the discrepancy is due to the different energy band used for the fit, indicating a spectral softening at MeV energies. The catalog of \fermi\ sources above 10 GeV \citep[1FHL,][]{1FHL} differs indeed from the results included in the 2FGL. Alternatively, the difference could be related to contamination from the Milky Way foreground which dominates at lower energies, as discussed for example in \citet{101015}.  \\
    For 1ES\, 0347-121 our result agrees with the \fermilat\ analysis recently presented by \citet{Tanaka14}.\\
    In the following, we include these new \fermilat\ spectra in our SEDs, with the exception of RGB\, J0710+591, for which, given the GeV variability, we used the spectrum estimated by \citep{0710_veritas} that is simultaneous with the \veritas\ observations. \\
     
   For the five sources, we find two distinct sets of solutions, corresponding to the proton-synchrotron scenario and a lepto-hadronic scenario, dominated by emission from secondary particles from p-$\gamma$ interactions. In the latter case, the SSC contribution is at a level comparable to the emission from hadronic processes, and justifies the ''\textit{lepto-hadronic}'' classification.  
   When scanning the parameter space, we compute for each model curve the $\chi^2$ deviation with respect to the observational data. This information is used to determine the solution which provides the best description of the SED, as well as the 1-$\sigma$ region in the $B$-$R$ plane (which corresponds to $\Delta \chi^2 \leq 2.3$, i.e. to a
   significance of 1$\sigma$ for two free parameters). The evaluation of the $\chi^2$ ranges was done separately for the two sets of solutions. Given the degeneracy
   of the problem and the difficulty in accounting correctly for different statistical and systematic uncertainties between the data sets, an actual $\chi^2$ minimisation
   was not attempted. \\
    
    In the top panel of Figure \ref{fig2} we show the two families of solutions for the case of RGB\, J0710+591, which is the only source for which there is a \fermilat\ spectrum simultaneous with VHE observations. The corresponding plots for the other four sources are provided in Appendix \ref{appa} (Figure \ref{figa2} for 1ES\, 0229+200, Figure \ref{figa1} for 1ES\, 0347-121, Figure \ref{figa3} for 1ES\, 1101-232, and Figure \ref{figa4} for 1ES\, 1218+304).\\
   The location of the two solutions in the parameter space is represented in the bottom panel of Figure \ref{fig2}, in which we show the contour-plot in the $B$-$R$ plane, as well as the ratio between the particle and magnetic energy density versus the total kinetic plus magnetic power of the emitting material. The total power of the emitting region is evaluated as 
   \begin{equation}
   \label{luminosityequation}
   L=\pi R^2c\Gamma_{bulk}^2(u_B+u_e+u_p)
   \end{equation}
    where $ \Gamma_{bulk}=\delta /2$, and $u_B$, $u_e$ and $u_p$ represent the energy densities of the magnetic field, the electrons, and the protons, respectively. The resulting ranges of all the parameter values are reported in Tables \ref{table1a} and \ref{table1b} for the five sources and for each of the two scenarios. \\  It can be seen from Figure~\ref{fig2} that the ``proton-synchrotron'' solutions consist principally of the electron-synchrotron component at low energies and the proton-synchrotron 
component at high energies. Only very small contributions from proton-synchrotron induced cascades and from muon-synchrotron emission are visible at low and high energies, respectively.
For this scenario, the synchrotron-cooling-dominated regime is accessible for all the sources but RGB\, J0710+591, the only source for which we used simultaneous \fermilat\ measurements to constrain the models.\\
 In the lepto-hadronic solutions, the low-energy bump is still due to electron-synchrotron emission, while the high-energy bump is a combination of SSC and pion-induced
cascade emission. The proton-synchrotron component is strongly suppressed and shifted to intermediate energies.  
   
As can be seen in the SED plots, our models do not always describe the available data on the MeV-GeV emission correctly. While the results for 1ES\, 0229+200 and RGB\, J0710+591 match the \fermilat\ measurements (even though for 1ES\, 0229+200 the proton-synchrotron model systematically underestimates the data at $\simeq 100$ MeV), for the other three sources our models can reproduce either the spectral index, or the normalization, but not both of them. In general, the lepto-hadronic solutions predict flatter GeV spectra, more in agreement for example with the \fermilat\ spectrum of 1ES\, 1101-232.\\
   This inconsistency may well be attributable to the long term $\gamma$-ray variability of the five UHBLs \citep[as suggested for 1ES\, 0229+200 by \veritas, see][]{0229_Veritas}: the \fermilat\ bow-ties are estimations of the average $\gamma$-ray emission integrated over five years. Looking at the SEDs of 1ES\, 1101-232 and 1ES\, 1218+304, it is clear that there is indeed some conflict between the \fermi\ and the IACT measurements at $\simeq 100$ GeV.\\
   The \fermilat\ measurements should thus be considered as an estimation of the mean MeV-GeV emission, and not used to put strong constraints on the models. The exception is RGB\, J0710+591, for which we used the strictly simultaneous \fermi\ data provided by \citet{0710_veritas}. In this case our models correctly reproduce the \fermilat\ flux and spectral index. We conclude that the mismatch between our model and the average \fermi\ spectra might well be due to long- or medium-term variability of the source spectra.\\

      \section{Discussion}   
   \label{sec5}
   
      \begin{figure*}
      \smallskip
	   \centering
	   		\includegraphics[width=220pt]{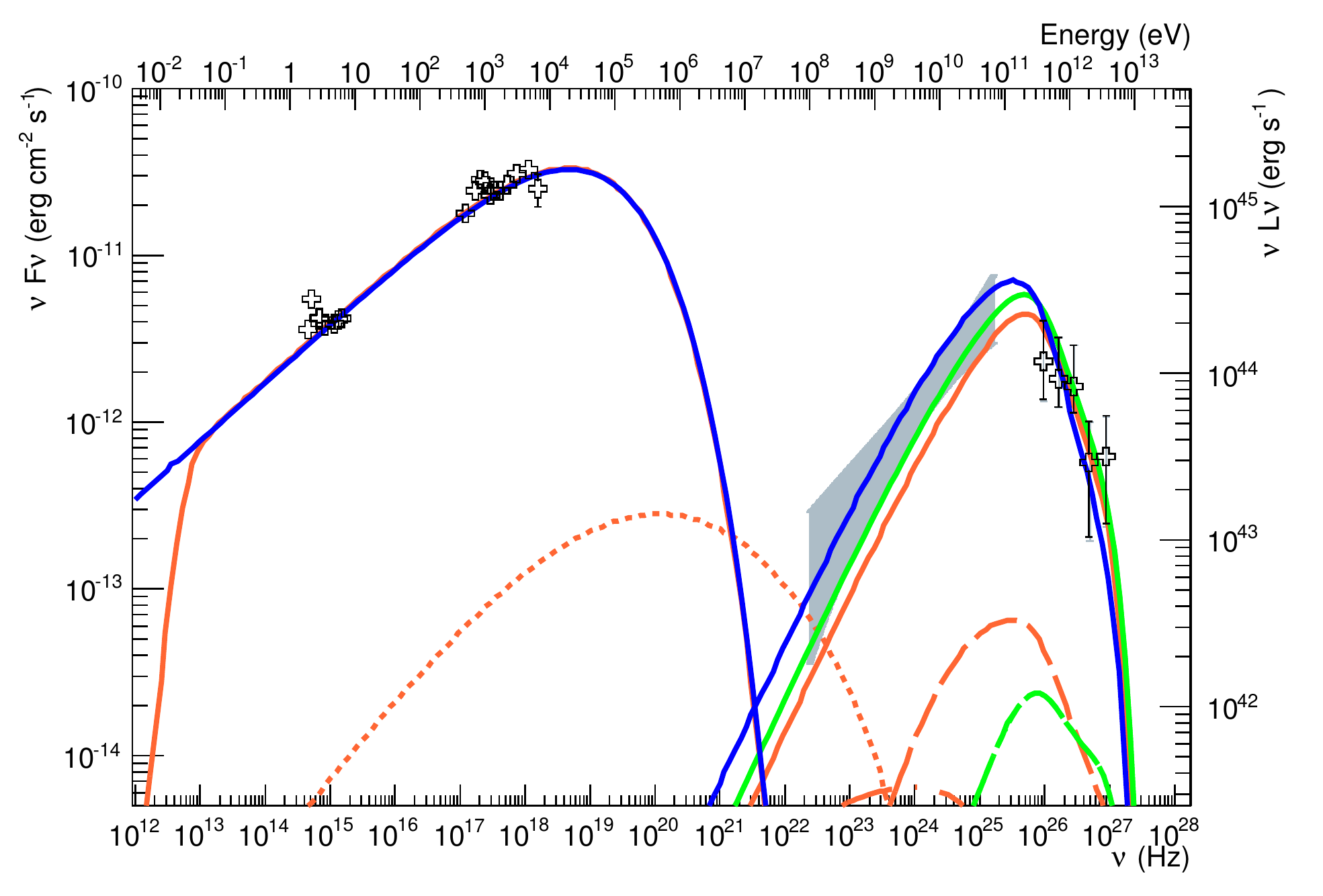}
		\includegraphics[width=220pt]{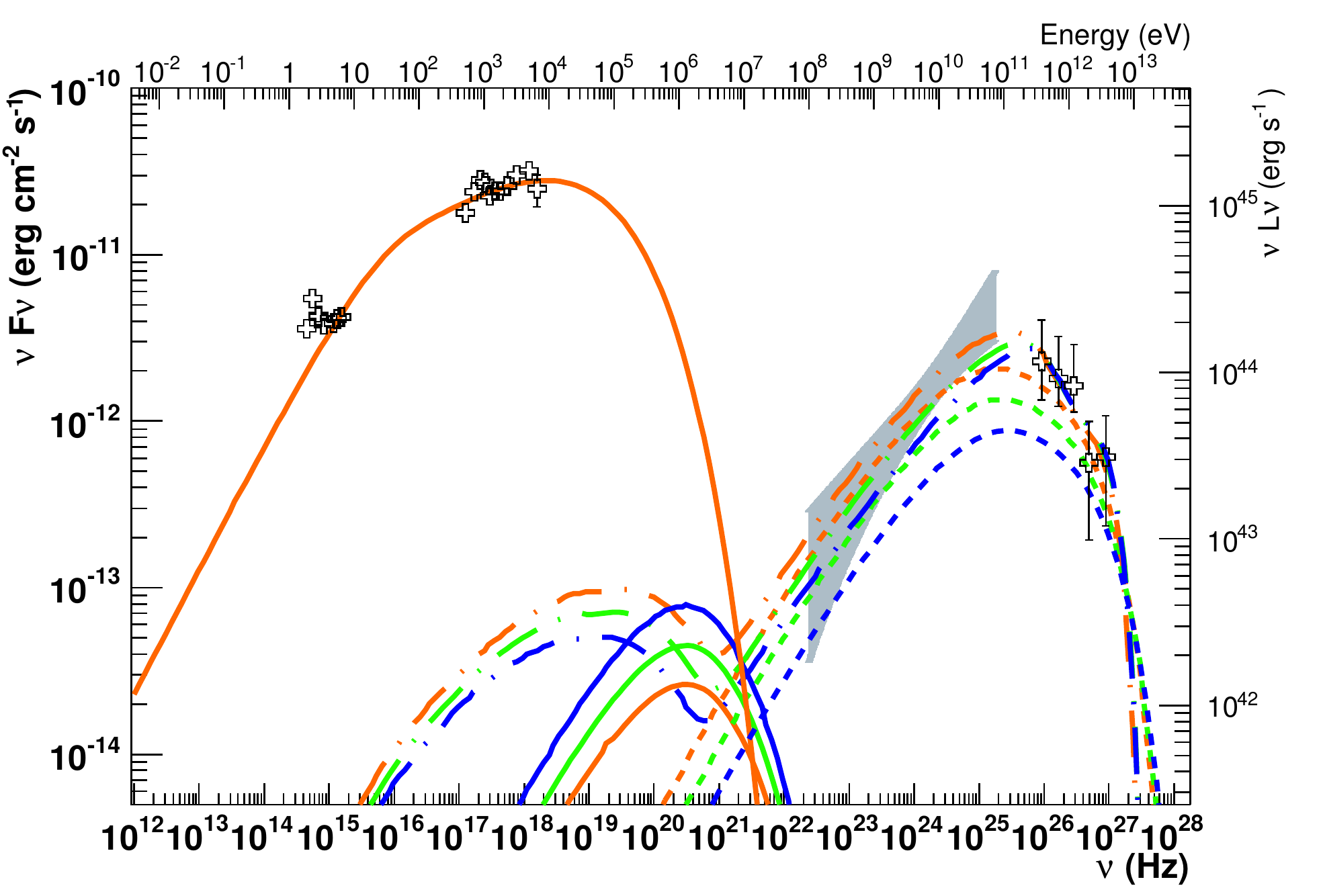}\\
		\smallskip
		\includegraphics[width=220pt]{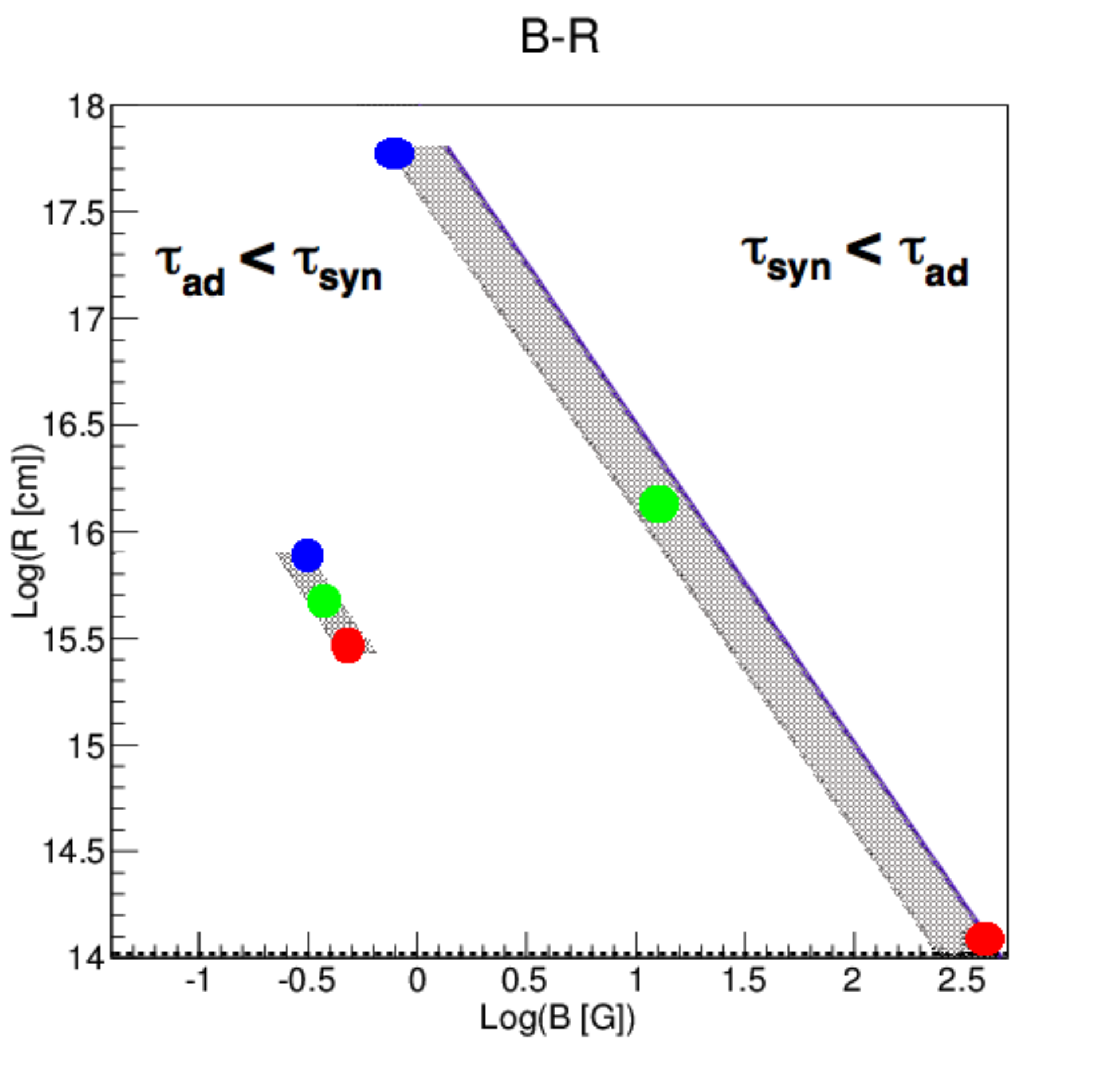}
		\includegraphics[width=220pt]{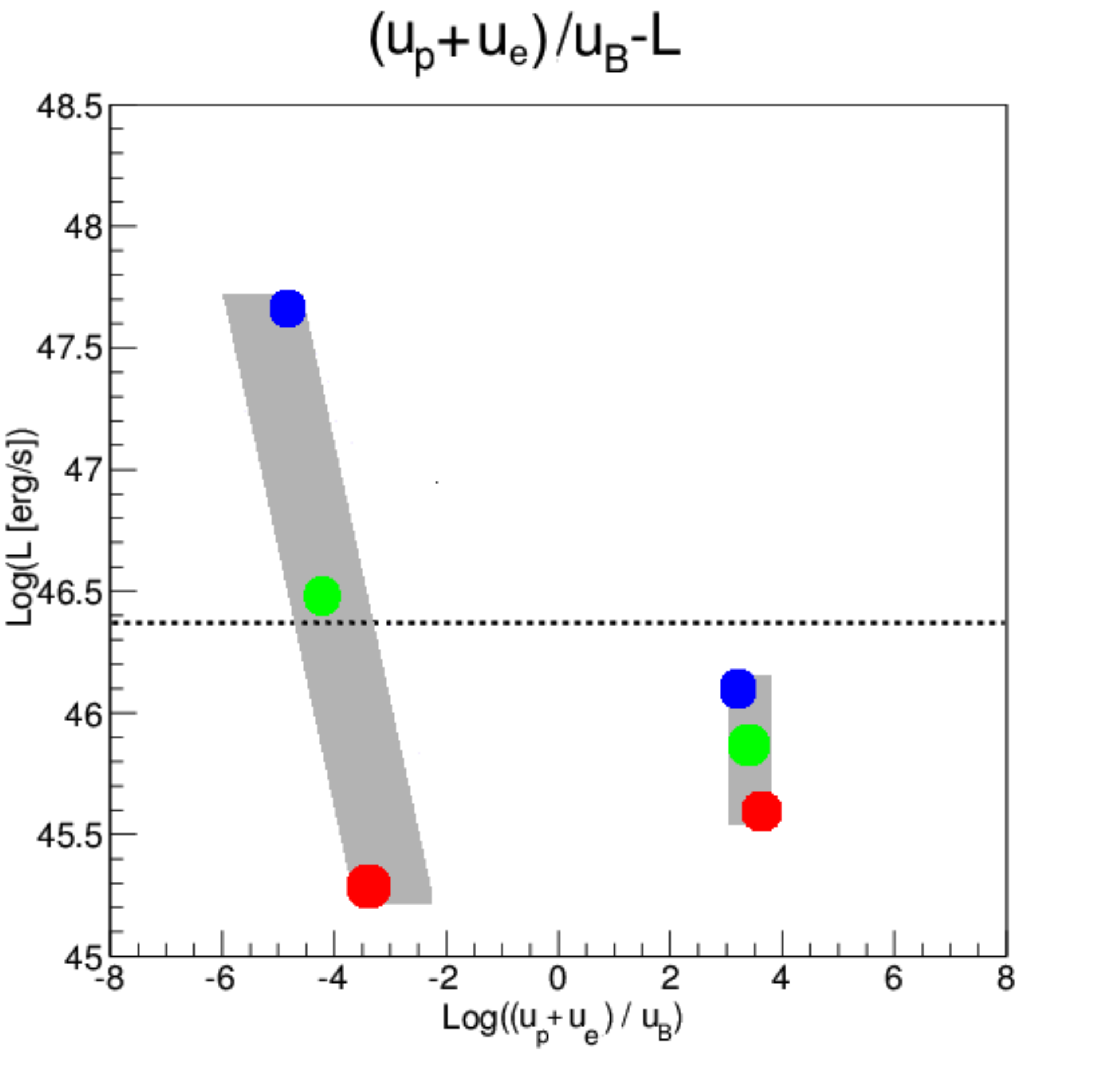}
	  \caption{\textit{Top:} modelling of the SED of RGB J0710+591, using data from \citet{0710_veritas}. \textit{left:} proton-synchrotron scenario for $(\textrm{B[G],R[cm]})=(1,6.7\times10^{17}),(21,8.2\times10^{15}),(446,1\times10^{14})$; \textit{right}: lepto-hadronic scenario for $(\textrm{B[G],R[cm]})=(0.3,8\times10^{15}),(0.4,5\times10^{15}),(0.6,3\times10^{15})$. Colours are used to identify the components corresponding to the same exemplary solutions in the $B$-$R$ parameter space. In the left plot, from lower to higher energies, the SED components are (with the same line-style as in Fig. \ref{figmrk}): electron synchrotron emission (solid lines), synchrotron emission from cascades associated with proton synchrotron emission (dotted lines), proton-synchrotron emission (solid line, high energies) and muon synchrotron emission (dashed lines). In the right plot, the visible components are:  electron synchrotron emission (solid line), proton synchrotron emission (solid lines at intermediate energies), SSC emission (dotted lines) and the sum of SSC emission and the synchrotron emission from $\pi^0$- and $\pi^\pm$-induced cascades (dashed lines). The negligible Bethe-Heitler component has not been computed to save CPU time.  For a more detailed view of all the secondary particles associated with p-$\gamma$ interactions, see Figure \ref{figmrk}. \textit{Bottom left}: representation in the $B$-$R$ plane of the two distinct regions of solutions. The solid violet line corresponds to the equality in Equations \ref{equation19}, and the nearby shadowed region represents the band of acceptable parameters for the proton-synchrotron scenario. The separate set of solutions in the bottom-left part of the plot represents the lepto-hadronic scenario. Solutions with $R \leq 10^{14}$ cm have been excluded. The three coloured dots correspond to the solutions shown in the top plots. \textit{Bottom right}: same as the previous plot, but in the $(u_p+u_e)/u_B-L$ plane. The horizontal dotted line represents the Eddington limit for $M_\bullet=10^{8.25} M_\odot$. 	  \label{fig2} }  
   \end{figure*}
   
        \begin{table*}
   \centering
   \caption{Parameters used for the hadronic modelling of our sources (proton-synchrotron scenario)}
   		\begin{tabular}{l c c c c c  }
		\hline
		\hline
		 \noalign{\smallskip}
		%&\multicolumn{5}{c}{Proton-synchrotron}&  \multicolumn{5}{c}{Lepto-hadronic}\\
		&1ES 0229+200 &1ES 0347-121  &RGB J0710+591 & 1ES 1101-232 & 1ES 1218+304\\
		 \noalign{\smallskip}
		\hline
		 \noalign{\smallskip}
	%	$\theta$ & $1^\circ$ & $1^\circ$& & & & & & & & \\
	z & 0.140 & 0.188 & 0.125 & 0.186 & 0.184  \\
		$\delta$ & $30$ & $30$&$30$  &$30$  &$30$   \\
%		$\Gamma_{bulk}$ & $16$ & $16$& & & & & & & & \\
$R_{src}$ [10$^{16}$ cm] & $0.1-68$& $0.03-65$ & $0.01-67$ & $0.1-66$ & $0.03-6.6$ \\
%		$\tau_{var} [h]$ & $0.35-240$ & $0.37-240$ & & &  \\
 \noalign{\smallskip}
		\hline
		 \noalign{\smallskip}
		$B$ [G] & $1.0-160$& $1.0-296$ & $1.0-446$ & $1.0-133$ & $3.4-454$ \\
		$^\star u_B\ [\textrm{erg cm}^{\textrm{-3}}\textrm{]}$ & $0.04-1017$& $0.04-3480$ & $0.04-7900$& $0.04-704$ &  $0.5-8210$\\
		 \noalign{\smallskip}
		\hline
		 \noalign{\smallskip}
		$\gamma_{e;min}\ [10^2]$& $1.6-20$ & $2.2-38$ & $0.01$ & $4.3-50$ &   $2.3-27$\\
		$\gamma_{e;break}\ [10^3]$& $\leq\gamma_{e,min}$ &$\leq\gamma_{e,min}$ & $0.001-0.03$&$\leq\gamma_{e,min}$ & $\leq\gamma_{e,min}$  \\
		$\gamma_{e;max}\ [10^5]$& $0.3-4.1$& $0.1-2.1$ & $0.2-3.7$ & $0.07-0.8$ & $0.04-0.5$  \\
		$\alpha_{e;1}=\alpha_{p;1}$ &$1.3$& $1.7$ & $1.35$ & $1.7$ &  $1.7$\\
		$\alpha_{e;2}=\alpha_{p;2}$ &$2.3$& $2.7$& $2.35$ & $2.7$ &  $2.7$\\
		$K_e\ \textrm{[cm}^{\textrm{-3}}\textrm{]}$ & $7.0\times10^{-8}-0.36$ & $0.05-1.2\times10^{5}$ & $7.3\times10^{-5}-1040$& $0.3-7.2\times10^{4}$ &  $3.2\times10^{-3}-8.0\times10^{4}$\\
		$^\star u_e\ [\textrm{erg cm}^{\textrm{-3}}\textrm{]}$ & $2.2\times10^{-11}-$ & $5.7\times10^{-7}$ & $1.6\times10^{-9}-$&  $2.6\times10^{-6}-$& $2.6\times10^{-8}-$\\
		& $3.2\times10^{-5}$ & $0.7$ & $2.8\times10^{-2}$& $0.4$& $0.4$\\
		 \noalign{\smallskip}
		\hline
		 \noalign{\smallskip}
		$\gamma_{p;min}$& 1& 1 &1 &1 &1 \\
		$\gamma_{p;break} [10^9]$& $2.6-57$& $2.4-56$  & $2.8-47$ & $3.7-56$ & $1.0-26$ \\
		$\gamma_{p;max} [10^9]$& $4.8-57$& $3.2-56$ & $2.8-47$ & $4.8-56$ &  $3.1-26$\\
		%$\alpha_{p,1}$ & $1.3$& $1.7$ & &  $1.7$ &  \\
		%$\alpha_{p,2}$ & $2.3$ & $2.7$  & & $2.7$ &  \\
		$\eta$ & $(9.7-19)\times10^{-6}$& $(0.7-18)\times10^{-6}$ & $(1.0-3.3)\times10^{-7}$ & $(0.2-2.6)\times10^{-6}$ & $2.1\times10^{-6}-0.02$ \\
		$^\star u_p\ \textrm{[erg cm}^{\textrm{-3}}\textrm{]}$ & $6.1\times10^{-8}-0.08$& $4.9\times10^{-7}-4.8$ & $3.0\times10^{-7}-6.4$ & $5.4\times10^{-7}-0.5$&  $5.9\times10^{-5}-3.5$ \\
		 \noalign{\smallskip}
		\hline
		 \noalign{\smallskip}
	%	$u_e/u_B$ & $35$ & & & & & & & & &  \\
   	$^\star (u_p+u_e)/u_B [10^{-5}]$ & $0.04-36$ & $0.8-540$ & $0.3-200$ & $2.5-430$& $0.9-2.2\times10^3$  \\
		$^\star L$ [10$^{45}$ erg s$^{\textrm{-1}}$ ]& $4.6-1670$& $2.1-1120$ & $1.7-460$ & $4.6-1120$  &$2.6-610$  \\
				$^\star min(\tau_{ad};\tau_{syn}(\gamma_{p;max}))\ [\textrm{hr}]$&  $12-1.2\times10^4$ &$4.3-1.2\times10^4$ &$1.9-1.2\times10^4$ & $19-1.2\times10^4$& $1.9-1.2\times10^3$\\
		%$L_{bol}$& & &\\
		 \noalign{\smallskip}
		\hline
		\hline		
		 \noalign{\bigskip}	
		\end{tabular}
	 \label{table1a}
	 \newline
		The luminosity of the emitting region has been calculated as $L=\pi R^2c\Gamma_{bulk}^2(u_B+u_e+u_p)$, where $ \Gamma_{bulk}=\delta /2$. In the last row we indicate the time-scale (in hours) which constrains $\gamma_{p;max}$, see Section \ref{phycons}. If the value of $\gamma_{e;min}$ required by the modelling is higher than the value of $\gamma_{e;break}$ estimated from synchrotron cooling, we do not report the value of the latter. The quantities flagged with a star are derived quantities and not model parameters.
		\end{table*}

    \begin{table*}
   \centering
   \caption{Parameters used for the hadronic modelling of our sources (lepto-hadronic scenario).}
   		\begin{tabular}{l c c c c c }
		\hline
		\hline
		 \noalign{\smallskip}
%		&\multicolumn{5}{c}{Proton-synchrotron}&  \multicolumn{5}{c}{Lepto-hadronic}\\
		&1ES 0229+200 &1ES 0347-121  &RGB J0710+591 & 1ES 1101-232 & 1ES 1218+304\\
		 \noalign{\smallskip}
		\hline
		 \noalign{\smallskip}
	%	$\theta$ & $1^\circ$ & $1^\circ$& & & & & & & & \\
	z & 0.140 & 0.188 & 0.125 & 0.186 & 0.184  \\
		$\delta$ & $30$ & $30$&$30$  &$30$  &$30$  \\
%		$\Gamma_{bulk}$ & $16$ & $16$& & & & & & & & \\
$R_{src}$ [10$^{16}$ cm] &  $0.1-3.2$ &$0.6-3.2$ & $0.3-0.8$ & $5.0-10$ & $0.2-0.9$\\
		%$\tau_{var} [h]$ &  $0.88-5.6$ & $0.92-5.8$ & & &  \\
		 \noalign{\smallskip}
		\hline
		 \noalign{\smallskip}
		$B$ [G] &  $0.1-1.4$ & $0.1-0.7$ & $0.3-0.6$ & $0.1-0.2$ & $0.2-1.8$ \\
		$^\star u_B\ [\textrm{erg cm}^{\textrm{-3}}\textrm{]}$ & $3.0\times 10^{-4}-0.08$ & $6.2\times 10^{-4}-0.02$ & $3.6\times 10^{-3}-0.01$ &  $(0.5-2.4)\times 10^{-3}$ & $3.4\times10^{-3}-0.1$\\
		 \noalign{\smallskip}
		\hline
		 \noalign{\smallskip}
		$\gamma_{e,min}\ [10^2]$&   $0.01$ & $47-108$ & $0.01$ & $100-150$ & $37-92$ \\
		$\gamma_{e,break}\ [10^3]$&  $3.9-31$  &  $4.7-15$ & $12-18$ & $10-15$ &  $3.7-9.2$\\
		$\gamma_{e,max}\ [10^5]$& $3.4-14$& $2.6-5.9$ & $4.6-6.5$ & $1.5-2.3$ & $0.6-1.5$\\
		$\alpha_{e,1}=\alpha_{p,1}$ &$1.3$ & $1.7$ & $1.5$ &  $1.7$ & $1.7$\\
		$\alpha_{e,2}=\alpha_{p,2}$ &$2.3$ & $2.7$ & $2.5$ & $2.7$ & $2.7$\\
		$K_e\ \textrm{[cm}^{\textrm{-3}}\textrm{]}$ & $5.6\times10^{-3}-1.5$& $2.9-135$ & $2.7-29$ & $0.5-3.5$ & $26-2190$ \\
		$^\star u_e\ [\textrm{erg cm}^{\textrm{-3}}\textrm{]}$ & $1.4\times10^{-5}-$& $4.2\times10^{-5}-$ & $1.0\times10^{-3}-$ & $7.3\times10^{-6}-$& $3.0\times10^{-4}-$\\
		&$2.2\times10^{-2}$ & $2.4\times10^{-3}$ & $9.2\times10^{-3}$& $4.9\times10^{-5}$ & $0.03$\\
		 \noalign{\smallskip}
		\hline
		 \noalign{\smallskip}
		$\gamma_{p,min}$ &1 &1 &1 &1 & 1\\
		$\gamma_{p,break} [10^9]$& $=\gamma_{p,max}$ &$=\gamma_{p,max}$ & $=\gamma_{p,max}$&$=\gamma_{p,max}$ & $=\gamma_{p,max}$ \\
		$\gamma_{p,max} [10^9]$&  $0.06-0.3$& $0.15-0.45$& $0.1-0.15$ & $0.6-1.0$ & $0.1-0.4$\\
		%$\alpha_{p,1}$ &  $1.3$ & $1.7$ & & $1.7$  & \\
		%$\alpha_{p,2}$ &  $2.3$ & $2.7$ & & $2.7$ & \\
		$\eta = K_p/K_e$ &  $0.1-0.8$ & $0.1-0.3$ & $0.04-0.1$ & $0.05-0.1$ & $0.1-0.4$\\
		$^\star u_p\ \textrm{[erg cm}^{\textrm{-3}}\textrm{]}$ &  $2.3-238$ &  $1.7-21$ & $9.3-28$ & $0.1-0.3$ & $15-270$\\
		 \noalign{\smallskip}
		\hline
		 \noalign{\smallskip}
	%	$u_e/u_B$ & $35$ & & & & & & & & &  \\
   	$^\star (u_p+u_e)/u_B  [10^{3}]$ &  $2.6-8.7$&  $0.9-4.0$ & $1.8-2.9$ & $0.1-0.2$ &  $1.1-8.5$\\
		$^\star L$ [10$^{45}$ erg s$^{\textrm{-1}}$ ]& $4-57$& $15-44$ & $3.9-12$& $17-27$ & $12-42$\\
		$^\star min(\tau_{ad};\tau_{syn}(\gamma_{p;max}))\  [\textrm{hr}]$& $19-590$ &$120-590$ &$56-150$ & $930-1.9\times10^3$& $33-180$\\
		 \noalign{\smallskip}
		\hline
		\hline
		 \noalign{\bigskip}			
		\end{tabular}
		\newline
		\label{table1b}
	         For a description, see Table \ref{table1a}.
		\end{table*}

    Lepto-hadronic modelling can reproduce well the VHE emission of the UHBLs under study, as well as the lower energy SEDs (with the caveat  on the non-simultaneous \fermilat\ data discussed in the previous Section). Compared to standard leptonic models, our solutions have the advantage of avoiding large values of the Doppler factor of the emitting region. The minimum Lorentz factor of the electron population $\gamma_{e,min}$ is of the order of $10^{2-3}$ or smaller in the proton-synchrotron scenario, in line with the values required in leptonic modelling of common HBLs. For the lepto-hadronic solutions, $\gamma_{e,min}$ can be as low as $1$ (for  1ES\, 0229+200 and RGB\, J0710+591), while it is of the order of $10^{3-4}$ for the other sources, due to constraints from the optical data. Only 1ES\, 1101-232 requires a $\gamma_{e;min}$ of 10$^{4}$ for all lepto-hadronic solutions.\\ 
   On the other hand, the only parameter that takes particularly non-trivial values for all our solutions is the spectral index of the particle population: for all sources, we need very hard injection functions ($\alpha_{p;1}=\alpha_{e;1} \in [1.3-1.5]$).    Such hard spectra are at odds with the simplest diffusive shock acceleration models, which predict injection indices close to 2.0 \citep[see e.g.][]{ODrury83, Bednarz98, Achterberg01, Protheroe04, Rieger07}. A particle index of 1.5 is usually considered as the lower limit still in agreement with simple acceleration scenarios \citep[see][]{Aha06}. However, harder values (even lower than 1.5) are still possible in certain acceleration scenarios \citep[see][]{Ellison90, Vainio03}.  Second-order Fermi acceleration may also produce hard ($\alpha<1.5$) spectra \citep[see][]{Virtanen05}. Another mechanism for efficiently accelerating particles is magnetic reconnection; in this case as well it is possible to obtain hard ($\alpha \simeq 1.5$) particle energy distributions \citep[see e.g.][]{Benoit1, Lorenzo14}.\\
    
    It should be stressed, however, that low values of  $\alpha_{e;1}$ and $\alpha_{p;1}$ are not directly constrained by observations and come from two physical assumptions detailed in Section \ref{phycons}: (i) the hypothesis that the leptonic particle population is simply cooled by synchrotron emission, and characterised by a break in the spectral index of $\Delta \Gamma=1.0$; and (ii) the hypothesis that protons and electrons are co-accelerated in magnetic fields with a single power-law spectrum of turbulence over a wide range of scales, sharing the same injection index.  Relaxing one or more of these assumptions can allow solutions with $\alpha_{p;1}\geq 1.5$ and $\alpha_{e;1}\geq 1.5$.\\
    In standard SSC modelling of HBLs, a spectral break of 1.0 is rarely observed \citep{Tavecchio10,constraints}, suggesting that acceleration and cooling in the emitting region is more complicated, with additional injection/escape terms. On the other hand, the magnetic turbulence spectrum may well be appreciably different for scales corresponding to gyroradii of TeV and EeV particles, resulting in $\alpha_{p;1} \not= \alpha_{e;1}$.\\
    Values of $\alpha_{p;1}$ softer than $1.3$-$1.5$ would provide as well a better fit of the \fermilat\ measurements, without affecting significantly the other parameters.  In summary, the hard injection spectrum does not represent a limitation of the hadronic model itself, but might point instead to the operation of mechanisms for acceleration and/or cooling beyond the simplest expectations.\\
     
     The hard particle spectra is also responsible for the difference between the hadronic modelling of UHBLs and the solution shown for \mrk\ in Section~\ref{sec3}  ($\alpha_{p;1}=1.9$): with larger values of $\alpha_{p;1}$ the number of p-$\gamma$ interactions increases, and the synchrotron emission from muons and cascades becomes more significant. In addition, the modelling of \mrk\ used a lower emission region size ($R=3.3\times10^{14}$ cm)  with a higher particle density, which also increases the number of p-$\gamma$ interactions. These two differences explain why we could not find a similar solution, dominated by muon synchrotron emission together with a very significant cascade component in the proton-synchrotron scenario for UHBLs.\\
   In the $B$-$R$ contour-plot, larger values of $\alpha_{p;1}$ would affect the proton-synchrotron parameter space by lowering the maximum value of $\nu_{peak, p}$ (see Equation \ref{equation22} and \ref{equation21b}) for given values of $R$ and $B$ and thus shrinking the width of the proton-synchrotron band of solutions. On the other hand, solutions dominated by secondary p-$\gamma$ particles would show up for higher values of $\nu_{peak,p}$ (i.e. for a larger size and stronger magnetic field, see Equation \ref{equation21}), moving the \textit{''island''} of lepto-hadronic solutions closer to the proton-synchrotron domain.\\
     
     The two main arguments generally put forward to disfavour hadronic blazar models are the energy budget of the emitting region and the constraints from short variability time-scales.\\   
    The total power of the emitting region (Equation \ref{luminosityequation}) need not necessarily be limited by the Eddington luminosity of the black hole, as is observed to be case for some narrow-line Seyfert 1 galaxies, and especially gamma-ray bursts. Nevertheless, for radio-loud AGN hosted by SMBHs with masses in the range 10$^{8-9} M_\odot$, observational estimates of jet powers generally do not show any evidence for highly super-Eddington values \citep[e.g.][]{Cavagnolo}. The Eddington luminosity for RGB\, J0710+591 (using the SMBH mass estimate provided in Table \ref{tab:uhbl}), is denoted with a dotted line in the bottom-right plot of Fig. \ref{fig2}, as well as in Appendix \ref{appa} for all the other sources. As can be seen, our solutions are all characterised by $\log(L) \in [45,48]$. While the solutions with the highest luminosities (i.e. lowest magnetic field and largest size) may be disfavored, a significant part of the hadronic solutions do not exceed the Eddington value. However, note that for the two UHBLs with the lowest SMBH mass estimates (1ES\, 0347-121 and 1ES\, 1218+304), the total luminosity for the lepto-hadronic models becomes comparable with the Eddington luminosity of the SMBH.\\
    The relatively low luminosities of our lepto-hadronic solutions are also related to the hard proton spectra: for $\alpha_{1;p} < 2.0$ the proton energy density is dominated by hadrons at $\gamma_{p;max}$. On the other hand, if $\alpha_{1;p} \geq 2.0$ the contribution of the low-energy part of the proton distribution becomes dominant in the evaluation of $u_p$, and thus $L$. Softer injection spectra, although more in agreement with simple shock acceleration scenarios, would thus have the disadvantage of significantly increasing the total power of our solutions.\\
     The energy budget of the emitting region is dominated by the magnetic field energy density $u_B$ for the proton-synchrotron scenario, and by the proton energy density $u_p$ in the lepto-hadronic scenario. The equipartition factor $(u_p+u_e)/u_B$ (which is $\simeq u_p/u_B$) is provided for all the sources in Tables \ref{table1a} and \ref{table1b}. Equipartition is often used to reduce the number of free parameters in blazar modelling, providing a specific solution characterised by an equipartition factor close to unity. An equipartition scenario is appealing in particular because it provides the minimum power of the emitting region, and has been successfully applied in leptonic blazar modelling, when adding additional
 external photon fields \citep[see e.g.][]{Chuck1}. None of our models is close to equipartition: the best cases are $(u_p+u_e)/u_B \simeq 0.02$ and $\simeq 100$ for the proton synchrotron and the lepto-hadronic scenario, respectively.  However, this is also the case for widely discussed SSC models for HBLs \citep[see e.g][]{constraints}.  \\
     
    Concerning variability, hadronic models are often disfavored when rapid flares are detected, in view of the typically long cooling times expected for the parent protons \citep[see however][]{Barkov12}. For the proton-synchrotron and the lepto-hadronic scenarios the associated variability time-scales (at VHE) are respectively the proton synchrotron cooling or the development of synchrotron-pair cascades, which depends on the time-scale of proton-photon interactions. The fastest variability time-scale may be achieved for the proton-synchrotron solutions with the highest B value: in this case we can have variability of the order of hours in the source frame. In general we have however time-scales of the order of days, which cannot account for minute-long flares, as shown by \citet{2155flare} or \citet{BLLacflare}. With the exception of 1ES\, 1218+304, none of the sources under study showed any significant $\gamma$-ray flare: they are thus not in conflict with a hadronic origin of the VHE emission. The flare of 1ES\, 1218+304
was characterized by a time-scale of a day, which is still in agreement with the expected time-scales from hadronic models. 

 In general, in the proton-synchrotron scenario, we would expect a different temporal behaviour between the low-energy and high-energy components, with time lags due to the 
different acceleration and cooling time-scales of the leptonic and hadronic particle populations. The lepto-hadronic scenarios could exhibit different variability patterns 
even within the (very) high-energy range, due to the different contributions from the SSC and cascade components. The SSC component, more dominant in the 
\fermilat\ energy range, would vary simultaneously with the electron-synchrotron emission.\\
 A closer evaluation of variability would require detailed time-dependent modelling of the hadronic blazar emission, which is a complex problem, requiring a Monte-Carlo study of the p-$\gamma$ interactions and evolution of the associated cascades. Only recently \citet{Dimi12,Mastichiadis13, Weidinger13} presented the first results of time-dependent hadronic codes.\\  
     
   One of the open questions in blazar physics is the location of the $\gamma$-ray emitting region. For FSRQs and LBLs, in which the external photon field is important and the high-energy emission is dominated by EIC components, it is possible to constrain the location of the emitting region with respect to the SMBH environment, in particular the broad-line region and the accretion disk \citep[see][]{chuck2, Chuck1}. For HBLs it is more difficult to estimate the location of the emitting region. One possibility is to assume a specific jet structure (conical) and that the emitting region fills the entire jet section. In this case $R$ can be expressed as a function of the distance $r$ from the central SMBH \citep[see][]{Ghisellini85, Moderski03, Potter12}. For the models presented here, assuming a conical jet with aperture angle $\psi \simeq 10^\circ$ and $r\simeq R/\tan{\psi}$, we can estimate $r\simeq 40-2000\ R_G$ for the lepto-hadronic solutions, while for the proton-synchrotron models the location of the $\gamma$-ray emitting region is constrained to an even lesser degree, between $40$ and $2.5\times10^4\ R_G$ (the lower values corresponding to lower power and closer to equipartition, with a limit at 40 $R_G$ due to our arbitrary choice of the minimum size of the emission region at 10$^{14-15}$\,cm).\\ 
   
    To reduce the number of free parameters of the models, we have related the maximum proton energy to the acceleration and cooling time-scales, and thus the magnetic field $B$ and the size $R$. The maximum electron energy, on the other hand, is constrained by the X-ray observations, which sample the peak of the low-energy component of the SED. An important information on the acceleration mechanism can indeed be extracted from the ratio $\gamma_{p;max}/\gamma_{e;max}$: if the acceleration takes place in the Bohm diffusion regime at all energies, we expect $\gamma_{p;max}/\gamma_{e;max} = m_p/m_e$ (or lower, if the maximum proton energy is determined by adiabatic losses instead of synchrotron losses). All our models are instead characterized by much higher ratios, which may indicate the presence of Kolmogorov or Kraichnan turbulence excited predominantly by the maximum energy protons. This is consistent with earlier studies, as discussed in \citet[][and references therein]{Biermann, Kolmo, Kraichnan, Mucke01}.\\
   
   One of the main goals of AGN hadronic modeling is to study a potential link between photons and cosmic-rays or neutrinos detected at Earth. In particular, the detection in blazar spectra of emission associated  with energetic protons would provide strong indication for AGN as the source of extra-galactic cosmic rays  or high-energy neutrinos. In all our models, the maximum proton energy is around 10$^{18-19}$ eV (for the 
   proton-synchrotron scenario, and one order of magnitude lower for the lepto-hadronic one). Under the simple hypothesis for the acceleration time-scale considered in this work, $\gamma_{p;max}$ would thus not be sufficient to explain the most energetic cosmic rays (up to 10$^{20}$ eV) measured on Earth. On the other hand, these energies may still suffice for generating neutrinos in the PeV energy range that were recently discovered by the IceCube observatory. Further discussions of such implications are beyond the scope of this paper.\\
   
   In recent years several authors \citep[e.g.][]{Essey10, Dermer12, Murase12, Tavecchio14}  have studied the possibility that the blazar $\gamma$-ray emission is not produced at the source but rather along the path from the AGN to the observer. Observed $\gamma$-rays could be due to the interactions of cosmic-rays with the EBL and the CMB, and explain the detection of hard and distant blazars with IACTs as long as the intergalactic magnetic fields along the cosmic-ray propagation path are sufficiently weak. In our work, we did not consider this component, and implicitly made the assumption that the emission at the source dominates over every other emission. Note that such contributions are expected to dominate for distant blazars, while all our sources are located at $z < 0.2$.  Interestingly, there are no models in the literature studying both components (hadronic emission in the source and UHECR cascade in the path towards the observer); a self-consistent model linking the two contribution would be a natural evolution of the present code, and will be the object of further studies.\\ 
   
   The observation of anomalous spectral components in the GeV-TeV energy band could represent  a signature to distinguish hadronic models from (one-zone) SSC scenarios. As can be seen in Fig. \ref{figmrk}, synchrotron emission by secondary cascades can produce a spectral hardening in the observed TeV spectrum, which would be hard to explain by standard SSC models. There is  already some evidence for a hardening of the TeV emission seen in the VHE spectra of BL Lac objects \citep{Horns12, Inoue13}. Even though the current generation of Cherenkov telescopes probably would not be able to confirm this hard excess, such a feature could be clearly measurable with the future Cherenkov Telescope Array (CTA) \citep{CTA, Sol}, as shown by \citet{ICRC2013}.  A systematic study of the parameter space to estimate the strength of this ''cascade-bump'' as a function of the model parameters and the perspectives for CTA observations is currently in preparation and will be discussed in a separate publication.\\
   
    Given the constraints described in Section~\ref{phycons}, the (lepto-) hadronic modelling of UHBLs does not result in universal spectral features, but the coexistence of 
several spectral components at high energies in the lepto-hadronic scenario could lead to discernible shapes in certain cases. The emergence of additional components at
intermediate energies --- from cascade emission and proton-synchrotron photons (for low $B$-fields) --- is another signature of these models, although difficult to detect due
to a lack of instrumental coverage with sufficient sensitivity at energies below the \fermilat\ band and above current X-ray telescopes. A combination of spectral and timing analysis with multi-wavelength data, including the large energy coverage at (very) high energies of CTA, will be the most promising approach to probe the
different emission scenarios of these sources.
   
   \section{Conclusions}
   \label{sec6}
  We propose an alternative interpretation of the SEDs of the currently known
UHBLs, also known as 
EHBLs. Within a  lepto-hadronic framework,  their SEDs can be explained
without the need for extreme Doppler factors (in all the models $\delta$ was fixed equal to $30$) or minimum Lorentz factors
of the radiating particle distribution (which is around $10^{2-3}$, two order of magnitudes lower than that required by SSC models). A caveat of this interpretation is the very hard particle injection spectra required under the assumption of simple schemes for electron cooling and co-acceleration of leptons and protons, which may point to the operation of non-trivial physical mechanisms for such processes. \\

In a systematic approach, we have identified, for the first time, the complete
parameter space in which hadronic interpretations of the high-energy bumps
for the known UHBLs can be found. These solutions are grouped into two 
distinct regions in the B-R parameter space corresponding to the dominance
of either proton-synchrotron or p-$\gamma$-induced cascade emission.\\

The hadronic solutions proposed here lead to values for the jet power that are generally sub-Eddington, in contrast to previous hadronic interpretations
for the emission from other classes of blazars that often required highly super-Eddington values. On the other hand, 
they are several orders of magnitude away from equipartition between the
proton energy density and magnetic field energy density. In our framework,
acceleration to the highest observed energies of ultra-high energy cosmic
rays is not expected for the sources under study, although they may possibly be of interest for the recently discovered high-energy neutrinos.\\
   
   \section*{Acknowledgments} 
   This work has made use of the computing capabilities of the Paris Observatory and the Harvard-Smithsonian Center for Astrophysics. The authors wish to thank Anita Reimer for providing the latest version of the \textit{SOPHIA} code, as well as H\'{e}l\`{e}ne Sol for fruitful discussions, and the referee, Michael Zacharias, for the detailed report and the useful comments and remarks. The preliminary results of this project have been presented at several conferences and workshops: all the inputs from colleagues have significantly improved the present work, and are warmly acknowledged.  
    
   \bibliographystyle{mn2e}
   \bibliography{leha_mnras}

\begin{thebibliography}{}
\makeatletter
\relax
\def\mn@urlcharsother{\let\do\@makeother \do\$\do\&\do\#\do\^\do\_\do\%\do\~}
\def\mn@doi{\begingroup\mn@urlcharsother \@ifnextchar [ {\mn@doi@}
  {\mn@doi@[]}}
\def\mn@doi@[#1]#2{\def\@tempa{#1}\ifx\@tempa\@empty \href
  {http://dx.doi.org/#2} {doi:#2}\else \href {http://dx.doi.org/#2} {#1}\fi
  \endgroup}
\def\mn@eprint#1#2{\mn@eprint@#1:#2::\@nil}
\def\mn@eprint@arXiv#1{\href {http://arxiv.org/abs/#1} {{\tt arXiv:#1}}}
\def\mn@eprint@dblp#1{\href {http://dblp.uni-trier.de/rec/bibtex/#1.xml}
  {dblp:#1}}
\def\mn@eprint@#1:#2:#3:#4\@nil{\def\@tempa {#1}\def\@tempb {#2}\def\@tempc
  {#3}\ifx \@tempc \@empty \let \@tempc \@tempb \let \@tempb \@tempa \fi \ifx
  \@tempb \@empty \def\@tempb {arXiv}\fi \@ifundefined
  {mn@eprint@\@tempb}{\@tempb:\@tempc}{\expandafter \expandafter \csname
  mn@eprint@\@tempb\endcsname \expandafter{\@tempc}}}

\bibitem[\protect\citeauthoryear{{Abdo} et~al.,}{{Abdo}
  et~al.}{2010}]{FermiSED}
{Abdo} A.~A.,  et~al., 2010, \mn@doi [ApJ] {10.1088/0004-637X/716/1/30}, \href
  {http://adsabs.harvard.edu/abs/2010ApJ...716...30A} {716, 30}

\bibitem[\protect\citeauthoryear{{Abdo} et~al.,}{{Abdo} et~al.}{2011}]{Mrk421}
{Abdo} A.~A.,  et~al., 2011, \mn@doi [ApJ] {10.1088/0004-637X/736/2/131}, \href
  {http://adsabs.harvard.edu/abs/2011ApJ...736..131A} {736, 131}

\bibitem[\protect\citeauthoryear{{Acciari} et~al.,}{{Acciari}
  et~al.}{2009}]{Mrk421VERITAS}
{Acciari} V.~A.,  et~al., 2009, \mn@doi [ApJ] {10.1088/0004-637X/703/1/169},
  \href {http://adsabs.harvard.edu/abs/2009ApJ...703..169A} {703, 169}

\bibitem[\protect\citeauthoryear{{Acciari} et~al.,}{{Acciari}
  et~al.}{2010a}]{1218_Veritas}
{Acciari} V.~A.,  et~al., 2010a, \mn@doi [ApJL] {10.1088/2041-8205/709/2/L163},
  \href {http://adsabs.harvard.edu/abs/2010ApJ...709L.163A} {709, L163}

\bibitem[\protect\citeauthoryear{{Acciari} et~al.,}{{Acciari}
  et~al.}{2010b}]{0710_veritas}
{Acciari} V.~A.,  et~al., 2010b, \mn@doi [ApJL] {10.1088/2041-8205/715/1/L49},
  \href {http://adsabs.harvard.edu/abs/2010ApJ...715L..49A} {715, L49}

\bibitem[\protect\citeauthoryear{{Achterberg}, {Gallant}, {Kirk}  \&
  {Guthmann}}{{Achterberg} et~al.}{2001}]{Achterberg01}
{Achterberg} A.,  {Gallant} Y.~A.,  {Kirk} J.~G.,   {Guthmann} A.~W.,  2001,
  \mn@doi [MNRAS] {10.1046/j.1365-8711.2001.04851.x}, \href
  {http://adsabs.harvard.edu/abs/2001MNRAS.328..393A} {328, 393}

\bibitem[\protect\citeauthoryear{{Ackermann} et~al.,}{{Ackermann}
  et~al.}{2013a}]{1FHL}
{Ackermann} M.,  et~al., 2013a, \mn@doi [ApJS] {10.1088/0067-0049/209/2/34},
  \href {http://adsabs.harvard.edu/abs/2013ApJS..209...34A} {209, 34}

\bibitem[\protect\citeauthoryear{{Ackermann} et~al.,}{{Ackermann}
  et~al.}{2013b}]{Ackermann13}
{Ackermann} M.,  et~al., 2013b, \mn@doi [Science] {10.1126/science.1231160},
  \href {http://adsabs.harvard.edu/abs/2013Sci...339..807A} {339, 807}

\bibitem[\protect\citeauthoryear{{Actis} et~al.,}{{Actis} et~al.}{2011}]{CTA}
{Actis} M.,  et~al., 2011, \mn@doi [Experimental Astronomy]
  {10.1007/s10686-011-9247-0}, \href
  {http://adsabs.harvard.edu/abs/2011ExA....32..193A} {32, 193}

\bibitem[\protect\citeauthoryear{{Aharonian}, {Atoian}  \&
  {Nagapetian}}{{Aharonian} et~al.}{1983}]{Aha83}
{Aharonian} F.~A.,  {Atoian} A.~M.,   {Nagapetian} A.~M.,  1983, Astrofizika,
  \href {http://adsabs.harvard.edu/abs/1983Afz....19..323A} {19, 323}

\bibitem[\protect\citeauthoryear{{Aharonian} et~al.,}{{Aharonian}
  et~al.}{2002}]{1426_hegra}
{Aharonian} F.,  et~al., 2002, \mn@doi [A\&A] {10.1051/0004-6361:20020206},
  \href {http://adsabs.harvard.edu/abs/2002A%26A...384L..23A} {384, L23}

\bibitem[\protect\citeauthoryear{{Aharonian} et~al.,}{{Aharonian}
  et~al.}{2005}]{Mrk421HESS}
{Aharonian} F.,  et~al., 2005, \mn@doi [A\&A] {10.1051/0004-6361:20053050},
  \href {http://adsabs.harvard.edu/abs/2005A%26A...437...95A} {437, 95}

\bibitem[\protect\citeauthoryear{{Aharonian} et~al.,}{{Aharonian}
  et~al.}{2006}]{Aha06}
{Aharonian} F.,  et~al., 2006, \mn@doi [Nature] {10.1038/nature04680}, \href
  {http://adsabs.harvard.edu/abs/2006Natur.440.1018A} {440, 1018}

\bibitem[\protect\citeauthoryear{{Aharonian} et~al.,}{{Aharonian}
  et~al.}{2007a}]{1101_HESS}
{Aharonian} F.,  et~al., 2007a, \mn@doi [A\&A] {10.1051/0004-6361:20077057},
  \href {http://adsabs.harvard.edu/abs/2007A%26A...470..475A} {470, 475}

\bibitem[\protect\citeauthoryear{{Aharonian} et~al.,}{{Aharonian}
  et~al.}{2007b}]{0347_HESS}
{Aharonian} F.,  et~al., 2007b, \mn@doi [A\&A] {10.1051/0004-6361:20078412},
  \href {http://adsabs.harvard.edu/abs/2007A%26A...473L..25A} {473, L25}

\bibitem[\protect\citeauthoryear{{Aharonian} et~al.,}{{Aharonian}
  et~al.}{2007c}]{2155flare}
{Aharonian} F.,  et~al., 2007c, \mn@doi [ApJL] {10.1086/520635}, \href
  {http://adsabs.harvard.edu/abs/2007ApJ...664L..71A} {664, L71}

\bibitem[\protect\citeauthoryear{{Aharonian}, {Khangulyan}  \&
  {Costamante}}{{Aharonian} et~al.}{2008}]{Aha08}
{Aharonian} F.~A.,  {Khangulyan} D.,   {Costamante} L.,  2008, \mn@doi [MNRAS]
  {10.1111/j.1365-2966.2008.13315.x}, \href
  {http://adsabs.harvard.edu/abs/2008MNRAS.387.1206A} {387, 1206}

\bibitem[\protect\citeauthoryear{{Ahn} et~al.,}{{Ahn} et~al.}{2012}]{Ahn12}
{Ahn} C.~P.,  et~al., 2012, \mn@doi [ApJS] {10.1088/0067-0049/203/2/21}, \href
  {http://adsabs.harvard.edu/abs/2012ApJS..203...21A} {203, 21}

\bibitem[\protect\citeauthoryear{{Albert} et~al.,}{{Albert}
  et~al.}{2007}]{Mrk421MAGIC}
{Albert} J.,  et~al., 2007, \mn@doi [ApJ] {10.1086/518221}, \href
  {http://adsabs.harvard.edu/abs/2007ApJ...663..125A} {663, 125}

\bibitem[\protect\citeauthoryear{{Aliu} et~al.,}{{Aliu}
  et~al.}{2014}]{0229_Veritas}
{Aliu} E.,  et~al., 2014, \mn@doi [ApJ] {10.1088/0004-637X/782/1/13}, \href
  {http://adsabs.harvard.edu/abs/2014ApJ...782...13A} {782, 13}

\bibitem[\protect\citeauthoryear{{Arlen} et~al.,}{{Arlen}
  et~al.}{2013}]{BLLacflare}
{Arlen} T.,  et~al., 2013, \mn@doi [ApJ] {10.1088/0004-637X/762/2/92}, \href
  {http://adsabs.harvard.edu/abs/2013ApJ...762...92A} {762, 92}

\bibitem[\protect\citeauthoryear{{Asano}, {Takahara}, {Kusunose}, {Toma}  \&
  {Kakuwa}}{{Asano} et~al.}{2014}]{Asano14}
{Asano} K.,  {Takahara} F.,  {Kusunose} M.,  {Toma} K.,   {Kakuwa} J.,  2014,
  \mn@doi [ApJ] {10.1088/0004-637X/780/1/64}, \href
  {http://adsabs.harvard.edu/abs/2014ApJ...780...64A} {780, 64}

\bibitem[\protect\citeauthoryear{{Barkov}, {Aharonian}  \&
  {Bosch-Ramon}}{{Barkov} et~al.}{2010}]{Barkov10}
{Barkov} M.~V.,  {Aharonian} F.~A.,   {Bosch-Ramon} V.,  2010, \mn@doi [ApJ]
  {10.1088/0004-637X/724/2/1517}, \href
  {http://adsabs.harvard.edu/abs/2010ApJ...724.1517B} {724, 1517}

\bibitem[\protect\citeauthoryear{{Barkov}, {Aharonian}, {Bogovalov}, {Kelner}
  \& {Khangulyan}}{{Barkov} et~al.}{2012}]{Barkov12}
{Barkov} M.~V.,  {Aharonian} F.~A.,  {Bogovalov} S.~V.,  {Kelner} S.~R.,
  {Khangulyan} D.,  2012, \mn@doi [ApJ] {10.1088/0004-637X/749/2/119}, \href
  {http://adsabs.harvard.edu/abs/2012ApJ...749..119B} {749, 119}

\bibitem[\protect\citeauthoryear{{Bednarz} \& {Ostrowski}}{{Bednarz} \&
  {Ostrowski}}{1998}]{Bednarz98}
{Bednarz} J.,  {Ostrowski} M.,  1998, \mn@doi [Physical Review Letters]
  {10.1103/PhysRevLett.80.3911}, \href
  {http://adsabs.harvard.edu/abs/1998PhRvL..80.3911B} {80, 3911}

\bibitem[\protect\citeauthoryear{{Benbow}}{{Benbow}}{2011}]{wystan}
{Benbow} W.,  2011, Proc. of the 32nd ICRC, astro-ph/1110.0038, \href
  {http://adsabs.harvard.edu/abs/2011arXiv1110.0038W} {}

\bibitem[\protect\citeauthoryear{{Biermann} \& {Strittmatter}}{{Biermann} \&
  {Strittmatter}}{1987}]{Biermann}
{Biermann} P.~L.,  {Strittmatter} P.~A.,  1987, \mn@doi [ApJ] {10.1086/165759},
  \href {http://adsabs.harvard.edu/abs/1987ApJ...322..643B} {322, 643}

\bibitem[\protect\citeauthoryear{{Blumenthal}}{{Blumenthal}}{1970}]{Blumenthal70}
{Blumenthal} G.~R.,  1970, \mn@doi [PhRD] {10.1103/PhysRevD.1.1596}, \href
  {http://adsabs.harvard.edu/abs/1970PhRvD...1.1596B} {1, 1596}

\bibitem[\protect\citeauthoryear{{B{\"o}ttcher}, {Reimer}, {Sweeney}  \&
  {Prakash}}{{B{\"o}ttcher} et~al.}{2013}]{Markus13}
{B{\"o}ttcher} M.,  {Reimer} A.,  {Sweeney} K.,   {Prakash} A.,  2013, \mn@doi
  [ApJ] {10.1088/0004-637X/768/1/54}, \href
  {http://adsabs.harvard.edu/abs/2013ApJ...768...54B} {768, 54}

\bibitem[\protect\citeauthoryear{{Cao} \& {Wang}}{{Cao} \&
  {Wang}}{2014}]{Cao14}
{Cao} G.,  {Wang} J.,  2014, \mn@doi [ApJ] {10.1088/0004-637X/783/2/108}, \href
  {http://adsabs.harvard.edu/abs/2014ApJ...783..108C} {783, 108}

\bibitem[\protect\citeauthoryear{{Cavagnolo}, {McNamara}, {Nulsen}, {Carilli},
  {Jones}  \& {B{\^i}rzan}}{{Cavagnolo} et~al.}{2010}]{Cavagnolo}
{Cavagnolo} K.~W.,  {McNamara} B.~R.,  {Nulsen} P.~E.~J.,  {Carilli} C.~L.,
  {Jones} C.,   {B{\^i}rzan} L.,  2010, \mn@doi [ApJ]
  {10.1088/0004-637X/720/2/1066}, \href
  {http://adsabs.harvard.edu/abs/2010ApJ...720.1066C} {720, 1066}

\bibitem[\protect\citeauthoryear{{Cerruti}, {Boisson}  \& {Zech}}{{Cerruti}
  et~al.}{2013a}]{constraints}
{Cerruti} M.,  {Boisson} C.,   {Zech} A.,  2013a, \mn@doi [A\&A]
  {10.1051/0004-6361/201220963}, \href
  {http://adsabs.harvard.edu/abs/2013A%26A...558A..47C} {558, A47}

\bibitem[\protect\citeauthoryear{{Cerruti}, {Dermer}, {Lott}, {Boisson}  \&
  {Zech}}{{Cerruti} et~al.}{2013b}]{chuck2}
{Cerruti} M.,  {Dermer} C.~D.,  {Lott} B.,  {Boisson} C.,   {Zech} A.,  2013b,
  \mn@doi [ApJL] {10.1088/2041-8205/771/1/L4}, \href
  {http://adsabs.harvard.edu/abs/2013ApJ...771L...4C} {771, L4}

\bibitem[\protect\citeauthoryear{{Cerutti}, {Uzdensky}  \&
  {Begelman}}{{Cerutti} et~al.}{2012}]{Benoit1}
{Cerutti} B.,  {Uzdensky} D.~A.,   {Begelman} M.~C.,  2012, \mn@doi [ApJ]
  {10.1088/0004-637X/746/2/148}, \href
  {http://adsabs.harvard.edu/abs/2012ApJ...746..148C} {746, 148}

\bibitem[\protect\citeauthoryear{{Chodorowski}, {Zdziarski}  \&
  {Sikora}}{{Chodorowski} et~al.}{1992}]{Chodorowski92}
{Chodorowski} M.~J.,  {Zdziarski} A.~A.,   {Sikora} M.,  1992, \mn@doi [ApJ]
  {10.1086/171984}, \href {http://adsabs.harvard.edu/abs/1992ApJ...400..181C}
  {400, 181}

\bibitem[\protect\citeauthoryear{{Costamante} et~al.,}{{Costamante}
  et~al.}{2001}]{Costamante01}
{Costamante} L.,  et~al., 2001, \mn@doi [A\&A] {10.1051/0004-6361:20010412},
  \href {http://cdsads.u-strasbg.fr/abs/2001A%26A...371..512C} {371, 512}

\bibitem[\protect\citeauthoryear{{Dermer} \& {Atoyan}}{{Dermer} \&
  {Atoyan}}{2001}]{Dermer01}
{Dermer} C.~D.,  {Atoyan} A.,  2001, Proc. of the 27th ICRC, astro-ph/0107200,
  \href {http://adsabs.harvard.edu/abs/2001astro.ph..7200D} {}

\bibitem[\protect\citeauthoryear{{Dermer}, {Cavadini}, {Razzaque}, {Finke},
  {Chiang}  \& {Lott}}{{Dermer} et~al.}{2011}]{Dermer11}
{Dermer} C.~D.,  {Cavadini} M.,  {Razzaque} S.,  {Finke} J.~D.,  {Chiang} J.,
  {Lott} B.,  2011, \mn@doi [ApJL] {10.1088/2041-8205/733/2/L21}, \href
  {http://adsabs.harvard.edu/abs/2011ApJ...733L..21D} {733, L21}

\bibitem[\protect\citeauthoryear{{Dermer}, {Murase}  \& {Takami}}{{Dermer}
  et~al.}{2012}]{Dermer12}
{Dermer} C.~D.,  {Murase} K.,   {Takami} H.,  2012, \mn@doi [ApJ]
  {10.1088/0004-637X/755/2/147}, \href
  {http://adsabs.harvard.edu/abs/2012ApJ...755..147D} {755, 147}

\bibitem[\protect\citeauthoryear{{Dermer}, {Cerruti}, {Lott}, {Boisson}  \&
  {Zech}}{{Dermer} et~al.}{2014}]{Chuck1}
{Dermer} C.~D.,  {Cerruti} M.,  {Lott} B.,  {Boisson} C.,   {Zech} A.,  2014,
  \mn@doi [ApJ] {10.1088/0004-637X/782/2/82}, \href
  {http://adsabs.harvard.edu/abs/2014ApJ...782...82D} {782, 82}

\bibitem[\protect\citeauthoryear{{Dimitrakoudis}, {Mastichiadis}, {Protheroe}
  \& {Reimer}}{{Dimitrakoudis} et~al.}{2012}]{Dimi12}
{Dimitrakoudis} S.,  {Mastichiadis} A.,  {Protheroe} R.~J.,   {Reimer} A.,
  2012, \mn@doi [A\&A] {10.1051/0004-6361/201219770}, \href
  {http://adsabs.harvard.edu/abs/2012A%26A...546A.120D} {546, A120}

\bibitem[\protect\citeauthoryear{{Djannati-Ata{\"i}}
  et~al.,}{{Djannati-Ata{\"i}} et~al.}{2002}]{Arache}
{Djannati-Ata{\"i}} A.,  et~al., 2002, \mn@doi [A\&A]
  {10.1051/0004-6361:20021034}, \href
  {http://adsabs.harvard.edu/abs/2002A%26A...391L..25D} {391, L25}

\bibitem[\protect\citeauthoryear{{Drury}}{{Drury}}{1983}]{ODrury83}
{Drury} L.~O.,  1983, \mn@doi [Reports on Progress in Physics]
  {10.1088/0034-4885/46/8/002}, \href
  {http://adsabs.harvard.edu/abs/1983RPPh...46..973D} {46, 973}

\bibitem[\protect\citeauthoryear{{Ellison}, {Reynolds}  \& {Jones}}{{Ellison}
  et~al.}{1990}]{Ellison90}
{Ellison} D.~C.,  {Reynolds} S.~P.,   {Jones} F.~C.,  1990, \mn@doi [ApJ]
  {10.1086/169156}, \href {http://adsabs.harvard.edu/abs/1990ApJ...360..702E}
  {360, 702}

\bibitem[\protect\citeauthoryear{{Essey} \& {Kusenko}}{{Essey} \&
  {Kusenko}}{2010}]{Essey10}
{Essey} W.,  {Kusenko} A.,  2010, \mn@doi [Astroparticle Physics]
  {10.1016/j.astropartphys.2009.11.007}, \href
  {http://adsabs.harvard.edu/abs/2010APh....33...81E} {33, 81}

\bibitem[\protect\citeauthoryear{{Falomo}, {Carangelo}  \& {Treves}}{{Falomo}
  et~al.}{2003}]{Falomo03}
{Falomo} R.,  {Carangelo} N.,   {Treves} A.,  2003, \mn@doi [MNRAS]
  {10.1046/j.1365-8711.2003.06690.x}, \href
  {http://cdsads.u-strasbg.fr/abs/2003MNRAS.343..505F} {343, 505}

\bibitem[\protect\citeauthoryear{{Franceschini}, {Rodighiero}  \&
  {Vaccari}}{{Franceschini} et~al.}{2008}]{Franceschini}
{Franceschini} A.,  {Rodighiero} G.,   {Vaccari} M.,  2008, \mn@doi [A\&A]
  {10.1051/0004-6361:200809691}, \href
  {http://adsabs.harvard.edu/abs/2008A%26A...487..837F} {487, 837}

\bibitem[\protect\citeauthoryear{{Ghisellini}, {Maraschi}  \&
  {Treves}}{{Ghisellini} et~al.}{1985}]{Ghisellini85}
{Ghisellini} G.,  {Maraschi} L.,   {Treves} A.,  1985, A\&A, \href
  {http://adsabs.harvard.edu/abs/1985A%26A...146..204G} {146, 204}

\bibitem[\protect\citeauthoryear{{H.E.S.S.~Collaboration}
  et~al.,}{{H.E.S.S.~Collaboration} et~al.}{2012}]{101015}
{H.E.S.S.~Collaboration} et~al., 2012, \mn@doi [A\&A]
  {10.1051/0004-6361/201218910}, \href
  {http://adsabs.harvard.edu/abs/2012A%26A...542A..94H} {542, A94}

\bibitem[\protect\citeauthoryear{{H.E.S.S.~Collaboration}
  et~al.,}{{H.E.S.S.~Collaboration} et~al.}{2013}]{HESSEBL}
{H.E.S.S.~Collaboration} et~al., 2013, \mn@doi [A\&A]
  {10.1051/0004-6361/201220355}, \href
  {http://adsabs.harvard.edu/abs/2013A%26A...550A...4H} {550, A4}

\bibitem[\protect\citeauthoryear{{Henri} \& {Saug{\'e}}}{{Henri} \&
  {Saug{\'e}}}{2006}]{Henri06}
{Henri} G.,  {Saug{\'e}} L.,  2006, \mn@doi [ApJ] {10.1086/500039}, \href
  {http://adsabs.harvard.edu/abs/2006ApJ...640..185H} {640, 185}

\bibitem[\protect\citeauthoryear{{Horan} et~al.,}{{Horan}
  et~al.}{2002}]{1426_whipple2}
{Horan} D.,  et~al., 2002, \mn@doi [ApJ] {10.1086/340019}, \href
  {http://adsabs.harvard.edu/abs/2002ApJ...571..753H} {571, 753}

\bibitem[\protect\citeauthoryear{{Horns} \& {Meyer}}{{Horns} \&
  {Meyer}}{2012}]{Horns12}
{Horns} D.,  {Meyer} M.,  2012, \mn@doi [JCAP] {10.1088/1475-7516/2012/02/033},
  \href {http://adsabs.harvard.edu/abs/2012JCAP...02..033H} {2, 33}

\bibitem[\protect\citeauthoryear{{IceCube Collaboration}}{{IceCube
  Collaboration}}{2013}]{IceCube}
{IceCube Collaboration} 2013, \mn@doi [Science] {10.1126/science.1242856},
  \href {http://adsabs.harvard.edu/abs/2013Sci...342E...1I} {342, 1}

\bibitem[\protect\citeauthoryear{{Inoue} \& {Takahara}}{{Inoue} \&
  {Takahara}}{1996}]{Susumu}
{Inoue} S.,  {Takahara} F.,  1996, \mn@doi [ApJ] {10.1086/177270}, \href
  {http://adsabs.harvard.edu/abs/1996ApJ...463..555I} {463, 555}

\bibitem[\protect\citeauthoryear{{Inoue}, {Inoue}, {Kobayashi}, {Makiya},
  {Niino}  \& {Totani}}{{Inoue} et~al.}{2013}]{Inoue13}
{Inoue} Y.,  {Inoue} S.,  {Kobayashi} M.~A.~R.,  {Makiya} R.,  {Niino} Y.,
  {Totani} T.,  2013, \mn@doi [ApJ] {10.1088/0004-637X/768/2/197}, \href
  {http://adsabs.harvard.edu/abs/2013ApJ...768..197I} {768, 197}

\bibitem[\protect\citeauthoryear{{Jones}}{{Jones}}{1968}]{Jones68}
{Jones} F.~C.,  1968, \mn@doi [Physical Review] {10.1103/PhysRev.167.1159},
  \href {http://adsabs.harvard.edu/abs/1968PhRv..167.1159J} {167, 1159}

\bibitem[\protect\citeauthoryear{{Katarzy{\'n}ski}, {Sol}  \&
  {Kus}}{{Katarzy{\'n}ski} et~al.}{2001}]{Kata01}
{Katarzy{\'n}ski} K.,  {Sol} H.,   {Kus} A.,  2001, \mn@doi [A\&A]
  {10.1051/0004-6361:20000538}, \href
  {http://adsabs.harvard.edu/abs/2001A%26A...367..809K} {367, 809}

\bibitem[\protect\citeauthoryear{{Katarzy{\'n}ski}, {Ghisellini}, {Tavecchio},
  {Gracia}  \& {Maraschi}}{{Katarzy{\'n}ski} et~al.}{2006}]{Katarzynski06}
{Katarzy{\'n}ski} K.,  {Ghisellini} G.,  {Tavecchio} F.,  {Gracia} J.,
  {Maraschi} L.,  2006, \mn@doi [MNRAS] {10.1111/j.1745-3933.2006.00156.x},
  \href {http://adsabs.harvard.edu/abs/2006MNRAS.368L..52K} {368, L52}

\bibitem[\protect\citeauthoryear{{Kaufmann}, {Wagner}, {Tibolla}  \&
  {Hauser}}{{Kaufmann} et~al.}{2011}]{Kaufmann11}
{Kaufmann} S.,  {Wagner} S.~J.,  {Tibolla} O.,   {Hauser} M.,  2011, \mn@doi
  [A\&A] {10.1051/0004-6361/201117215}, \href
  {http://adsabs.harvard.edu/abs/2011A%26A...534A.130K} {534, A130}

\bibitem[\protect\citeauthoryear{{Kelner} \& {Aharonian}}{{Kelner} \&
  {Aharonian}}{2008}]{Kelner}
{Kelner} S.~R.,  {Aharonian} F.~A.,  2008, \mn@doi [PhRD]
  {10.1103/PhysRevD.78.034013}, \href
  {http://adsabs.harvard.edu/abs/2008PhRvD..78c4013K} {78, 034013}

\bibitem[\protect\citeauthoryear{{Kolmogorov}}{{Kolmogorov}}{1991}]{Kolmo}
{Kolmogorov} A.~N.,  1991, \mn@doi [Royal Society of London Proceedings Series
  A] {10.1098/rspa.1991.0075}, \href
  {http://adsabs.harvard.edu/abs/1991RSPSA.434....9K} {434, 9}

\bibitem[\protect\citeauthoryear{{Konigl}}{{Konigl}}{1981}]{Konigl81}
{Konigl} A.,  1981, \mn@doi [ApJ] {10.1086/158638}, \href
  {http://adsabs.harvard.edu/abs/1981ApJ...243..700K} {243, 700}

\bibitem[\protect\citeauthoryear{{Kotera} \& {Olinto}}{{Kotera} \&
  {Olinto}}{2011}]{Kumiko11}
{Kotera} K.,  {Olinto} A.~V.,  2011, \mn@doi [ARA\&A]
  {10.1146/annurev-astro-081710-102620}, \href
  {http://adsabs.harvard.edu/abs/2011ARA%26A..49..119K} {49, 119}

\bibitem[\protect\citeauthoryear{{Kraichnan} \& {Montgomery}}{{Kraichnan} \&
  {Montgomery}}{1980}]{Kraichnan}
{Kraichnan} R.~H.,  {Montgomery} D.,  1980, \mn@doi [Reports on Progress in
  Physics] {10.1088/0034-4885/43/5/001}, \href
  {http://adsabs.harvard.edu/abs/1980RPPh...43..547K} {43, 547}

\bibitem[\protect\citeauthoryear{{Lefa}, {Rieger}  \& {Aharonian}}{{Lefa}
  et~al.}{2011}]{Lefa11}
{Lefa} E.,  {Rieger} F.~M.,   {Aharonian} F.,  2011, \mn@doi [ApJ]
  {10.1088/0004-637X/740/2/64}, \href
  {http://adsabs.harvard.edu/abs/2011ApJ...740...64L} {740, 64}

\bibitem[\protect\citeauthoryear{{Leonardo} et~al.,}{{Leonardo}
  et~al.}{2009}]{Leonardo09}
{Leonardo} E.,  et~al., 2009, Proc. of the 31st ICRC, astro-ph/0907.0959, \href
  {http://adsabs.harvard.edu/abs/2009arXiv0907.0959L} {}

\bibitem[\protect\citeauthoryear{{Lister} et~al.,}{{Lister}
  et~al.}{2013}]{Lister13}
{Lister} M.~L.,  et~al., 2013, \mn@doi [AJ] {10.1088/0004-6256/146/5/120},
  \href {http://adsabs.harvard.edu/abs/2013AJ....146..120L} {146, 120}

\bibitem[\protect\citeauthoryear{{Mannheim}}{{Mannheim}}{1993}]{Mannheim93}
{Mannheim} K.,  1993, A\&A, \href
  {http://adsabs.harvard.edu/abs/1993A%26A...269...67M} {269, 67}

\bibitem[\protect\citeauthoryear{{Mastichiadis}, {Petropoulou}  \&
  {Dimitrakoudis}}{{Mastichiadis} et~al.}{2013}]{Mastichiadis13}
{Mastichiadis} A.,  {Petropoulou} M.,   {Dimitrakoudis} S.,  2013, \mn@doi
  [MNRAS] {10.1093/mnras/stt1210}, \href
  {http://adsabs.harvard.edu/abs/2013MNRAS.434.2684M} {434, 2684}

\bibitem[\protect\citeauthoryear{{Meyer}, {Raue}, {Mazin}  \& {Horns}}{{Meyer}
  et~al.}{2012a}]{Meyer12}
{Meyer} M.,  {Raue} M.,  {Mazin} D.,   {Horns} D.,  2012a, \mn@doi [A\&A]
  {10.1051/0004-6361/201118284}, \href
  {http://adsabs.harvard.edu/abs/2012A%26A...542A..59M} {542, A59}

\bibitem[\protect\citeauthoryear{{Meyer}, {Fossati}, {Georganopoulos}  \&
  {Lister}}{{Meyer} et~al.}{2012b}]{EyleenMeyer12}
{Meyer} E.~T.,  {Fossati} G.,  {Georganopoulos} M.,   {Lister} M.~L.,  2012b,
  \mn@doi [ApJL] {10.1088/2041-8205/752/1/L4}, \href
  {http://adsabs.harvard.edu/abs/2012ApJ...752L...4M} {752, L4}

\bibitem[\protect\citeauthoryear{{Moderski}, {Sikora}  \&
  {B{\l}a{\.z}ejowski}}{{Moderski} et~al.}{2003}]{Moderski03}
{Moderski} R.,  {Sikora} M.,   {B{\l}a{\.z}ejowski} M.,  2003, \mn@doi [A\&A]
  {10.1051/0004-6361:20030794}, \href
  {http://adsabs.harvard.edu/abs/2003A%26A...406..855M} {406, 855}

\bibitem[\protect\citeauthoryear{{Moore} \& {Stockman}}{{Moore} \&
  {Stockman}}{1981}]{Moore81}
{Moore} R.~L.,  {Stockman} H.~S.,  1981, \mn@doi [ApJ] {10.1086/158567}, \href
  {http://adsabs.harvard.edu/abs/1981ApJ...243...60M} {243, 60}

\bibitem[\protect\citeauthoryear{{M{\"u}cke} \& {Protheroe}}{{M{\"u}cke} \&
  {Protheroe}}{2001}]{Mucke01}
{M{\"u}cke} A.,  {Protheroe} R.~J.,  2001, \mn@doi [Astroparticle Physics]
  {10.1016/S0927-6505(00)00141-9}, \href
  {http://adsabs.harvard.edu/abs/2001APh....15..121M} {15, 121}

\bibitem[\protect\citeauthoryear{{M{\"u}cke}, {Engel}, {Rachen}, {Protheroe}
  \& {Stanev}}{{M{\"u}cke} et~al.}{2000}]{Sophia}
{M{\"u}cke} A.,  {Engel} R.,  {Rachen} J.~P.,  {Protheroe} R.~J.,   {Stanev}
  T.,  2000, \mn@doi [Computer Physics Communications]
  {10.1016/S0010-4655(99)00446-4}, \href
  {http://adsabs.harvard.edu/abs/2000CoPhC.124..290M} {124, 290}

\bibitem[\protect\citeauthoryear{{Murase}, {Dermer}, {Takami}  \&
  {Migliori}}{{Murase} et~al.}{2012}]{Murase12}
{Murase} K.,  {Dermer} C.~D.,  {Takami} H.,   {Migliori} G.,  2012, \mn@doi
  [ApJ] {10.1088/0004-637X/749/1/63}, \href
  {http://adsabs.harvard.edu/abs/2012ApJ...749...63M} {749, 63}

\bibitem[\protect\citeauthoryear{{Nolan} et~al.,}{{Nolan} et~al.}{2012}]{2FGL}
{Nolan} P.~L.,  et~al., 2012, \mn@doi [ApJS] {10.1088/0067-0049/199/2/31},
  \href {http://adsabs.harvard.edu/abs/2012ApJS..199...31N} {199, 31}

\bibitem[\protect\citeauthoryear{{Padovani} \& {Giommi}}{{Padovani} \&
  {Giommi}}{1995}]{Padovani95}
{Padovani} P.,  {Giommi} P.,  1995, \mn@doi [ApJ] {10.1086/175631}, \href
  {http://adsabs.harvard.edu/abs/1995ApJ...444..567P} {444, 567}

\bibitem[\protect\citeauthoryear{{Petry} et~al.,}{{Petry}
  et~al.}{2002}]{1426_whipple}
{Petry} D.,  et~al., 2002, \mn@doi [ApJ] {10.1086/343102}, \href
  {http://adsabs.harvard.edu/abs/2002ApJ...580..104P} {580, 104}

\bibitem[\protect\citeauthoryear{{Potter} \& {Cotter}}{{Potter} \&
  {Cotter}}{2012}]{Potter12}
{Potter} W.~J.,  {Cotter} G.,  2012, \mn@doi [MNRAS]
  {10.1111/j.1365-2966.2012.20918.x}, \href
  {http://adsabs.harvard.edu/abs/2012MNRAS.423..756P} {423, 756}

\bibitem[\protect\citeauthoryear{{Protheroe} \& {Clay}}{{Protheroe} \&
  {Clay}}{2004}]{Protheroe04}
{Protheroe} R.~J.,  {Clay} R.~W.,  2004, \mn@doi [PASA] {10.1071/AS03047},
  \href {http://adsabs.harvard.edu/abs/2004PASA...21....1P} {21, 1}

\bibitem[\protect\citeauthoryear{{Protheroe} \& {Johnson}}{{Protheroe} \&
  {Johnson}}{1996}]{Protheroe96}
{Protheroe} R.~J.,  {Johnson} P.~A.,  1996, \mn@doi [Astroparticle Physics]
  {10.1016/0927-6505(95)00039-9}, \href
  {http://adsabs.harvard.edu/abs/1996APh.....4..253P} {4, 253}

\bibitem[\protect\citeauthoryear{{Punch} et~al.,}{{Punch}
  et~al.}{1992}]{Punch92}
{Punch} M.,  et~al., 1992, \mn@doi [Nature] {10.1038/358477a0}, \href
  {http://adsabs.harvard.edu/abs/1992Natur.358..477P} {358, 477}

\bibitem[\protect\citeauthoryear{{Rachen}}{{Rachen}}{2000}]{Rachen00}
{Rachen} J.~P.,  2000, in {Dingus} B.~L.,  {Salamon} M.~H.,   {Kieda} D.~B.,
  eds,  American Institute of Physics Conference Series Vol. 515, American
  Institute of Physics Conference Series. pp 41--52 (\mn@eprint {}
  {astro-ph/0003282}), \mn@doi{10.1063/1.1291342}

\bibitem[\protect\citeauthoryear{{Remillard}, {Tuohy}, {Brissenden}, {Buckley},
  {Schwartz}, {Feigelson}  \& {Tapia}}{{Remillard} et~al.}{1989}]{Remillard89}
{Remillard} R.~A.,  {Tuohy} I.~R.,  {Brissenden} R.~J.~V.,  {Buckley} D.~A.~H.,
   {Schwartz} D.~A.,  {Feigelson} E.~D.,   {Tapia} S.,  1989, \mn@doi [ApJ]
  {10.1086/167888}, \href {http://adsabs.harvard.edu/abs/1989ApJ...345..140R}
  {345, 140}

\bibitem[\protect\citeauthoryear{{Reynoso}, {Medina}  \& {Romero}}{{Reynoso}
  et~al.}{2011}]{Reynoso11}
{Reynoso} M.~M.,  {Medina} M.~C.,   {Romero} G.~E.,  2011, \mn@doi [A\&A]
  {10.1051/0004-6361/201014998}, \href
  {http://adsabs.harvard.edu/abs/2011A%26A...531A..30R} {531, A30}

\bibitem[\protect\citeauthoryear{{Rieger}, {Bosch-Ramon}  \& {Duffy}}{{Rieger}
  et~al.}{2007}]{Rieger07}
{Rieger} F.~M.,  {Bosch-Ramon} V.,   {Duffy} P.,  2007, \mn@doi [ApSS]
  {10.1007/s10509-007-9466-z}, \href
  {http://adsabs.harvard.edu/abs/2007Ap%26SS.309..119R} {309, 119}

\bibitem[\protect\citeauthoryear{{Romero}, {Torres}, {Kaufman Bernad{\'o}}  \&
  {Mirabel}}{{Romero} et~al.}{2003}]{Romero03}
{Romero} G.~E.,  {Torres} D.~F.,  {Kaufman Bernad{\'o}} M.~M.,   {Mirabel}
  I.~F.,  2003, \mn@doi [A\&A] {10.1051/0004-6361:20031314-1}, \href
  {http://adsabs.harvard.edu/abs/2003A%26A...410L...1R} {410, L1}

\bibitem[\protect\citeauthoryear{{R{\"u}ger}, {Spanier}  \&
  {Mannheim}}{{R{\"u}ger} et~al.}{2010}]{Rueger10}
{R{\"u}ger} M.,  {Spanier} F.,   {Mannheim} K.,  2010, \mn@doi [MNRAS]
  {10.1111/j.1365-2966.2009.15738.x}, \href
  {http://adsabs.harvard.edu/abs/2010MNRAS.401..973R} {401, 973}

\bibitem[\protect\citeauthoryear{{Rybicki} \& {Lightman}}{{Rybicki} \&
  {Lightman}}{1979}]{Rybicki}
{Rybicki} G.~B.,  {Lightman} A.~P.,  1979, {Radiative processes in
  astrophysics, Ed. Wiley-VCH}

\bibitem[\protect\citeauthoryear{{Salamon} \& {Stecker}}{{Salamon} \&
  {Stecker}}{1998}]{Salamon98}
{Salamon} M.~H.,  {Stecker} F.~W.,  1998, \mn@doi [ApJ] {10.1086/305134}, \href
  {http://adsabs.harvard.edu/abs/1998ApJ...493..547S} {493, 547}

\bibitem[\protect\citeauthoryear{{Sikora}, {Begelman}  \& {Rees}}{{Sikora}
  et~al.}{1994}]{Sikora94}
{Sikora} M.,  {Begelman} M.~C.,   {Rees} M.~J.,  1994, \mn@doi [ApJ]
  {10.1086/173633}, \href {http://adsabs.harvard.edu/abs/1994ApJ...421..153S}
  {421, 153}

\bibitem[\protect\citeauthoryear{{Sikora}, {Stawarz}, {Moderski}, {Nalewajko}
  \& {Madejski}}{{Sikora} et~al.}{2009}]{Sikora}
{Sikora} M.,  {Stawarz} {\L}.,  {Moderski} R.,  {Nalewajko} K.,   {Madejski}
  G.~M.,  2009, \mn@doi [ApJ] {10.1088/0004-637X/704/1/38}, \href
  {http://adsabs.harvard.edu/abs/2009ApJ...704...38S} {704, 38}

\bibitem[\protect\citeauthoryear{{Sironi} \& {Spitkovsky}}{{Sironi} \&
  {Spitkovsky}}{2014}]{Lorenzo14}
{Sironi} L.,  {Spitkovsky} A.,  2014, \mn@doi [ApJL]
  {10.1088/2041-8205/783/1/L21}, \href
  {http://adsabs.harvard.edu/abs/2014ApJ...783L..21S} {783, L21}

\bibitem[\protect\citeauthoryear{{Sol} et~al.,}{{Sol} et~al.}{2013}]{Sol}
{Sol} H.,  et~al., 2013, \mn@doi [Astroparticle Physics]
  {10.1016/j.astropartphys.2012.12.005}, \href
  {http://adsabs.harvard.edu/abs/2013APh....43..215S} {43, 215}

\bibitem[\protect\citeauthoryear{{Stein}, {Odell}  \& {Strittmatter}}{{Stein}
  et~al.}{1976}]{Stein76}
{Stein} W.~A.,  {Odell} S.~L.,   {Strittmatter} P.~A.,  1976, \mn@doi [ARA\&A]
  {10.1146/annurev.aa.14.090176.001133}, \href
  {http://adsabs.harvard.edu/abs/1976ARA%26A..14..173S} {14, 173}

\bibitem[\protect\citeauthoryear{{Tanaka} et~al.,}{{Tanaka}
  et~al.}{2014}]{Tanaka14}
{Tanaka} Y.~T.,  et~al., 2014, astro-ph/1404.3727, \href
  {http://adsabs.harvard.edu/abs/2014arXiv1404.3727T} {}

\bibitem[\protect\citeauthoryear{{Tavecchio}}{{Tavecchio}}{2014}]{Tavecchio14}
{Tavecchio} F.,  2014, \mn@doi [MNRAS] {10.1093/mnras/stt2437}, \href
  {http://adsabs.harvard.edu/abs/2014MNRAS.438.3255T} {438, 3255}

\bibitem[\protect\citeauthoryear{{Tavecchio}, {Maraschi}  \&
  {Ghisellini}}{{Tavecchio} et~al.}{1998}]{Tavecchio98}
{Tavecchio} F.,  {Maraschi} L.,   {Ghisellini} G.,  1998, \mn@doi [ApJ]
  {10.1086/306526}, \href {http://adsabs.harvard.edu/abs/1998ApJ...509..608T}
  {509, 608}

\bibitem[\protect\citeauthoryear{{Tavecchio}, {Ghisellini}, {Ghirlanda},
  {Costamante}  \& {Franceschini}}{{Tavecchio} et~al.}{2009}]{Tavecchio09}
{Tavecchio} F.,  {Ghisellini} G.,  {Ghirlanda} G.,  {Costamante} L.,
  {Franceschini} A.,  2009, \mn@doi [MNRAS] {10.1111/j.1745-3933.2009.00724.x},
  \href {http://adsabs.harvard.edu/abs/2009MNRAS.399L..59T} {399, L59}

\bibitem[\protect\citeauthoryear{{Tavecchio}, {Ghisellini}, {Ghirlanda},
  {Foschini}  \& {Maraschi}}{{Tavecchio} et~al.}{2010}]{Tavecchio10}
{Tavecchio} F.,  {Ghisellini} G.,  {Ghirlanda} G.,  {Foschini} L.,   {Maraschi}
  L.,  2010, \mn@doi [MNRAS] {10.1111/j.1365-2966.2009.15784.x}, \href
  {http://adsabs.harvard.edu/abs/2010MNRAS.401.1570T} {401, 1570}

\bibitem[\protect\citeauthoryear{{Urry} \& {Padovani}}{{Urry} \&
  {Padovani}}{1995}]{Urry95}
{Urry} C.~M.,  {Padovani} P.,  1995, \mn@doi [PASP] {10.1086/133630}, \href
  {http://adsabs.harvard.edu/abs/1995PASP..107..803U} {107, 803}

\bibitem[\protect\citeauthoryear{{Vainio}, {Virtanen}  \&
  {Schlickeiser}}{{Vainio} et~al.}{2003}]{Vainio03}
{Vainio} R.,  {Virtanen} J.~J.~P.,   {Schlickeiser} R.,  2003, \mn@doi [A\&A]
  {10.1051/0004-6361:20034038}, \href
  {http://adsabs.harvard.edu/abs/2003A%26A...409..821V} {409, 821}

\bibitem[\protect\citeauthoryear{{Virtanen} \& {Vainio}}{{Virtanen} \&
  {Vainio}}{2005}]{Virtanen05}
{Virtanen} J.~J.~P.,  {Vainio} R.,  2005, \mn@doi [ApJ] {10.1086/427324}, \href
  {http://adsabs.harvard.edu/abs/2005ApJ...621..313V} {621, 313}

\bibitem[\protect\citeauthoryear{{Vovk}, {Taylor}, {Semikoz}  \&
  {Neronov}}{{Vovk} et~al.}{2012}]{Vovk12}
{Vovk} I.,  {Taylor} A.~M.,  {Semikoz} D.,   {Neronov} A.,  2012, \mn@doi
  [ApJL] {10.1088/2041-8205/747/1/L14}, \href
  {http://adsabs.harvard.edu/abs/2012ApJ...747L..14V} {747, L14}

\bibitem[\protect\citeauthoryear{{Weidinger} \& {Spanier}}{{Weidinger} \&
  {Spanier}}{2010}]{Weidinger10}
{Weidinger} M.,  {Spanier} F.,  2010, \mn@doi [A\&A]
  {10.1051/0004-6361/201014299}, \href
  {http://adsabs.harvard.edu/abs/2010A%26A...515A..18W} {515, A18}

\bibitem[\protect\citeauthoryear{{Weidinger} \& {Spanier}}{{Weidinger} \&
  {Spanier}}{2013}]{Weidinger13}
{Weidinger} M.,  {Spanier} F.,  2013, in European Physical Journal Web of
  Conferences. p.~5009, \mn@doi{10.1051/epjconf/20136105009}

\bibitem[\protect\citeauthoryear{{Woo}, {Urry}, {van der Marel}, {Lira}  \&
  {Maza}}{{Woo} et~al.}{2005}]{Woo05}
{Woo} J.-H.,  {Urry} C.~M.,  {van der Marel} R.~P.,  {Lira} P.,   {Maza} J.,
  2005, \mn@doi [ApJ] {10.1086/432681}, \href
  {http://adsabs.harvard.edu/abs/2005ApJ...631..762W} {631, 762}

\bibitem[\protect\citeauthoryear{{Zech} \& {Cerruti}}{{Zech} \&
  {Cerruti}}{2013}]{ICRC2013}
{Zech} A.,  {Cerruti} M.,  2013, Proc. of the 33rd ICRC, astro-ph/1307.3038,
  \href {http://adsabs.harvard.edu/abs/2013arXiv1307.3038Z} {}

\bibitem[\protect\citeauthoryear{{{\c S}ent{\"u}rk}, {Errando}, {B{\"o}ttcher}
  \& {Mukherjee}}{{{\c S}ent{\"u}rk} et~al.}{2013}]{Gunes13}
{{\c S}ent{\"u}rk} G.~D.,  {Errando} M.,  {B{\"o}ttcher} M.,   {Mukherjee} R.,
  2013, \mn@doi [ApJ] {10.1088/0004-637X/764/2/119}, \href
  {http://adsabs.harvard.edu/abs/2013ApJ...764..119S} {764, 119}

\makeatother
\end{thebibliography}
 
\appendix
 \section{Figures}
 \label{appa}

          \begin{figure*}
	   \centering
	   	\includegraphics[width=220pt]{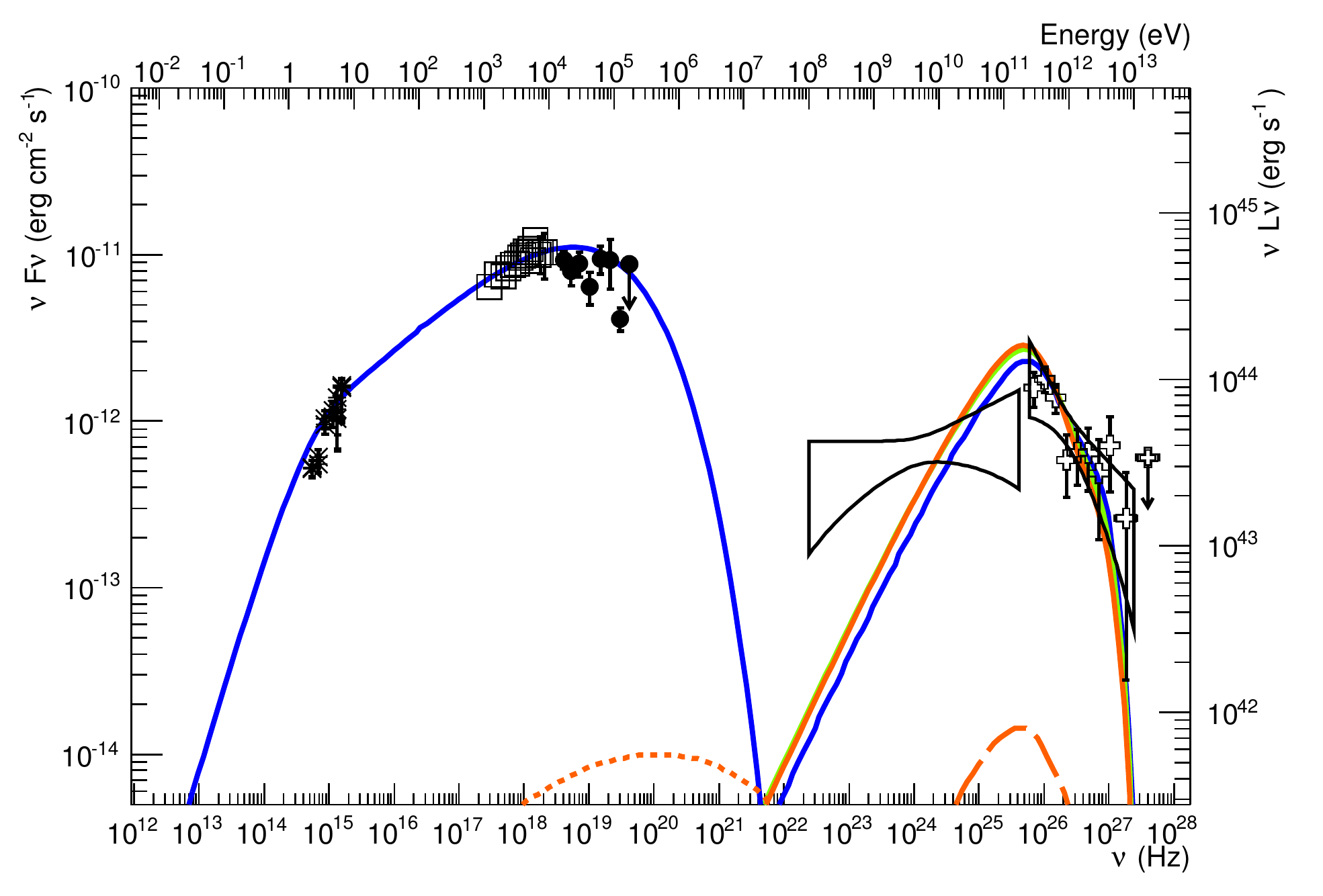}
	   	\includegraphics[width=220pt]{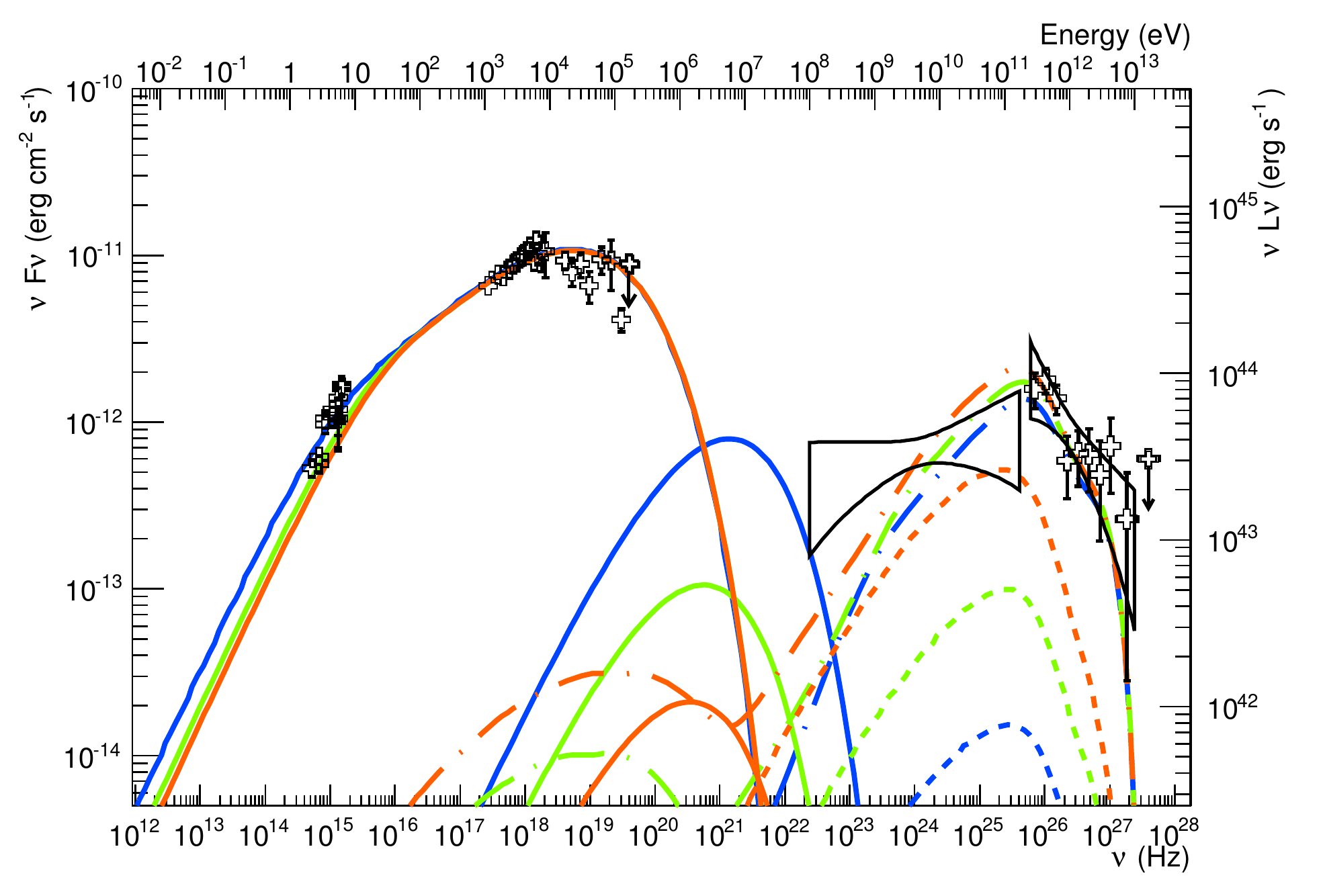}	
		\includegraphics[width=220pt]{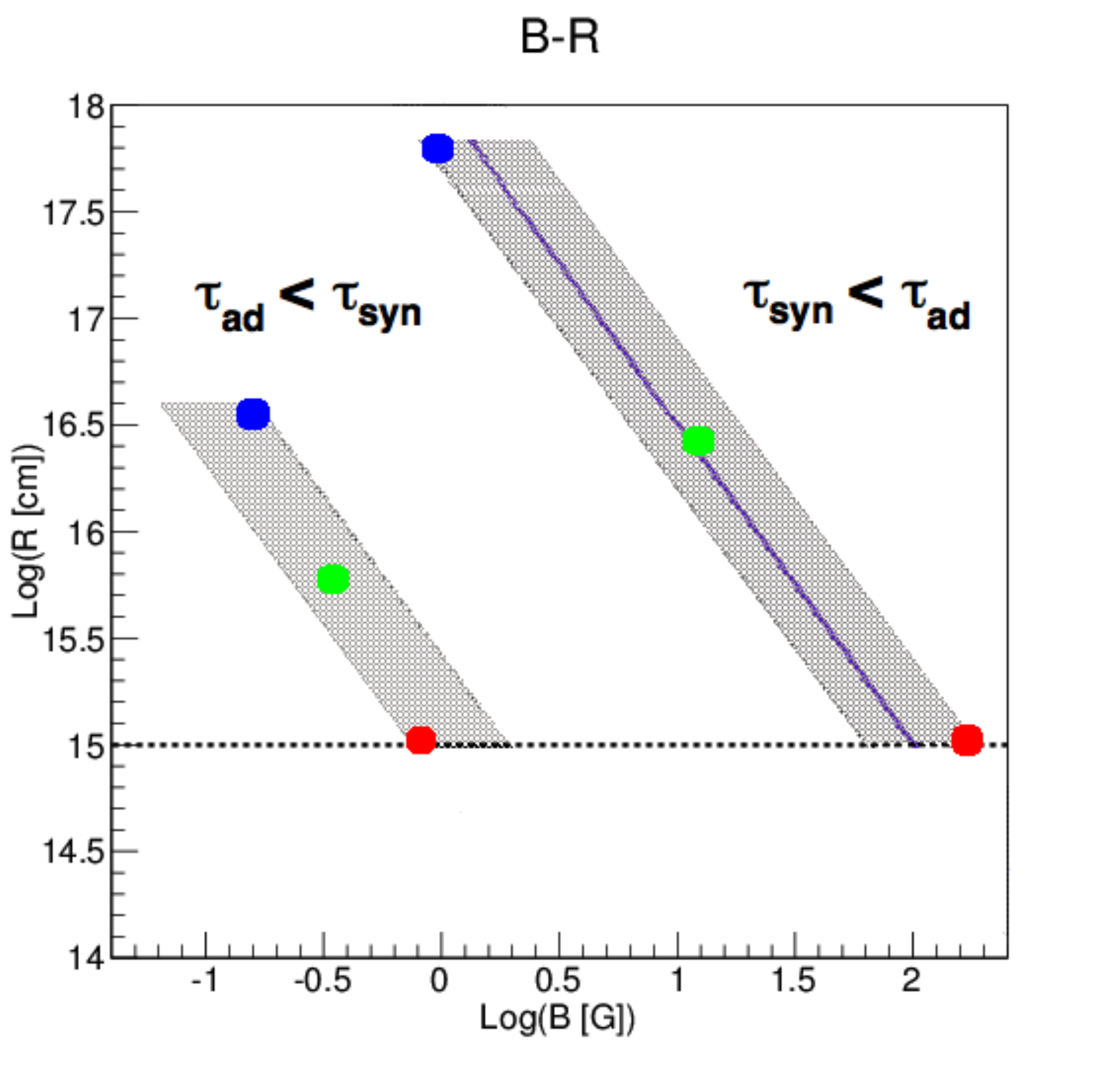}
		\includegraphics[width=220pt]{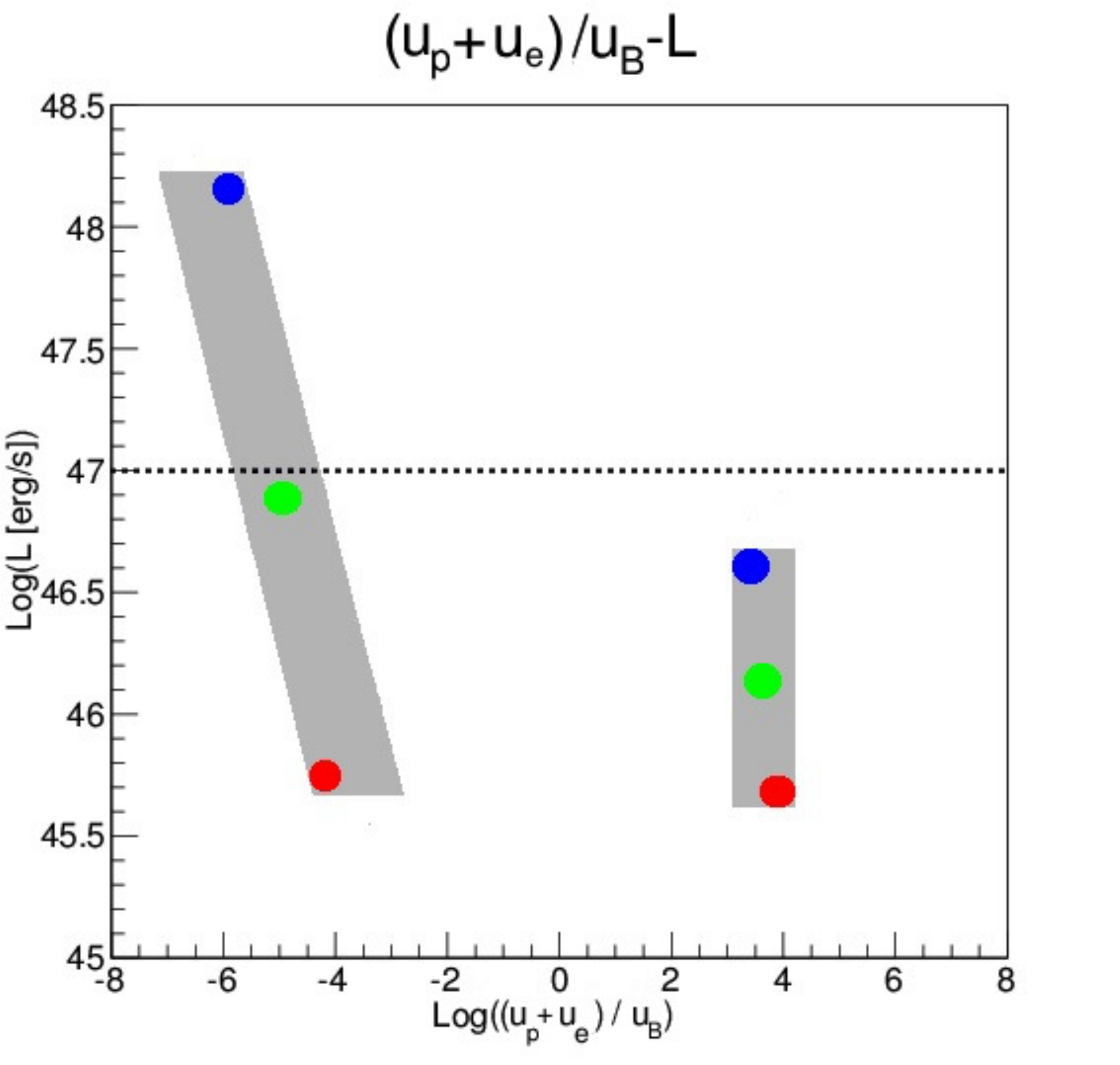} 
	  \caption{Same as Figure \ref{fig2}, for 1ES\, 0229+200, using data from \citet{0229_Veritas}.  The values of the magnetic field and the emitting region size are $(\textrm{B[G],R[cm]})=(1,6.8\times10^{17}),(13,2.6\times10^{16}),(160,1\times10^{15})$, for the proton-synchrotron scenario, and $(\textrm{B[G],R[cm]})=(0.2,3.2\times10^{16}),(0.3,5.6\times10^{15}),(0.8,1\times10^{15})$ for the lepto-hadronic scenario. \label{figa2}}
   \end{figure*}

  \newpage
        \begin{figure*}
	   \centering
		\includegraphics[width=220pt]{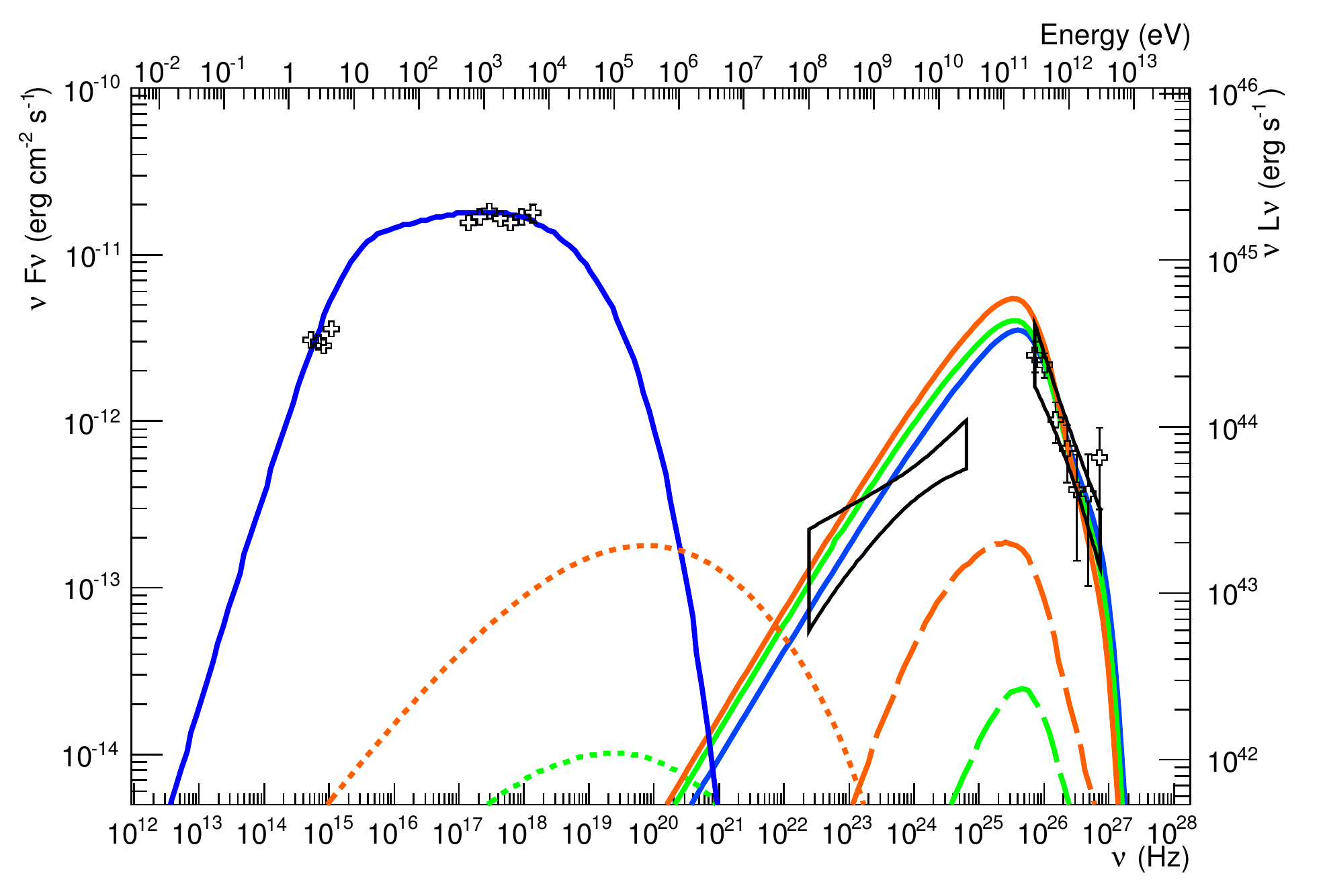}
		\includegraphics[width=220pt]{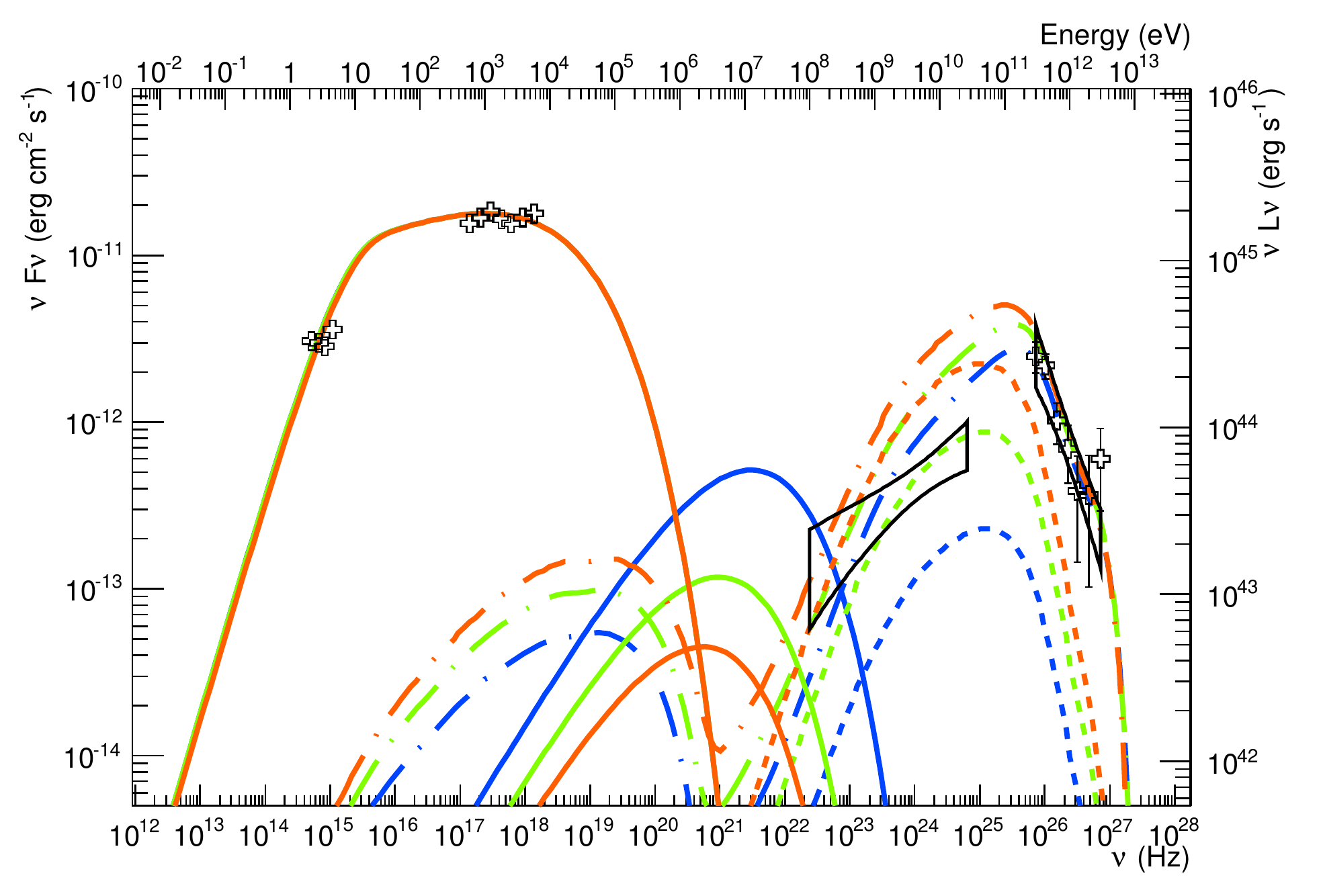}
		\includegraphics[width=220pt]{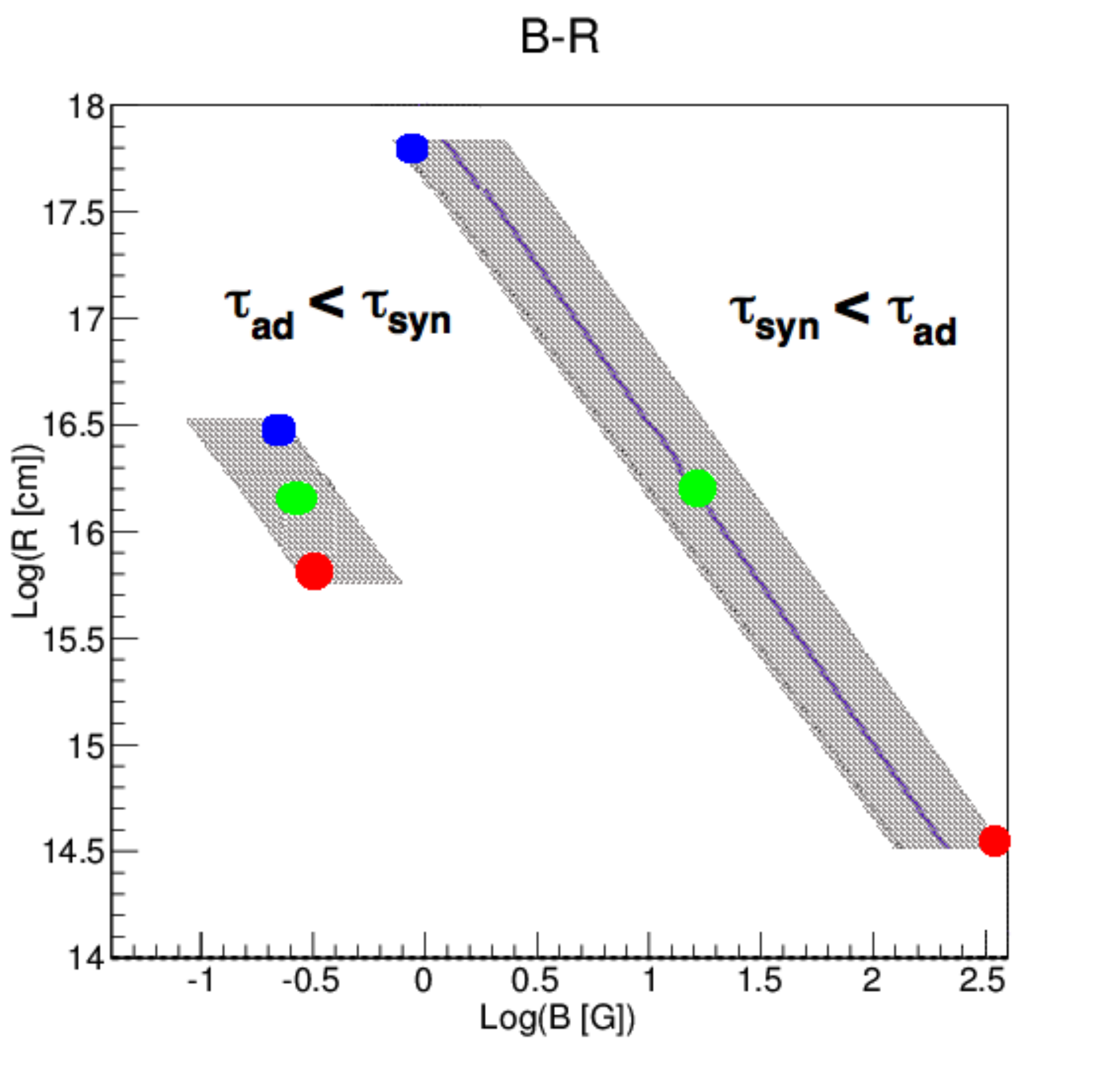}
		\includegraphics[width=220pt]{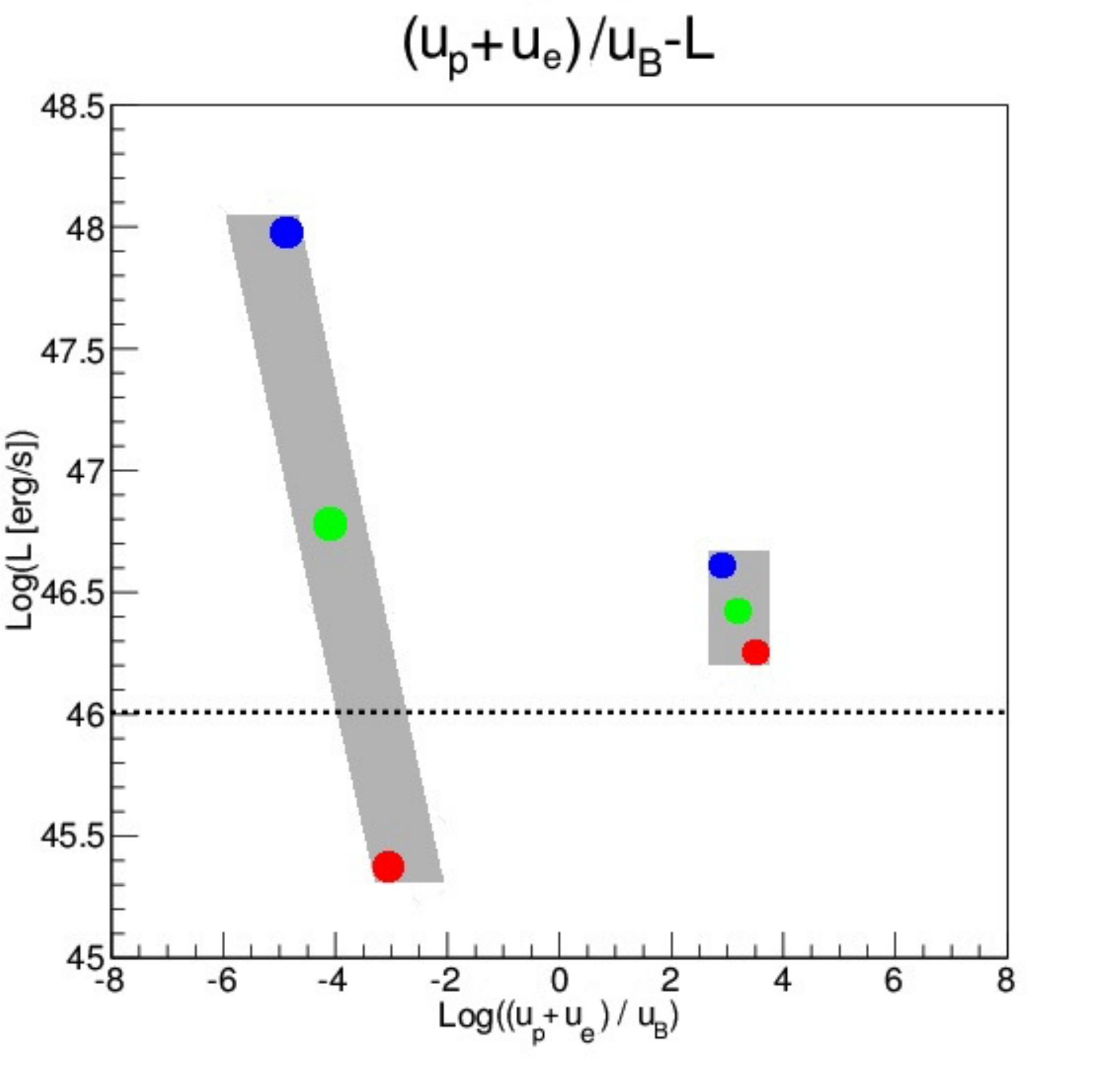}
	  \caption{Same as Figure \ref{fig2}, for 1ES\,0347-121, using data from \citet{0347_HESS}. The values of the magnetic field and the emitting region size are $(\textrm{B[G],R[cm]})=(1,6.5\times10^{17}),(17,1.4\times10^{16}),(296,3\times10^{14})$, for the proton-synchrotron scenario, and $(\textrm{B[G],R[cm]})=(0.25,3.2\times10^{16}),(0.3,1.4\times10^{16}),(0.4,6\times10^{15})$ for the lepto-hadronic scenario. \label{figa1}}
   \end{figure*}

 \newpage 
          \begin{figure*}
	   \centering
		\includegraphics[width=220pt]{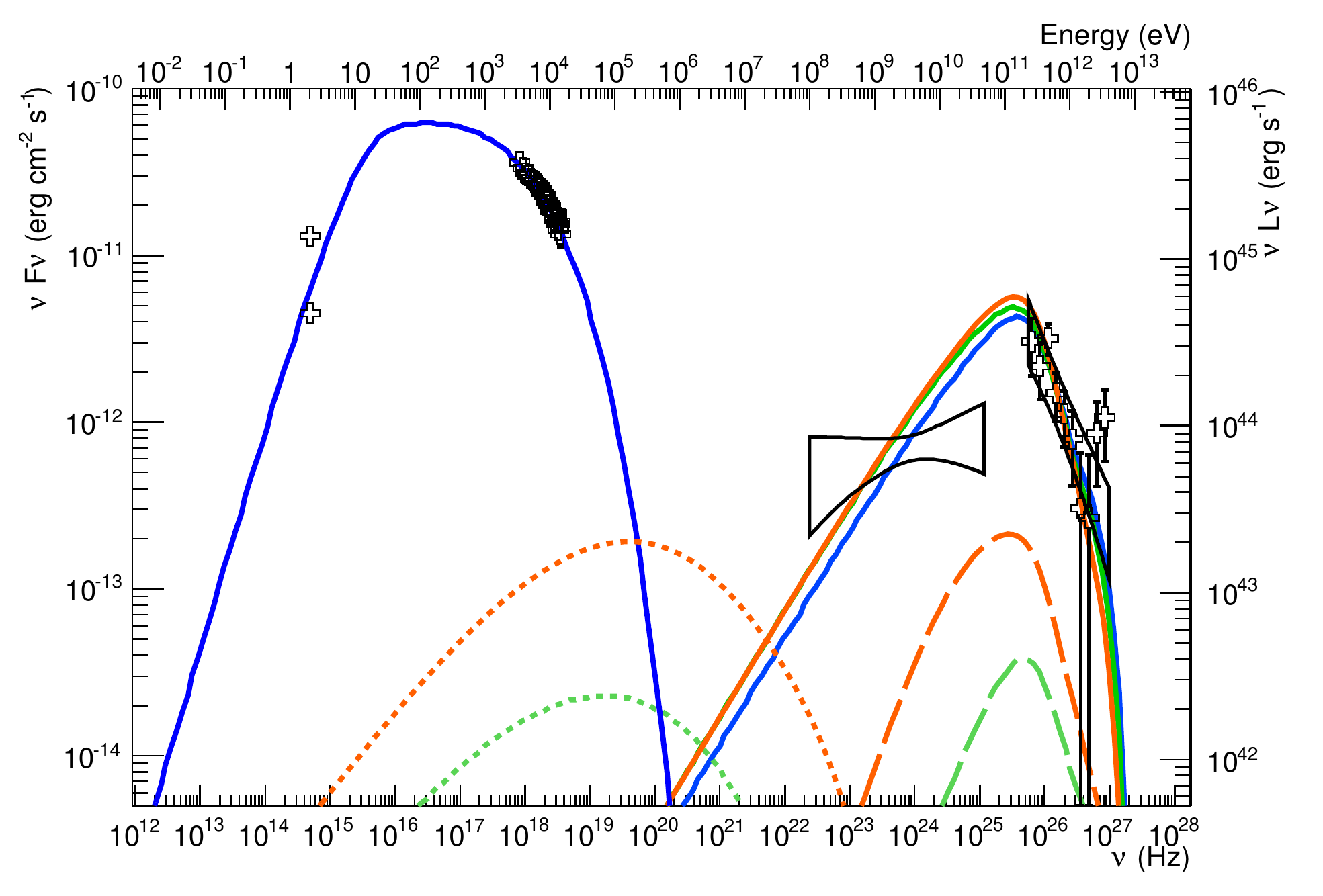}
		\includegraphics[width=220pt]{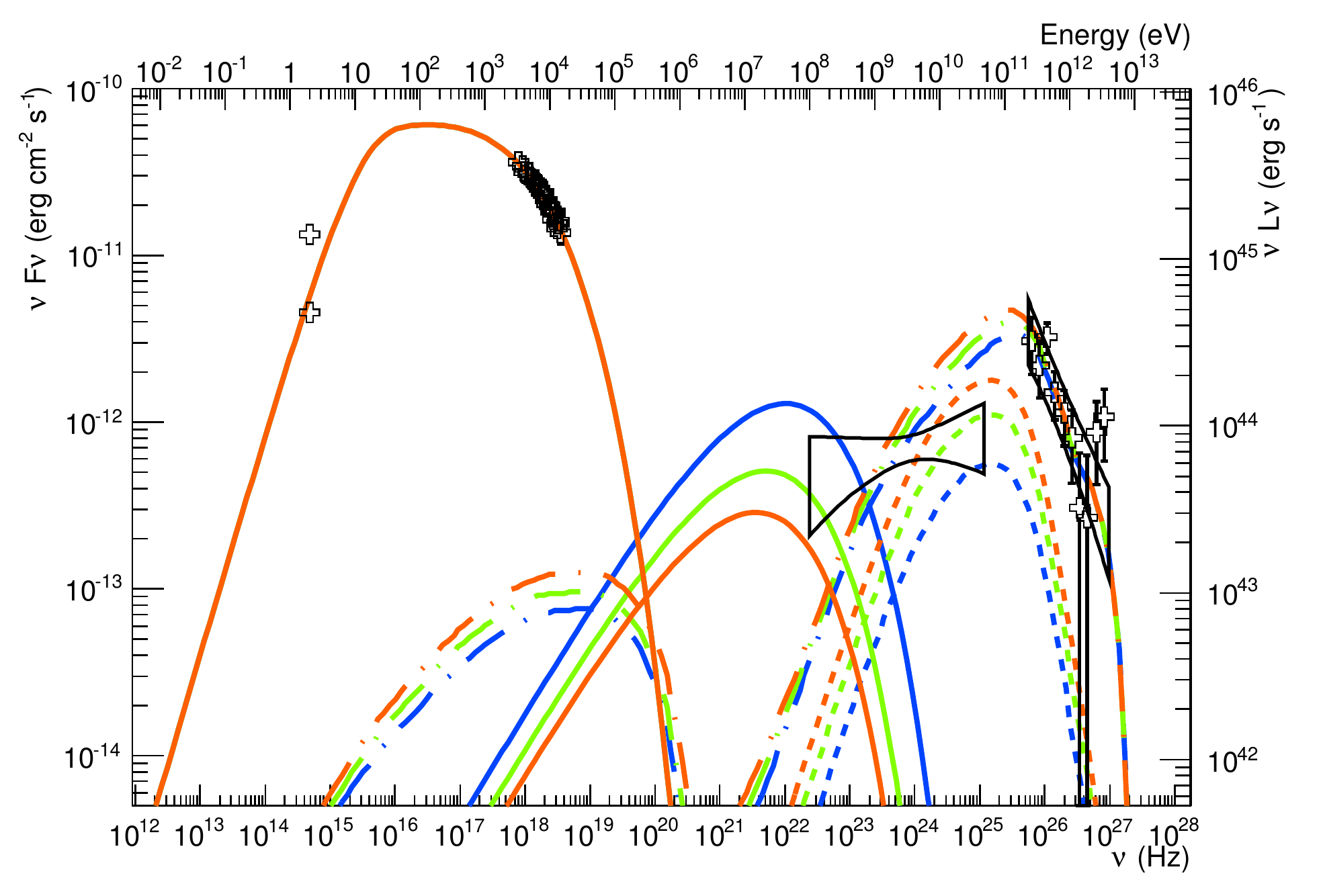}
		\includegraphics[width=220pt]{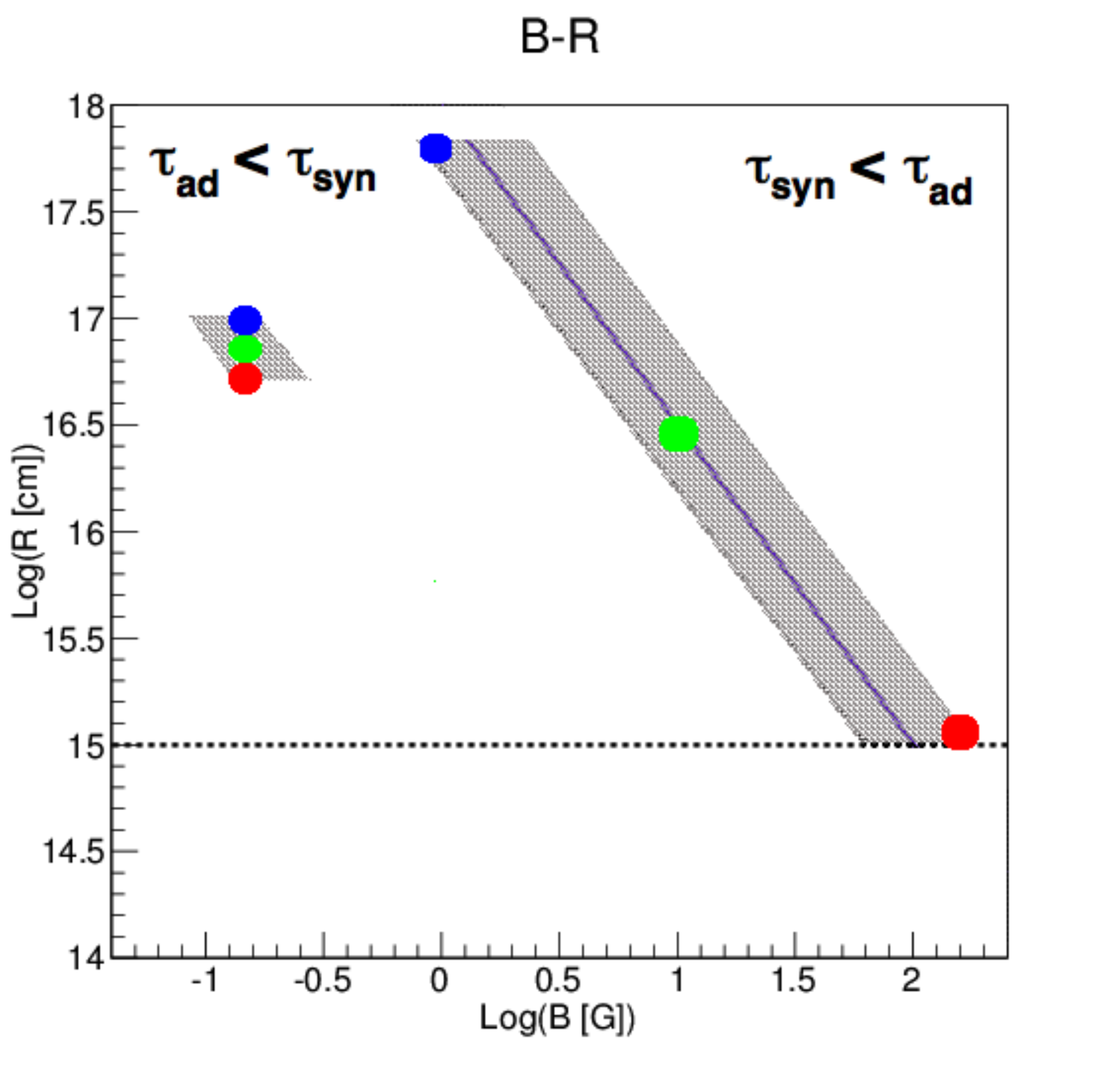}
		\includegraphics[width=220pt]{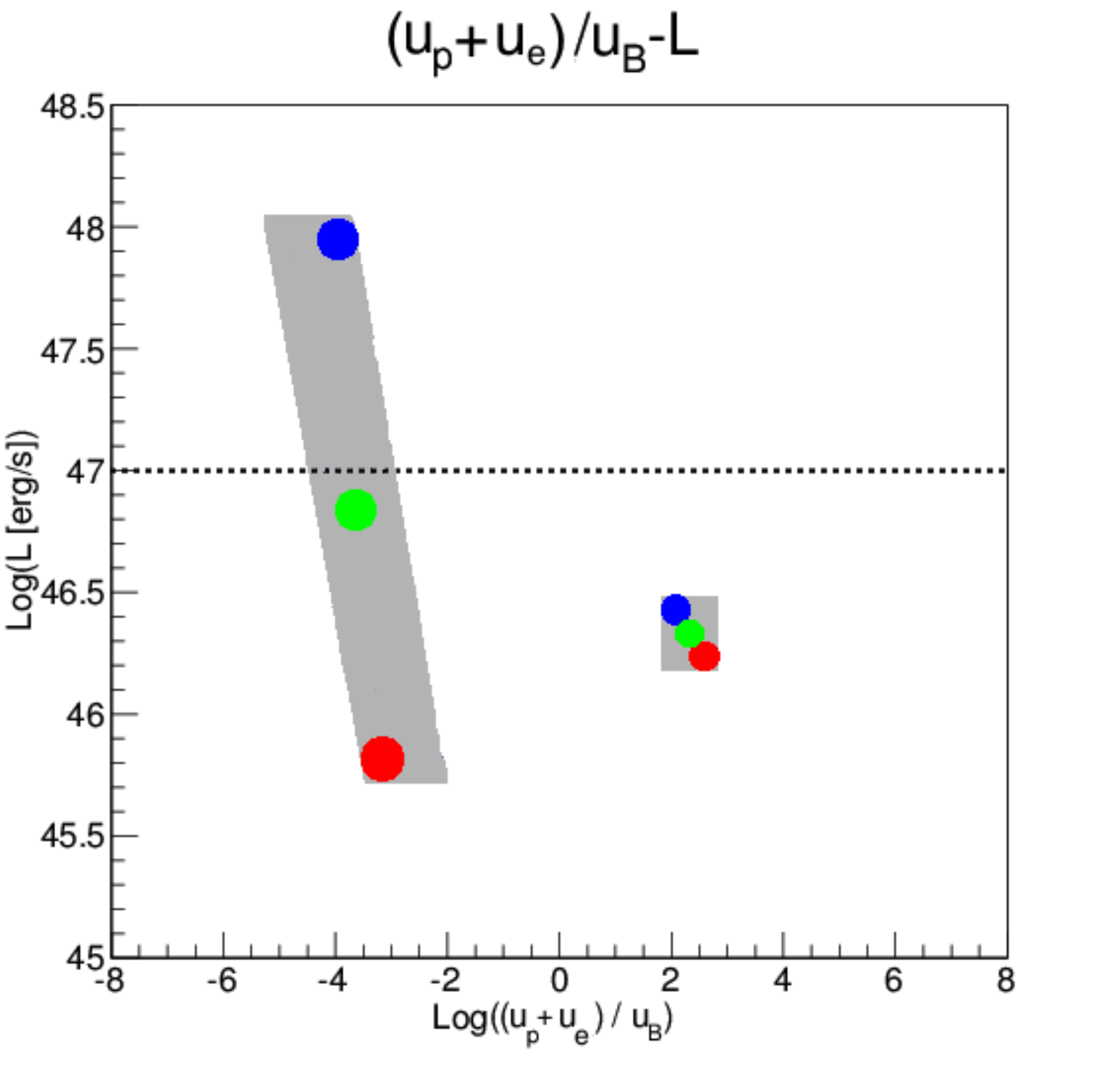}
	  \caption{Same as Figure \ref{fig2}, for 1ES\, 1101-232, using data from \citet{1101_HESS}. The two flux measurements in optical represent an estimation of the minimum and maximum flux from the AGN. The values of the magnetic field and the emitting region size are $(\textrm{B[G],R[cm]})=(1,6.6\times10^{17}),(12,2.6\times10^{16}),(133,1\times10^{15})$, for the proton-synchrotron scenario, and $(\textrm{B[G],R[cm]})=(0.15,1\times10^{17}),(0.15,7\times10^{16}),(0.15,5\times10^{16})$ for the lepto-hadronic scenario. \label{figa3}}
   \end{figure*}
  
  \newpage
          \begin{figure*}
	   \centering
		\includegraphics[width=220pt]{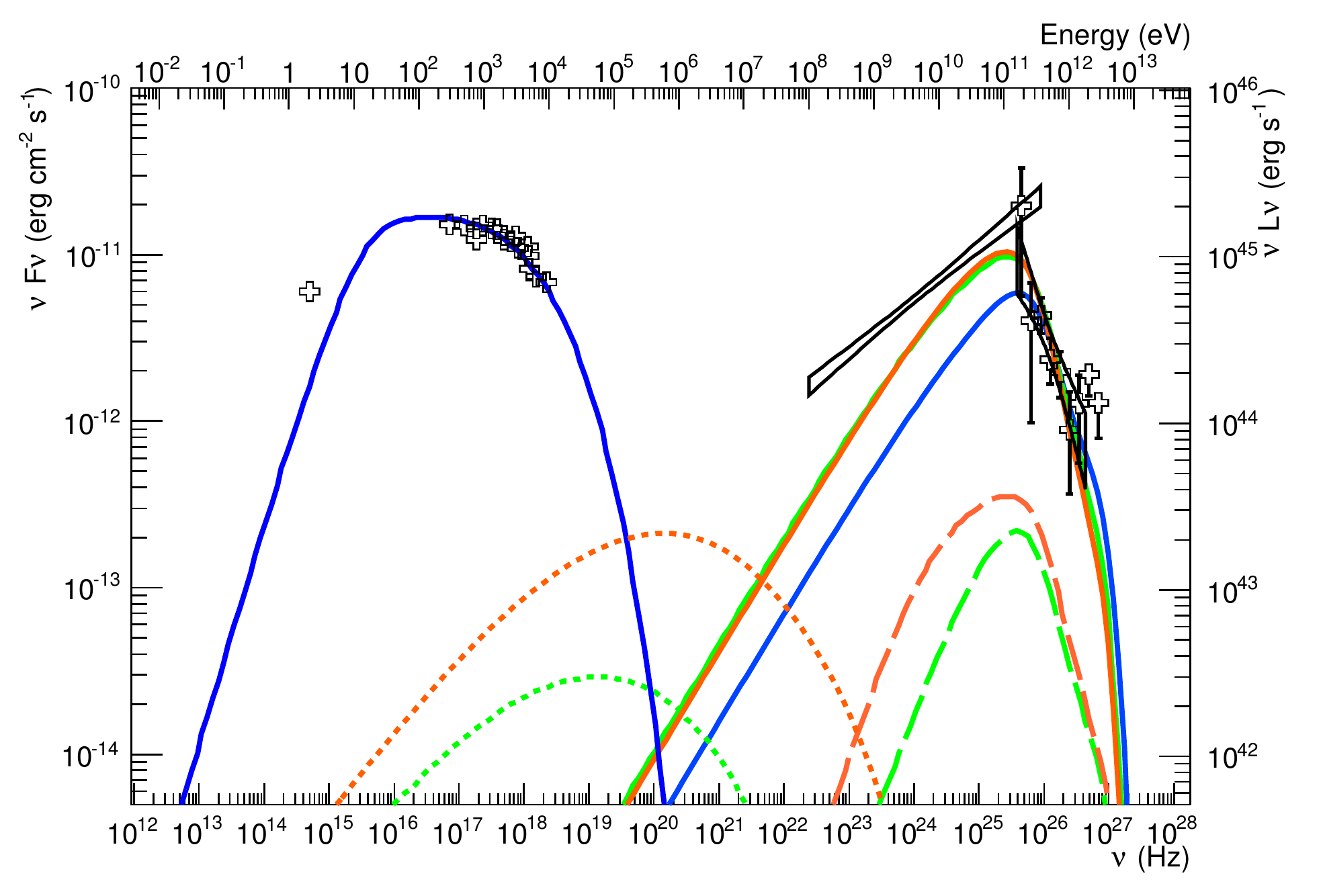}
		\includegraphics[width=220pt]{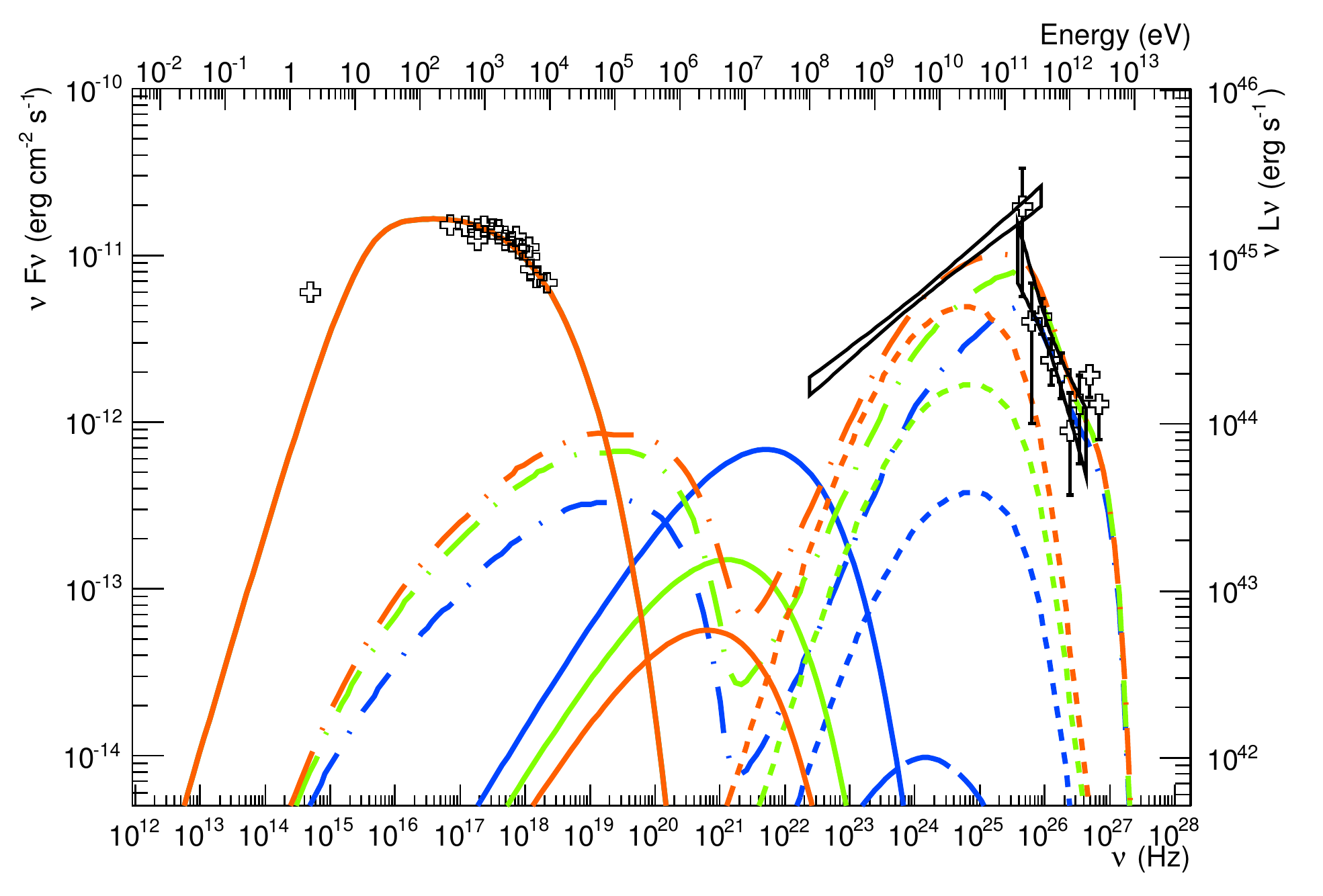}
				\includegraphics[width=220pt]{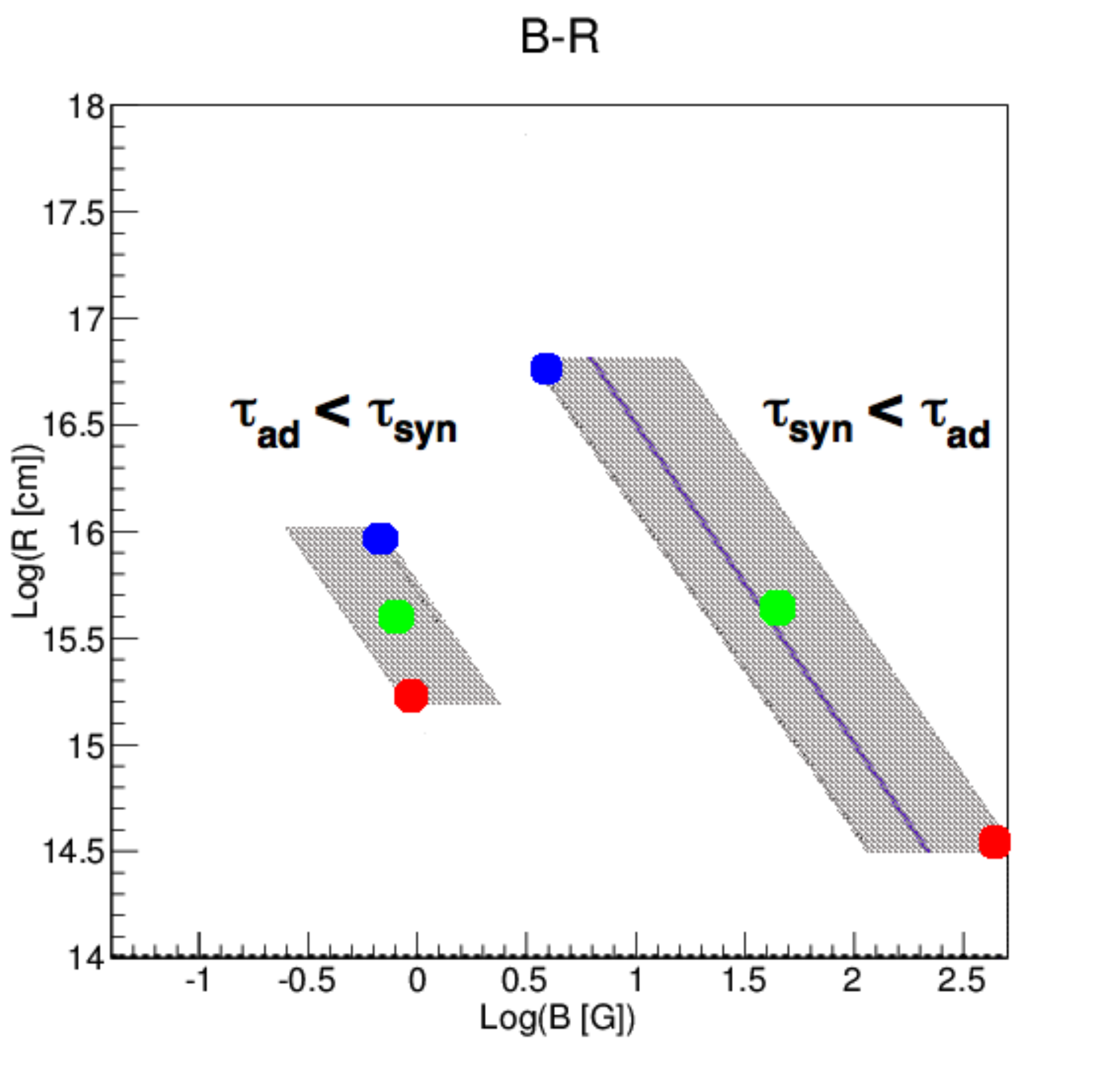}
		\includegraphics[width=220pt]{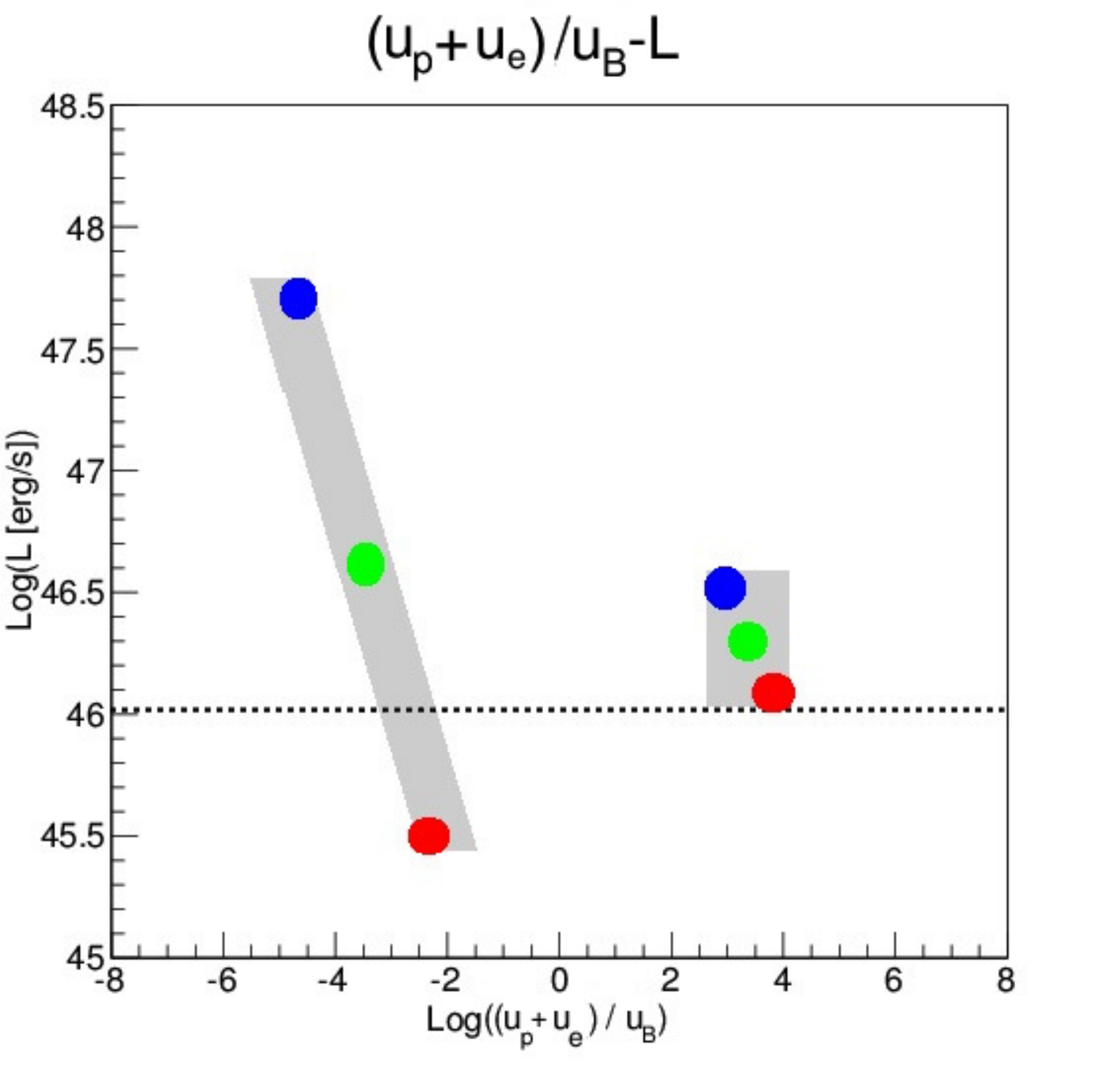}
	  \caption{Same as Figure \ref{fig2}, for 1ES\, 1218+304, using data from \citet{Rueger10}. The optical measurement includes the contamination from the host galaxy, and is considered as an upper limit of the AGN flux. The values of the magnetic field and the emitting region size are $(\textrm{B[G],R[cm]})=(3,6.6\times10^{16}),(39,4.5\times10^{15}),(454,3\times10^{14})$, for the proton-synchrotron scenario, and $(\textrm{B[G],R[cm]})=(0.5,9\times10^{15}),(0.8,4.2\times10^{15}),(1,2\times10^{15})$ for the lepto-hadronic scenario. \label{figa4}}
   \end{figure*}

   \end{document}